\documentclass[12pt,letterpaper]{article}
\usepackage[utf8]{inputenc}
\usepackage{geometry}
\usepackage{graphicx} 
\usepackage{bm}
\usepackage{amsfonts,amssymb}
\usepackage[dvipsnames]{xcolor}
\usepackage{hyperref}
\usepackage[round]{natbib}
\usepackage{bbding}
\usepackage{amsmath}
\usepackage{pifont}
\usepackage{mathtools}
\usepackage{bbm}
\usepackage{setspace}
\usepackage{tabularx}
\usepackage{pdfpages}
\usepackage{siunitx}
\usepackage{booktabs} 
\usepackage{multirow}
\usepackage{adjustbox}
\usepackage{makecell}
\usepackage{threeparttable}
\usepackage{appendix}
\usepackage{rotating}
\usepackage{graphicx}
\usepackage{subcaption}
\usepackage{actuarialsymbol}
\usepackage{float}
\usepackage{eurosym}
\usepackage{indentfirst}
\usepackage{titlesec}
\usepackage[ruled,vlined]{algorithm2e}
\usepackage{algpseudocode}
\usepackage{dcolumn}
\usepackage{xr}
\usepackage{fix-cm}
\usepackage{comment}

\definecolor{darkblue}{rgb}{0.0, 0.0, 0.55}
\hypersetup{colorlinks=true,linkcolor=darkblue,filecolor=darkblue,urlcolor=darkblue,citecolor=darkblue}

\newcommand{\Title}{Assessing the impact of climate change and \\ rising temperatures on life insurance portfolios}

\newcommand{\Authors}{Jiacheng Min\footnote{Centre for Actuarial Studies, Department of Economics, The University of Melbourne, Parkville, VIC 3010, Australia, email: \href{mailto:jiacheng.min@student.unimelb.edu.au}{jiacheng.min@student.unimelb.edu.au}.} \qquad Han Li\footnote{Centre for Actuarial Studies, Department of Economics, The University of Melbourne, Parkville, VIC 3010, Australia, email: \href{mailto:han.li@unimelb.edu.au}{han.li@unimelb.edu.au}.}\qquad Jean-Fran\c cois B\'egin\footnote{Corresponding author. Department of Statistics and Actuarial Science, Simon Fraser University, Burnaby, British Columbia, Canada, V5A 1S6, email: \href{mailto:jbegin@sfu.ca}{jbegin@sfu.ca}.}\qquad Shuanming Li\footnote{Centre for Actuarial Studies, Department of Economics, The University of Melbourne, Parkville, VIC 3010, Australia, email: \href{mailto:shli@unimelb.edu.au}{shli@unimelb.edu.au}.}}

\newcommand{\Abstract}{Climate change may materially affect long-term life insurance liabilities by altering both the level and seasonal pattern of mortality. This article develops a multi-population mortality framework that combines a Hermite spline model with a distributed lag non-linear model to capture age-, region-, and temperature-specific mortality effects. We apply the framework to mortality and temperature data from 15 Spanish NUTS-2 regions and project future mortality under three shared socioeconomic pathway (SSP) scenarios. We then assess the implications for a hypothetical whole life insurance portfolio through expected death-benefit payments and portfolio profit and loss. The results reveal an important seasonal offset: warmer conditions reduce expected payoffs during winter periods but increase them during summer periods, with these effects becoming more pronounced under more severe climate scenarios and for policies issued in later years. Over longer horizons, adverse summer mortality effects become increasingly important. The portfolio analysis further shows that climate-related mortality risk can materially increase the dispersion and downside risk of portfolio outcomes, particularly under SSP5-8.5. These findings highlight the importance of incorporating temperature-related mortality effects into long-term life insurance liability projections and risk assessment.}

\newcommand{\Keywords}{Mortality modeling; life insurance; multi-population; climate risk; Hermite splines; distributed lag non-linear model}

\newcommand{\JEL}{G22; J11; Q54.}

\SetKwInput{KwInput}{Input}               
\SetKwInput{KwOutput}{Output}            
\SetKwInput{KwProcedure}{Procedure}
\definecolor{DarkGreen}{RGB}{1,50,32}

\newcolumntype{d}[1]{D{.}{.}{#1}}
\newcolumntype{Y}{>{\raggedleft\arraybackslash}X}
\newcolumntype{C}{>{\centering\arraybackslash}X}

\numberwithin{equation}{section}

\title{\Title}
\author{\Authors} 
\date{\vspace{0.2cm}}

\titleformat{\section}{\normalfont\bfseries}{\thesection}{1em}{}
\titleformat{\subsection}{\normalfont\bfseries}{\thesubsection}{1em}{}
\titleformat{\subsubsection}{\normalfont\bfseries}{\thesubsubsection}{1em}{}

\titlespacing*{\section}{0pt}{6pt}{6pt}
\titlespacing*{\subsection}{0pt}{6pt}{6pt}
\titlespacing*{\subsubsection}{0pt}{6pt}{6pt}

\begin{document}

\maketitle
\vspace{-0.85cm}

\begin{abstract}
\Abstract \\ 


\noindent {\it JEL Classification:} \JEL
\end{abstract}
\vspace{-0.25cm}

\newpage 

\section{Introduction}

Climate change poses a growing challenge for life insurers and pension schemes, as its effects on population health introduce additional uncertainty into the long-term mortality assumptions used to value their liabilities \citep{miljkovic2018examining,boudreault2023changing,ofosuhene2026modelling}. Climate-related hazards, including extreme heat, air pollution, and flooding, can substantially increase mortality and morbidity risks, particularly among vulnerable populations \citep{romanello2024countdown}. Broader regulatory initiatives, including the Bank of England's \emph{2021 Climate Biennial Exploratory Scenario}, also reflect the growing attention to climate-related financial risks \citep{bankofengland2021a}. Despite this attention and recent methodological advances, translating climate exposures into mortality assumptions over the long projection horizons relevant to life insurance portfolios remains a central challenge. A targeted survey of practitioners in UK pension consultancies, insurers, and reinsurers highlights both the data limitations and methodological barriers to incorporating climate-related health risks into actuarial practice, as well as the need for climate scenario frameworks tailored to actuarial applications \citep{ofosuhene2026modelling}.

A particular challenge arises from the mismatch between the data needed to identify climate-related mortality effects and those required for actuarial valuation. Temperature--mortality relationships are inherently high-frequency, nonlinear, and lagged, and their magnitude may vary across ages and regions. At the same time, high-frequency mortality data are often available only for broad age groups, whereas life insurance pricing and valuation require mortality projections at individual ages over long horizons. Motivated by this mismatch, we develop a modeling framework that links climate scenarios to mortality projections and quantifies their implications for expected life insurance payoffs and portfolio risk. Focusing on temperature-related mortality, we augment a multi-population Hermite spline model with a distributed lag nonlinear model (DLNM). The Hermite component captures age-specific mortality patterns, long-term trends, and regular seasonality, while the DLNM captures the common nonlinear and lagged effects of heat and cold. Age-region-specific loadings allow the magnitude of these temperature effects to vary across populations. The resulting framework therefore draws on pooled death counts to estimate temperature effects while providing region-specific mortality forecasts at single-year ages for life insurance valuation.

Our framework builds on two closely related methodological strands of literature. The first concerns the estimation of temperature--mortality relationships. The introduction of the DLNM framework by \citet{gasparrini2010distributed} marked a major methodological advance in estimating associations between environmental exposures and mortality through a cross-basis structure that jointly captures nonlinear exposure--response relationships and delayed effects. Related epidemiological research examines mortality associations with a range of environmental exposures, including temperature \citep{armstrong2006models,madaniyazi2022assessing,madaniyazi2024seasonality}, air pollution \citep{dominici2002air}, relative humidity \citep{armstrong2019role}, and heat waves \citep{gasparrini2011impact}. We build on and extend this literature through a specification that estimates common temperature effects using data pooled across age groups and regions. Separate heat and cold loadings specific to each age-group--region combination allow the magnitude of these effects to vary across populations while retaining the benefits of pooled estimation. To accommodate the different temporal resolutions of mortality and temperature observations, we also adopt the modified mixed-frequency DLNM formulation of \citet{min2026mortality}, linking weekly mortality data with daily temperature data.

The second methodological strand concerns mortality modeling with seasonal and subannual data. Within the extensive actuarial literature on mortality modeling, growing attention has been devoted to models using subannual observations \citep{richards2020modelling, li2023pricing,begin2025modelling, li2025beyond,arik2025measuring,hanebeck2026dependence,miao2026gradient}. A practical challenge is that these observations may be available for age groups rather than single-year ages, creating a mismatch between the age resolution of the data and that required for insurance pricing and valuation. We address this challenge through a Hermite spline model that provides single-year-age mortality estimates and forecasts from grouped observations \citep{richards2020hermite, richards2020modelling, tang2023hermite}. Regular seasonality is incorporated through sine and cosine terms, allowing the model to retain seasonal variation beyond that explained by temperature. The proposed framework therefore combines the larger age-group death counts useful for estimating temperature effects with the single-year-age mortality forecasts needed for insurance applications.

Alongside these methodological developments, the insurance implications of climate-informed mortality projections have also received increasing attention. 
Recent studies have incorporated climate information into mortality models \citep{li2022joint, dong2022air,begin2025modelling, robben2025penalized,guibert2025impact,li2026modeling, min2026mortality, so2026climate} and examined its implications for insurance pricing and valuation \citep{wangevaluating,arandjelovic2026impact,begin2026modelling,kouton2026pricing}. These developments motivate further work connecting granular mortality estimation with the financial consequences of climate-related mortality risk. In particular, relatively less attention has been devoted to how climate change may alter both the seasonal timing of life insurance cash flows and the distribution of long-horizon liability outcomes. Our analysis addresses this issue by combining climate-sensitive mortality projections with an empirical analysis of a hypothetical whole life insurance portfolio under alternative climate scenarios.

This article advances the literature through two main contributions.
First, we develop a temperature-augmented multi-population Hermite mortality framework that bridges the gap between grouped high-frequency mortality observations and the single-year-age mortality projections required for actuarial valuation. The model combines age-specific mortality patterns, long-term improvement, regular seasonality, and nonlinear and lagged heat and cold effects. It pools information across populations while allowing temperature responses to vary by age group and region, and uses an expectation--maximization (EM) procedure to recover single-year-age mortality estimates from weekly age-group data. Second, from a life insurance risk-management perspective, we propagate climate-scenario and mortality-model uncertainty into long-duration insurance liabilities. This allows us to assess not only changes in expected death-benefit payments, but also how temperature-related mortality dynamics affect the seasonal timing, dispersion, and downside risk of portfolio outcomes.

We apply the proposed framework to mortality and temperature data from 15 Spanish NUTS-2 regions over 2000--2019. We then project mortality under SSP2-4.5, SSP3-7.0, and SSP5-8.5 and assess hypothetical whole-life portfolios issued in 2025 and 2050 to individuals aged 65. The results reveal an important interaction between rising temperatures and the seasonality of mortality. Warmer conditions reduce cold-related mortality during winter but increase heat-related mortality during summer, creating a winter compensation effect whose relative importance changes over the projection horizon. Relative to a fitted Hermite baseline without explicit future temperature information, discounted portfolio death-benefit payoffs are therefore generally lower during colder periods and higher during warmer periods, with these seasonal deviations becoming more pronounced under higher-warming scenarios and for policies issued in 2050. For policies issued in 2025, reductions in winter mortality offset much of the adverse summer effect and the profits and losses (P\&L) distributions overlap substantially across scenarios. For policies issued in 2050, the adverse effects of higher summer temperatures become increasingly important, with SSP5-8.5 producing substantially greater dispersion and larger losses at the reported lower percentiles. These findings highlight the importance of assessing climate-related mortality risk beyond mean projections and accounting for changes in both the timing and uncertainty of long-term insurance liabilities.

The remainder of the article is organized as follows. Section~\ref{sec:Data} describes the temperature and mortality data and presents visualizations and preliminary analyses. The multi-population temperature-augmented Hermite modeling framework is introduced in Section~\ref{sec:Modeling-framework}. Section \ref{sec:Empirical-results} presents the empirical results, focusing on estimated mortality trends, seasonal patterns, and regional heterogeneity in temperature--mortality relationships. Section \ref{sec:Scenario-projection} projects future mortality under SSP2-4.5, SSP3-7.0, and SSP5-8.5. The implications for a hypothetical whole life insurance portfolio are examined in Section \ref{sec:Pricing} through analyses of expected payoffs and the distribution of portfolio P\&L outcomes. Finally, Section \ref{sec:Conclusion} concludes.

\section{Data}\label{sec:Data}

\subsection{Notation}

Let $d_{r}(x,t)$ denote the weekly realized death counts for individuals aged $x \in \mathcal{X}$ in week $t \in \mathcal{T}$ and region $r \in \mathcal{R}$, where $\mathcal{X} = \{x_{\min}, ..., x_{\max}\}$, $\mathcal{T} = \{1, ..., T_{\!\max}\}$, and $\mathcal{R} = \{r_1, ..., r_R\}$. Similarly, let $E_{r}(x,t)$ represent the number of individuals aged $x$ during week $t$ in region $r$. The weekly empirical mortality rate for individual age $x$ in region $r$ is defined as
\begin{equation}
     \widehat{m}_{r}(x,t) = \frac{d_{r}(x,t)}{E_{r}(x,t)}.
\end{equation}

In our application, however, mortality data are typically available only in age-group aggregates, rather than at single-year ages, necessitating the use of grouped quantities. Therefore, we further define $d_{r}(\mathcal{G},t)$ as the weekly death counts for age group $\mathcal{G}$ in week $t$ and region $r$. The age groups are collected in the set $\mathcal{A}=\{\mathcal{G}_1,...,\mathcal{G}_G\}$, and form a partition of $\mathcal{X}$ such that $\bigcup_{\mathcal{G}\in\mathcal{A}}\mathcal{G}=\mathcal{X}$.

For age group $\mathcal{G}$, the corresponding death count is obtained by summing the individual-age death counts over all ages $x$ belonging to the group. That is, $d_{r}(\mathcal{G},t) = \sum_{x \in \mathcal{G}} d_{r}(x,t)$. The exposure to risk $E_{r}(\mathcal{G},t)$ denotes the number of individuals whose ages fall within group $\mathcal{G}$ during week $t$ in region $r$. The corresponding weekly mortality rate for age group $\mathcal{G}$ is defined as
\begin{equation}
    \widehat{m}_{r}(\mathcal{G},t) = \frac{d_{r}(\mathcal{G},t)}{E_{r}(\mathcal{G},t)}.
\end{equation}

Throughout the article, we distinguish between random death counts and their observed realizations. Specifically, let $D_r(x,t)$ denote the random number of deaths at age $x$, in week $t$, and region $r$. For an age group $\mathcal{G}$, we define
\begin{equation*}
D_r(\mathcal{G},t) = \sum_{x \in \mathcal{G}} D_r(x,t).
\end{equation*}

We also define $\mu_r(x,t)$ and $\mu_r(\mathcal{G}, t)$ as the model-implied  weekly mortality hazard at age $x$ and within age group $\mathcal{G}$, respectively, during week $t$ in region $r$. 

\subsection{Mortality data}

We collect weekly death counts and annual population data from the Eurostat database over the period 2000--2019, for five age groups: 65--69, 70--74, 75--79, 80--84, and 85 and older. These data are available for 15 NUTS-2 regions in mainland Spain. We focus on older age groups because we believe these segments are most vulnerable to climate change's impact on mortality. According to \cite{romanello2024countdown}, heat-related deaths among people over 65 are estimated to have increased by 167\% since the 1990s, while populations worldwide experience an additional 50 days of dangerous heat each year.

\begin{figure}[ht!]
    \centering
\includegraphics[width=0.95\linewidth]{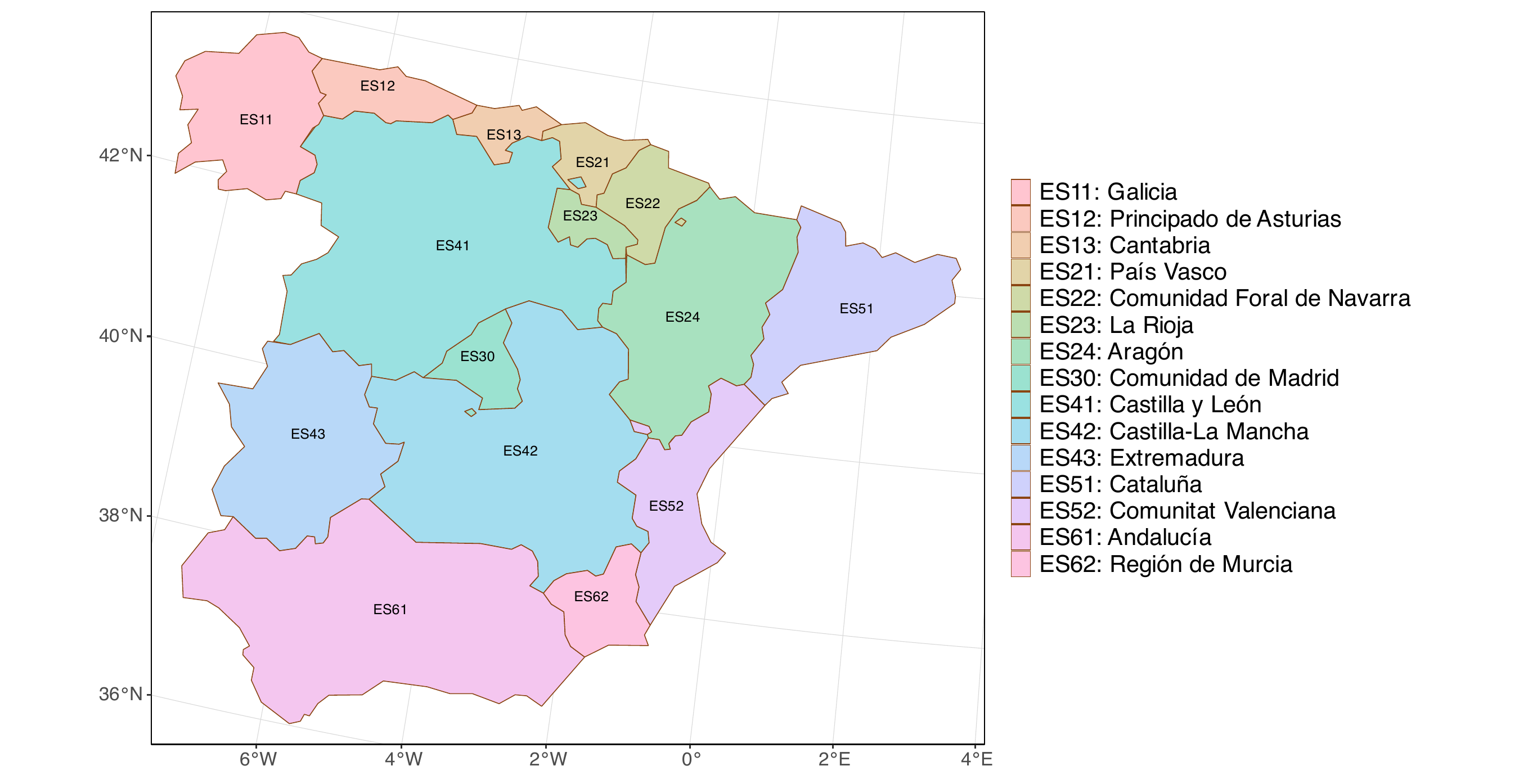}
    \caption{\textbf{Selected mainland Spanish NUTS-2 regions} \newline \footnotesize 
    \emph{Notes}: This figure reports the 15 mainland Spanish NUTS-2 regions considered in this study.}
    \label{fig:spain_nuts}
\end{figure}

Figure~\ref{fig:spain_nuts} shows the NUTS-2 regions included in the analysis.\footnote{Note that the NUTS-2 regions ES53, ES63, ES64, and ES70 are excluded, as they are island or geographically remote regions outside mainland Spain.} The 15 mainland Spanish NUTS-2 regions span the northern Atlantic coast, the Mediterranean coast, the central inland plateau, and the southern region of Andalusia. Among these, Andalucía (ES61), Cataluña (ES51), and Comunidad de Madrid (ES30) are the three most populous regions, whereas Comunidad Foral de Navarra (ES22),  Cantabria (ES13), and  La Rioja (ES23) are the three least populous regions. The geographical diversity of the selected regions allows our analysis to capture heterogeneity in climatic conditions and demographic characteristics.

Following \citet{li2023pricing} and \citet{min2026mortality}, we assume that the population size remains constant within each year and that each year consists of 52 weeks. The weekly exposure to risk is therefore approximated by $E_{r}(\mathcal{G},t) = P_{r}(\mathcal{G},t)/52$, where $P_{r}(\mathcal{G},t)$ denotes the annual population size in that year. The weekly mortality rate for age group $\mathcal{G}$ in week $t$ and region $r$ is thus computed as
\begin{equation}
    \widehat{m}_r(\mathcal{G},t) = \frac{d_{r}(\mathcal{G},t)}{E_{r}(\mathcal{G},t)} =\frac{d_{r}(\mathcal{G},t)}{P_{r}(\mathcal{G},t)/52}.
\end{equation}

\begin{figure}[ht!]
    \centering
\includegraphics[width=1\linewidth]{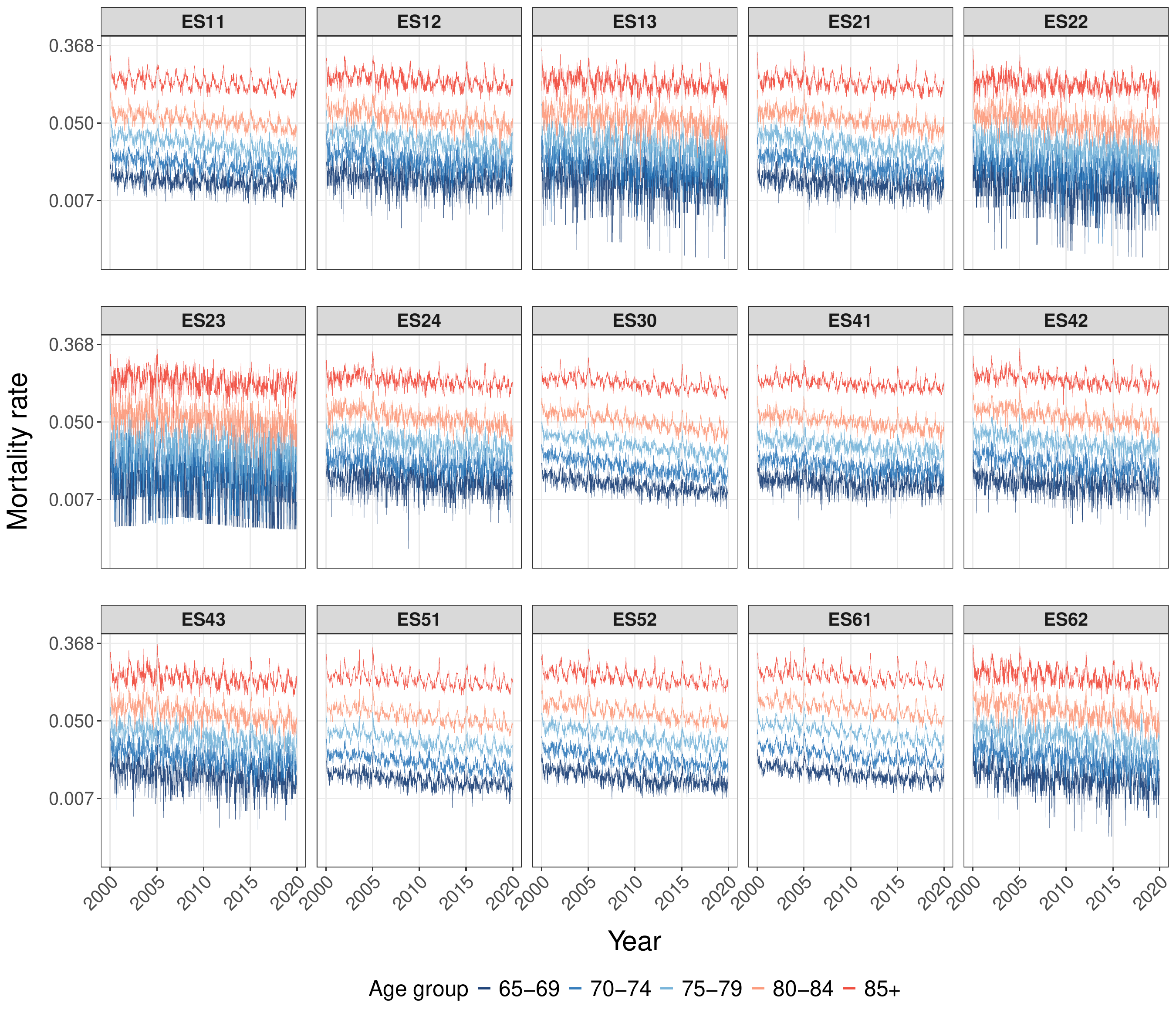}
\vspace{-0.1in}
    \caption{\textbf{Weekly mortality rates across different age groups and for selected mainland Spanish NUTS-2 regions} \newline \footnotesize 
    \emph{Notes}: This figure reports weekly mortality rates for selected mainland Spanish NUTS-2 regions over the period 2000--2019. Each panel corresponds to one region, identified by its NUTS-2 code. Lines represent age-specific mortality rates for five age groups: 65--69, 70--74, 75--79, 80--84, and 85 and older. 
    Mortality rates are displayed on a logarithmic scale to facilitate comparison across age groups.}
    \label{fig:mort_plot}
\end{figure}

Figure~\ref{fig:mort_plot} illustrates the computed weekly mortality rate for five different age groups across all fifteen regions. Each plot displays the oldest age groups at the top and the youngest at the bottom. Overall, we observe a slight downward trend in mortality rates over the period 2000--2019, although this trend is relatively modest compared with the pronounced short-term fluctuations in the data. Clear seasonal patterns are evident across the regions, particularly among older age groups, with mortality rates typically higher in winter and lower in summer. At the same time, several regions exhibit temporary increases in mortality during summer periods, which are likely associated with extreme heat events. It is also important to note that the time series display substantial variability, especially in regions with smaller populations,
where the data are comparatively noisy. This is particularly pronounced for the younger age groups considered in this study (i.e., 65--69 and 70--74), for which random fluctuations tend to obscure underlying patterns more strongly than in older age groups.

Table~\ref{tab:mortality-summary} reports average annual mortality rates over 2000--2019 across the five age groups considered in this study. While differences in mortality across the 15 regions are not large, a clear association with GDP per capita remains evident. In general, regions with higher GDP per capita tend to exhibit lower mortality rates, and this gradient is more pronounced at older ages. For example, Comunidad de Madrid (ES30), which has the highest GDP per capita among the regions considered, also records one of the lowest mortality levels.

\begin{table}[ht!]
\caption{\textbf{GDP per capita and average annual mortality rates across selected mainland Spanish NUTS-2 regions}}
{\fontsize{9.5}{10}\selectfont
\centering
\begin{tabularx}{\linewidth}{lX d{3.1} *{5}{d{1.4}}}
\toprule
& 
& & \multicolumn{5}{c}{Average annual mortality rates} \\
\cmidrule(lr){4-8}
NUTS-2 & {Region name} 
& \multicolumn{1}{c}{GDP per capita} &  \multicolumn{1}{c}{65--69} & \multicolumn{1}{c}{70--74} & \multicolumn{1}{c}{75--79} & \multicolumn{1}{c}{80--84} & \multicolumn{1}{c}{85+} \\
\midrule
ES11 & Galicia & 81 & 0.0112 & 0.0174 & 0.0298 & 0.0551 & 0.1439 \\
ES12 & Principado de Asturias & 79 & 0.0119 & 0.0189 & 0.0325 & 0.0600 & 0.1523 \\
ES13 & Cantabria & 83 & 0.0113 & 0.0175 & 0.0303 & 0.0571 & 0.1448 \\
ES21 & Pa\'is Vasco & 114 & 0.0107 & 0.0172 & 0.0297 & 0.0554 & 0.1438 \\
ES22 & Comunidad Foral de Navarra & 109 & 0.0099 & 0.0160 & 0.0285 & 0.0537 & 0.1405 \\
ES23 & La Rioja & 95 & 0.0100 & 0.0164 & 0.0290 & 0.0548 & 0.1446 \\
ES24 & Arag\'on & 99 & 0.0108 & 0.0172 & 0.0304 & 0.0572 & 0.1468 \\
ES30 & Comunidad de Madrid & 124 & 0.0099 & 0.0162 & 0.0281 & 0.0523 & 0.1369 \\
ES41 & Castilla y Le\'on & 84 & 0.0104 & 0.0160 & 0.0273 & 0.0507 & 0.1366 \\
ES42 & Castilla-La Mancha & 72 & 0.0102 & 0.0171 & 0.0307 & 0.0588 & 0.1512 \\
ES43 & Extremadura & 67 & 0.0124 & 0.0201 & 0.0343 & 0.0636 & 0.1536 \\
ES51 & Catalu\~na & 106 & 0.0110 & 0.0177 & 0.0309 & 0.0579 & 0.1479 \\
ES52 & Comunitat Valenciana & 79 & 0.0116 & 0.0192 & 0.0344 & 0.0646 & 0.1575 \\
ES61 & Andaluc\'ia & 67 & 0.0130 & 0.0216 & 0.0384 & 0.0714 & 0.1632 \\
ES62 & Regi\'on de Murcia & 75 & 0.0115 & 0.0196 & 0.0347 & 0.0671 & 0.1636 \\
\bottomrule
\end{tabularx}%
}

\par\smallskip\footnotesize 
\emph{Notes}: This table reports the gross domestic product (GDP) per capita for each selected NUTS-2 region, measured in 2019 purchasing power standards (PPS) and expressed relative to the average of the 27 European Union member states (the average is set to 100), available from the Eurostat database. Annual mortality rates are calculated as the total weekly deaths in each year divided by the annual population, then averaged over 2000--2019.
\label{tab:mortality-summary}
\end{table}

\subsection{Temperature data}

\begin{figure}[ht!]
    \centering
    \includegraphics[width=0.9\linewidth]{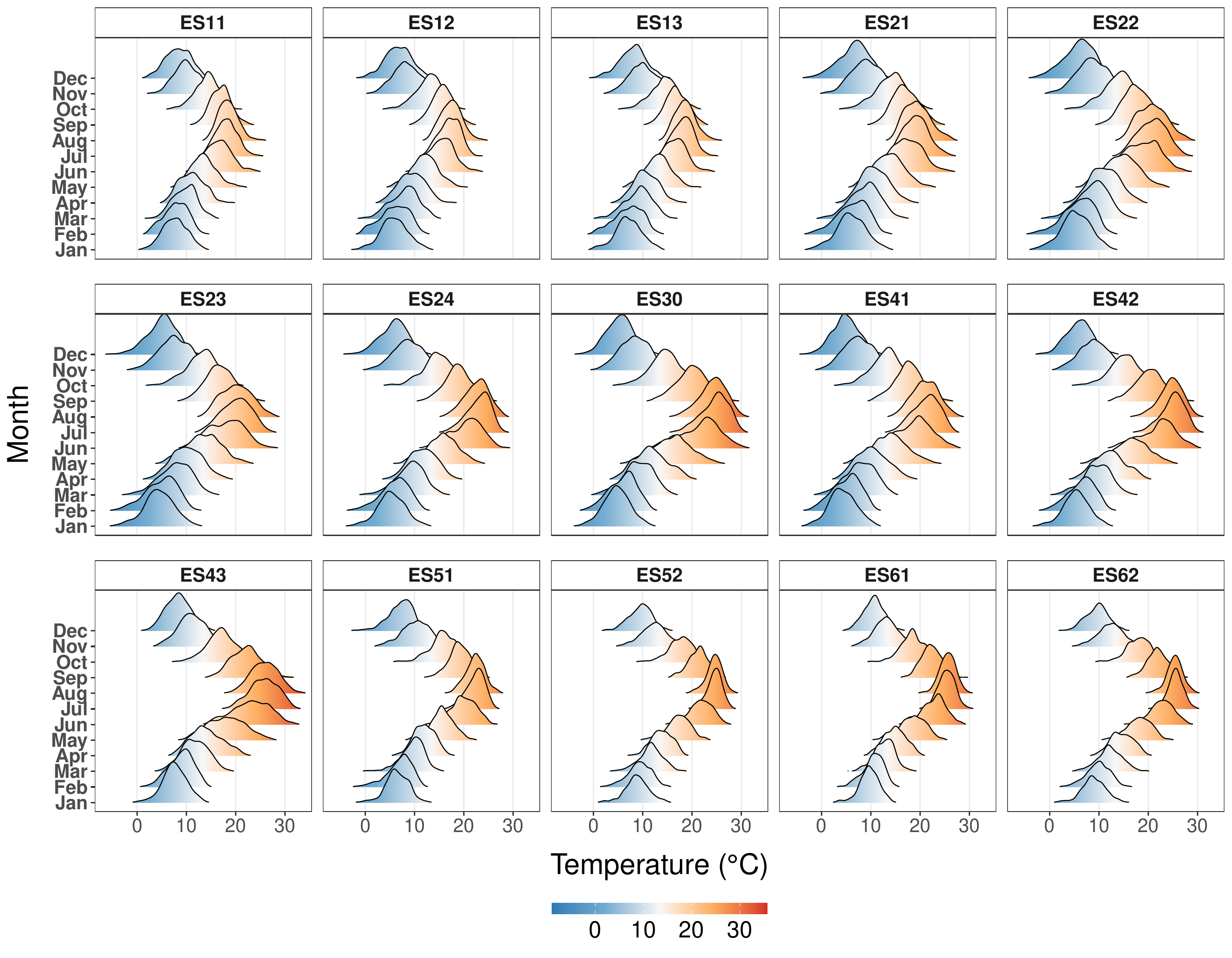}
    \vspace{-0.2in}
    \caption{\textbf{Monthly distributions of daily mean temperature across selected mainland Spanish NUTS-2 regions} \newline \footnotesize 
    \emph{Notes}: This figure reports monthly ridgeline distributions of daily mean temperatures for selected mainland Spanish NUTS-2 regions over 2000--2019. Each panel corresponds to one region, identified by its NUTS-2 code. Months are ordered from January at the bottom to December at the top. Colors indicate temperature in degrees Celsius, with colder values in blue and warmer in orange.}
    \label{fig:temp_plot}
\end{figure}

We obtain daily mean temperature data from the fifth generation European Centre for medium-range weather forecasts (ECMWF) reanalysis (ERA5), available in the \cite{era5_2025}. This dataset provides hourly temperature data with a spatial resolution of $0.1^{\circ} \times 0.1^{\circ}$. For each NUTS-2 region, temperature values are extracted from all grid points within the regional boundary, and their average is used to represent the regional temperature. We collect temperature data for the period 2000--2019 to align with the mortality data. Figure~\ref{fig:temp_plot} plots the monthly distributions of daily mean temperature for the fifteen NUTS-2 regions during the investigation period.

For each region, the monthly ridgeline is constructed using all daily mean temperature observations from the corresponding month over 2000--2019. The plots demonstrate pronounced seasonal patterns in daily mean temperature across all regions, together with notable differences in the spread of monthly temperature distributions across NUTS-2 regions. Table~\ref{tab:temp-summary} reports the minimum, mean, and maximum daily temperatures during the summer months (June, July, and August) and the winter months (December, January, and February) for the 15 NUTS-2 regions over the period 2000--2019. As expected, geographical variation is evident. The northern Atlantic regions, such as Galicia (ES11), Principado de Asturias (ES12), and Cantabria (ES13), experience relatively mild summers, whereas inland and southern regions, including Comunidad de Madrid (ES30), Extremadura (ES43), Andalucía (ES61), and Región de Murcia (ES62), record overall higher summer temperatures. Winter temperatures also exhibit a clear spatial gradient, with inland regions, including Aragón (ES24), Castilla y León (ES41), La Rioja (ES23), and Comunidad de Madrid (ES30), generally experiencing colder conditions than coastal regions, including Andalucía (ES61), Región de Murcia (ES62), Comunitat Valenciana (ES52), and Cataluña (ES51). For conciseness, we generally refer to regions by their NUTS-2 codes in the following sections.

\begin{table}[ht!]
\caption{\textbf{Summary statistics of daily mean temperature across mainland Spanish NUTS-2 regions}}
{\fontsize{9.5}{10}\selectfont
\centering
\begin{tabularx}{\linewidth}{lX *{6}{d{2.2}}}
\toprule
& & \multicolumn{3}{c}{Winter months} & \multicolumn{3}{c}{Summer months} \\
\cmidrule(lr){3-5} \cmidrule(lr){6-8}
NUTS-2 & {Region name} 
& \multicolumn{1}{c}{Minimum} 
& \multicolumn{1}{c}{Mean} 
& \multicolumn{1}{c}{Maximum} 
& \multicolumn{1}{c}{Minimum} 
& \multicolumn{1}{c}{Mean} 
& \multicolumn{1}{c}{Maximum} \\
\midrule
ES11 & Galicia & 0.51 & 7.92 & 14.92 & 10.51 & 17.76 & 25.60 \\
ES12 & Principado de Asturias & -4.07 & 6.13 & 13.58 & 8.74 & 16.71 & 24.15 \\
ES13 & Cantabria & -2.15 & 7.36 & 15.11 & 10.37 & 17.81 & 26.18 \\
ES21 & Pa\'is Vasco & -3.44 & 6.23 & 15.02 & 10.81 & 18.71 & 28.48 \\
ES22 & Comunidad Foral de Navarra & -3.96 & 5.66 & 14.37 & 10.87 & 20.25 & 29.31 \\
ES23 & La Rioja & -6.61 & 4.72 & 13.19 & 9.39 & 19.46 & 28.00 \\
ES24 & Arag\'on & -3.80 & 5.62 & 13.46 & 10.89 & 22.05 & 28.93 \\
ES30 & Comunidad de Madrid & -3.55 & 5.28 & 12.60 & 11.24 & 23.57 & 31.00 \\
ES41 & Castilla y Le\'on & -4.89 & 4.69 & 11.71 & 9.23 & 20.42 & 28.17 \\
ES42 & Castilla-La Mancha & -3.64 & 5.93 & 13.23 & 12.16 & 23.91 & 30.95 \\
ES43 & Extremadura & -0.24 & 8.22 & 14.99 & 13.95 & 24.94 & 33.47 \\
ES51 & Catalu\~na & -2.50 & 6.81 & 12.98 & 12.62 & 21.56 & 27.36 \\
ES52 & Comunitat Valenciana & 0.04 & 9.20 & 16.89 & 14.83 & 23.67 & 29.00 \\
ES61 & Andaluc\'ia & 1.35 & 10.22 & 16.43 & 15.92 & 24.61 & 30.33 \\
ES62 & Regi\'on de Murcia & -0.15 & 9.42 & 17.26 & 15.41 & 24.61 & 30.20 \\
\bottomrule
\end{tabularx}%
}
\par\smallskip\footnotesize 
\emph{Notes}: This table reports summary statistics for daily mean temperature, in degrees Celsius, across the mainland Spanish NUTS-2 regions included in the analysis. For each region, the table shows the minimum, mean, and maximum observed daily mean temperature during winter and summer months. Winter months are defined as December, January, and February, while summer months are defined as June, July, and August. Statistics are calculated using daily observations pooled over the period 2000--2019.
\label{tab:temp-summary}
\end{table}

\section{Modeling framework}
\label{sec:Modeling-framework}
This section introduces the temperature-augmented Hermite mortality model in a multi-population setting. The Hermite component serves as a baseline that captures age structure, long-term mortality trends, regular seasonal variation, and regional differences in mortality experience, whereas the temperature component captures age- and region-specific temperature effects on mortality, including nonlinear and lagged effects. We first outline the basic assumptions of our modeling framework and introduce the Hermite baseline mortality model without temperature effects. We then extend the DLNM by incorporating age- and region-specific loadings and by decomposing the cross-basis matrix into heat and cold components. Finally, we combine the Hermite baseline mortality model with temperature effects using an extended age- and region-specific DLNM. This combined framework is referred to as the temperature-augmented Hermite mortality model throughout this article.

\subsection{Distributional assumptions}\label{sec:assupmtions}

Suppose that the death count for age $x$ on week $t$ in region $r$ is a Poisson-distributed random variable such that 
\begin{equation*}
    D_{r}(x,t) \sim \text{Poi}\left(E_{r}(x,t) \, \mu_{r}(x,t) \right),
\end{equation*}
similar to \citet{brouhns2002poisson} and \citet{renshaw2006cohort}, among others. By assuming that $D_{r}(x,t)$ is conditionally independent across age, time, and regions, the death count for age group $\mathcal{G}$ in week $t \in T$ in region $r \in \mathcal{R}$ also follows a Poisson random variable. For simplicity, and because we do not have exposures at individual ages, we further assume that exposure to risk is evenly distributed across ages within each age group. Specifically, for any $x \in \mathcal{G}$ in week $t$, we set ${E}_{r}(x,t) = \frac{{E}_{r}(\mathcal{G},t)}{\mathrm{card}\left(\mathcal{G}\right)}$ and have
\begin{equation}
    D_{r}(\mathcal{G},t) = \sum_{x \in \mathcal{G}} D_{r}(x,t) \sim  \text{Poi}\left({E}_{r}(x,t) \sum_{x \in \mathcal{G} }\mu_{r}(x,t) \right),
\end{equation}
because $E_{r}(\mathcal{G},t)\, \mu_{r}(\mathcal{G},t) = \sum_{x\in\mathcal{G}} E_{r}(x,t)\, \mu_{r}(x,t) = {E}_{r}(x,t) \sum_{x \in \mathcal{G}}\mu_{r}(x,t)$. 

Under the Poisson mean specification, the weekly mortality hazard for age $x$ and age group $\mathcal{G}$ are expressed as 
\begin{equation}
    \mu_r(x,t) = \frac{\mathbb{E}[D_r(x,t)]}{E_r(x,t)} \quad \text{and} \quad \mu_r(\mathcal{G},t) = \frac{\mathbb{E}[D_r(\mathcal{G},t)]}{E_r(\mathcal{G},t)},
\end{equation}
respectively, where $E_r(x,t)$ and $E_r(\mathcal{G},t)$ are treated as known and enter the corresponding count-scale models through the offset term $\log \left(E_r(x,t)\right)$ and $\log \left(E_r(\mathcal{G},t)\right)$. In the following sections, we approximate weekly mortality hazards using their corresponding weekly mortality rates.

\subsection{Multi-population Hermite spline mortality model}\label{sec:hermite-model}

\begin{figure}[ht!]
    \centering
    \includegraphics[width=0.95\linewidth]{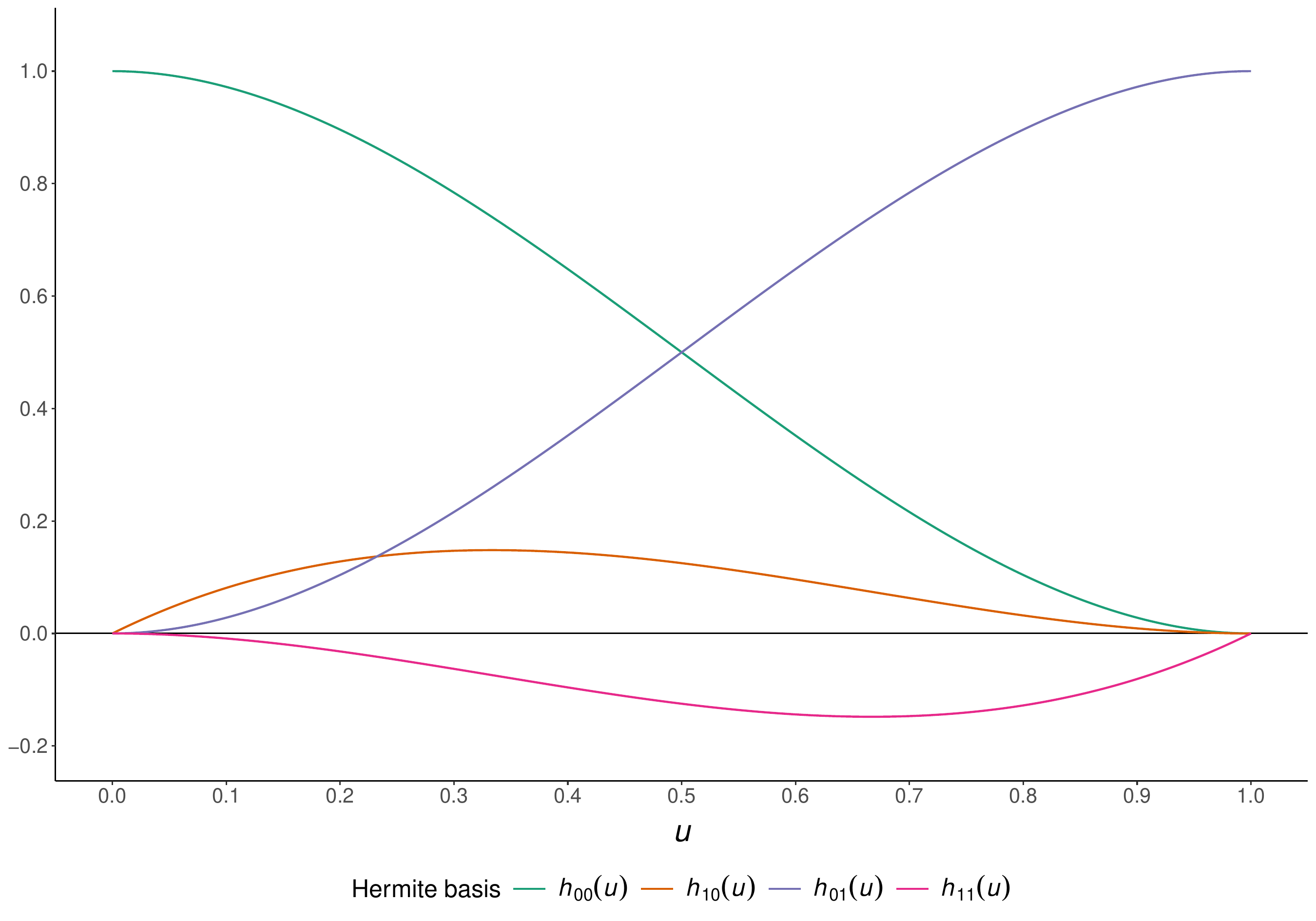}
    \caption{\textbf{Cubic Hermite basis functions on the unit interval} \newline \footnotesize
    \emph{Notes}: This figure shows the four cubic Hermite basis functions on $u \in [0,1]$. These functions determine the interpolation between two endpoints. The basis functions $h_{00}(u)=(1+2u)(1-u)^2$ and $h_{01}(u)=u^2(3-2u)$ weight the function values at the left and right endpoints, respectively. The functions $h_{10}(u)=u(1-u)^2$ and $h_{11}(u)=u^2(u-1)$ weight the corresponding endpoint slopes.}
    \label{fig:hermite_basis}
\end{figure}

The underlying mathematical concept of the \citet{hermite1878sur} spline builds on the foundational work of Charles Hermite on Hermite interpolation. Building on this groundwork, \cite{schoenberg1973cardinal} established the foundational B-spline theory for cardinal Hermite interpolation, which evolved into standard modern implementations. Specifically, as detailed by \cite{kreyszig1999advanced}, the one-dimensional Hermite spline is constructed using a basis collection of four cubic polynomial functions, as shown in Figure~\ref{fig:hermite_basis}.

The Hermite spline provides a smooth path between the start point $(0, \alpha)$ and the end point $(1, \gamma)$ via a linear combination of the Hermite basis functions, which can be expressed as 
\begin{equation}
    p(u) = \alpha \, h_{00}(u) +  m_0 \, h_{10}(u) + \gamma \, h_{01}(u) + m_1 \, h_{11}(u), \label{eq:hermitegen}
\end{equation}
with four basis functions as follows: 
\begin{align*}
    h_{00}(u) = & (1+2u)(1-u)^2, & h_{10}(u) = u(1-u)^2, \\
    h_{01}(u) = & u^2(3-2u),  & h_{11}(u) =  u^2 (u-1), 
\end{align*}
where $m_0$ and $m_1$ denote the gradients at the points $(0, \alpha)$ and $(1, \gamma)$, respectively. Varying the values of $m_0$ and $m_1$ yields a wide range of shapes, which provides sufficient flexibility to capture smooth curves in practice.

The Hermite spline functions were first applied to modeling mortality rates by \cite{richards2020hermite} and \cite{richards2020modelling}, and were extended by \cite{tang2023hermite} to incorporate time-varying parameters. In particular, \cite{richards2020hermite} modeled the log mortality curve as a smooth function of age through the Hermite-spline approach, which provides greater flexibility in capturing mortality patterns at older ages while maintaining smoothness and interpretability. The approach was subsequently extended in \cite{richards2020modelling} to model seasonal mortality patterns using individual-level data. These developments demonstrate the flexibility of Hermite splines in capturing complex mortality dynamics across age and time.

This article applies the general Hermite spline structure of Equation~\eqref{eq:hermitegen} with time trends and seasonality terms to model the weekly mortality hazard $\mu_{r}(x,t)$. Let $u = (x - x_{\min})/(x_{\max} - x_{\min})$ for age $x \in [x_{\min}, x_{\max}]$, our proposed multi-population Hermite model is defined as
\begin{align}
\mathcal{H}_r(x,t) = \log\left(\mu_r(x,t)\right) =&\, \underbrace{\alpha_r \, h_{00}(u) + m_{0,r} \, h_{10}(u) +\gamma_r \, h_{01}(u) + m_{1,r} \, h_{11}(u)}_{\text{Region-specific Hermite term}} \nonumber \\ 
&\,+ \underbrace{(\phi_r \, h_{00}(u) + \kappa_r \, h_{01}(u)) \frac{t}{52}}_{\text{Region-specific trend}} \nonumber \\ 
&\,+ \underbrace{\sum_{j=1}^2 \rho_{1,j}\sin\left(\frac{2\pi j t}{52}\right) + \sum_{k=1}^2 \rho_{2,k}\cos\left(\frac{2\pi k t}{52}\right)}_{\text{Common seasonal pattern}}\,\,,
\label{hermite-baseline}
\end{align}
where the parameters $\alpha_r$, $m_{0,r}$, $\gamma_r$, and $m_{1,r}$ determine the region-specific baseline age profile through the Hermite basis functions. The parameters $\phi_r$ and $\kappa_r$ capture the region-specific time trend at the left and right endpoints, respectively, thereby allowing the mortality trend to vary smoothly
with age through the Hermite basis. Finally, $\rho_{1,j}$ and $\rho_{2,k}$ govern the common seasonal pattern, represented by annual and semi-annual Fourier terms.\footnote{The specification in Equation~\eqref{hermite-baseline} corresponds to the preferred baseline model selected using the \citet{akaike1973information} information criteria (AIC) and  Bayesian information criteria (BIC) of \citet{schwarz1978estimating}. Additional details on the model-selection procedure are provided in Appendix~\ref{sec:model_selection}.}

\subsection{The distributed lag nonlinear model}\label{sec:extension-dlnm}

The DLNM approach was first proposed by \citet{gasparrini2010distributed} to analyze the nonlinear and lagged effects of temperature on mortality. Climate variables such as temperature \citep{cheng2014temperature, wen2023new, madaniyazi2022assessing,madaniyazi2024seasonality}, air pollution \citep{dominici2002air,gasparrini2011distributed}, humidity 
\citep{armstrong2019role}, and heat-wave and cold-spell indicators \citep{gasparrini2011impact, guo2018quantifying, min2026mortality} are commonly used in the DLNM. 

To introduce the standard DLNM formulation, we temporarily suppress the region and age indices and write the mortality hazard as $\mu(t)$; this foundation will then be used in our framework. The DLNM employs a quasi-Poisson regression model with a logarithmic link function to capture the nonlinear and lagged effects of temperature as
\begin{equation}
\label{original_dlnm}
\log \left( \mu (t) \right) = \beta_0 + \sum_{\ell = 0}^L f\big(y(t-\ell), \ell; \bm{\varsigma}\big) \, \bm{\eta} +\bm{z}_{t}^{\top}  \bm{\beta}, 
\end{equation}
where $\beta_0$ is the intercept, $y{(t-\ell)}$ is the exposure variable (e.g., temperature) at lag $\ell = 0, ..., L$, and $L$ represents the maximum lag considered. The vector $\bm{z}_t$ includes the confounders (e.g., the time trend and seasonality), with $\bm{\beta}$ being the vector of the corresponding coefficients. The function $f$ is a two-dimensional exposure--lag function defined as
\begin{equation}
    f\big(y{(t-\ell)}, \ell; \bm{\varsigma}\big) = \Big(\bm{s}_y\big(y{(t-\ell)}; \varsigma_y\big) \otimes \bm{s}_l(\ell; \varsigma_\ell)\Big)^{\!\top},
\end{equation}
where $\bm{\varsigma} = \left[ \,\, \varsigma_y \quad  \varsigma_\ell\,\,\right]$ denotes the degrees of freedom for the basis functions $\bm{s}_y$ and $\bm{s}_l$, representing the exposure--response and lag--response relationships, respectively. 
These two basis functions are combined through the Kronecker product, denoted by $\otimes$.
More specifically, we have 
$\bm{s}_y(y(t-\ell); \varsigma_y) = \left[\,\, \bm{s}_{y,1}(y(t-\ell)) \quad ... \quad\bm{s}_{y,\varsigma_y}(y(t-\ell)) \,\, \right]^{\top} \in \mathbb{R}^{\varsigma_y}$ and 
$\bm{s}_l(\ell; \varsigma_\ell) = \left[\,\,\bm{s}_{\ell, 1}(\ell) \quad ... \quad \bm{s}_{\ell, \varsigma_\ell}(\ell)\,\,\right]^{\top} \in \mathbb{R}^{\varsigma_\ell}$. The vector $\bm{\eta} \in \mathbb{R}^{\varsigma_y \varsigma_\ell}$ denotes the cross-basis coefficient vector. 
For notational simplicity, we define $\bm{C}(t)$ as
\begin{equation}
    \label{cross_basis_matrix}
    \bm{C}(t) = \sum_{\ell = 0}^L f\big(y{(t-\ell)}, \ell; \bm{\varsigma}\big).
\end{equation} 

In the epidemiological literature, DLNMs are typically applied to total deaths aggregated across all age groups to estimate the overall temperature--mortality association \citep[see, e.g.,][]{guo2018quantifying, madaniyazi2022assessing,madaniyazi2024seasonality}. However, this approach may overlook age differences in vulnerability to temperature exposure. It is widely acknowledged that older age groups are generally more sensitive to both extreme heat and extreme cold and therefore experience higher mortality risks under such conditions \citep{tillett1983excess,aastrom2011heat,li2022joint}. A straightforward way to uncover age-specific temperature--mortality associations is to apply DLNMs separately to different age groups. This approach has become increasingly common in recent actuarial studies, where understanding age-specific vulnerability to temperature-related mortality risk is of particular interest \citep[see, e.g.,][]{wangevaluating,guibert2025impact,min2026mortality}. It is worth noting that this approach may not be ideal, as it does not borrow strength across age groups through data pooling. In particular, age-specific death counts may be small, especially at finer temporal resolutions, leading to inefficient estimation and potentially unstable or unreliable DLNM fits. This can result in noisy estimates of the temperature--mortality relationship and poorer model performance. Yet there is limited information sharing across populations within regions. To the best of our knowledge, very few studies have considered these issues, including \cite{robben2025penalized} and \cite{begin2026modelling}. Therefore, addressing the limitations of current practice and extending the modeling framework to a multi-population setting are key motivations for our proposed approach.

\subsection{The temperature-augmented Hermite mortality model}

Building on recent work by \cite{robben2025penalized} and \cite{begin2026modelling}, we make two extensions to the classic DLNM. First, we distinguish between the effects of heat and cold temperatures on mortality. Instead of employing a single cross-basis function to model both effects simultaneously, we construct separate cross-basis functions for heat and cold exposure, defined with respect to selected temperature thresholds. Second, we introduce age-region-specific loadings linked to a common cross-basis function estimated from pooled data. Given the geographical proximity of the regions, it is reasonable to assume that they share similar underlying responses to extreme temperatures. By estimating a common cross-basis function, we reduce estimation error and model uncertainty through information pooling. Importantly, heterogeneity across age groups and regions is retained through the loadings, which allow the magnitude of temperature effects to vary by age group and region.

The proposed temperature-augmented Hermite mortality model is specified as 
\begin{equation}
    \label{hermite_delta_dlnm}
    \log \left(\mu_r(x,t)\right) = \mathcal{H}_r(x,t) + \delta_{r}^{\text{H}}(\mathcal{G}) \, \bar{\bm{C}}_{r}^{\text{H}}(t) \, \bm{\eta}^{\text{H}} + \delta_{r}^{\text{C}}(\mathcal{G}) \, \bar{\bm{C}}_{r}^{\text{C}}(t) \, \bm{\eta}^{\text{C}},
\end{equation}
where $\mathcal{H}_r(x,t)$ denotes the Hermite baseline component as defined in Equation \eqref{hermite-baseline}. For age $x \in \mathcal{G}$, $\delta_{r}^{\text{H}}(\mathcal{G})$ and $\delta_{r}^{\text{C}}(\mathcal{G})$ denote the age-region-specific loadings associated with heat- and cold-related cross-basis matrices $\bar{\bm{C}}^{\text{H}}_{r}(t)$ and $\bar{\bm{C}}^{\text{C}}_{r}(t)$, respectively. The vectors $\bm{\eta}^{\text{H}}$ and $\bm{\eta}^{\text{C}}$ are coefficient vectors, estimated by pooling information across regions and age groups, which define the common heat and cold exposure--lag response structures associated with $\bar{\bm{C}}^{\text{H}}_{r}(t)$ and $\bar{\bm{C}}^{\text{C}}_{r}(t)$, respectively.

To formally define $\bar{\bm{C}}^{\text{H}}_{r}(t)$ and $\bar{\bm{C}}^{\text{C}}_{r}(t)$, we first introduce the heat- and cold-adjusted temperature series $y_{r}^{\text{H}}(t)$ and $y_{r}^{\text{C}}(t)$ as
\begin{equation}
    \label{temperature_adjustment}
    y_{r}^{\text{H}}(t) = \max\left(y_{r}(t), y_{\text{ref}}^{\text{H}}\right), 
    \quad 
    y_{r}^{\text{C}}(t) = \min\left(y_{r}(t), y_{\text{ref}}^{\text{C}}\right), 
\end{equation}
where $y_{\text{ref}}^{\text{H}}$ and $y_{\text{ref}}^{\text{C}}$ denote the reference temperatures for heat and cold, respectively. Then, for $a \in \{\text{H},\text{C}\}$, we construct the heat- and cold-related cross-basis matrices as
\begin{align}
    \bar{\bm{C}}_{r}^{a}(t) &= \bm{C}_{r}^{a}(t) - \bm{C}_{\text{ref}}^{a} = \sum_{\ell = 0}^L f\big(y_{r}^{a}(t-\ell),\ell; \bm{\varsigma}\big) - \sum_{\ell = 0}^Lf(y_{\text{ref}}^{a},\ell; \bm{\varsigma})  \nonumber \\ 
    &= \sum_{\ell = 0}^L \Big(\bm{s}_y\big(y_{r}^{a}(t-\ell); \varsigma_y\big) - \bm{s}_y (y_{\text{ref}}^{a};\varsigma_y)\Big)^{\top} \otimes \bm{s}_{\ell}\big(\ell ; \varsigma_\ell \big)^{\top}.
    \label{adjusted_cross_basis_matrix}
\end{align}

\subsection{Model estimation}\label{sec:estimation}

We estimate the proposed model, as specified in Equation~\eqref{hermite_delta_dlnm}, using an iterative estimation procedure with refinement rounds. 
For clarity, we separate the parameters estimated in the first and second stages. The first-stage parameter set is defined as 
\begin{equation} 
    \bm{\theta} = \left[ \,\, \bm{\alpha} \quad \bm{\gamma} \quad \bm{m}_0 \quad \bm{m}_1 \quad \bm{\phi} \quad \bm{\kappa} \quad \bm{\rho}_1 \quad \bm{\rho}_2 \quad \bm{\eta}^{\text{H}} \quad \bm{\eta}^{\text{C}} \,\, \right], 
    \label{para_theta}
\end{equation}
where $\bm{\alpha} = \left\{\alpha_r\right\}_{r \in \mathcal{R}}, \bm{\gamma} = \left\{\gamma_r\right\}_{r \in \mathcal{R}}, \bm{m}_0 = \left\{m_{0,r}\right\}_{r \in \mathcal{R}}, \bm{m}_1 = \left\{m_{1,r}\right\}_{r \in \mathcal{R}}, \bm{\phi} = \left\{\phi_r\right\}_{r \in \mathcal{R}}$, and $\bm{\kappa} = \left\{\kappa_r\right\}_{r \in \mathcal{R}}$ denote region-specific parameter vectors in the Hermite baseline component, whereas $\bm{\rho}_1=\left[\,\,\rho_{1,1} \quad \rho_{1,2}\,\,\right]$, and $\bm{\rho}_2=\left[\,\,\rho_{2,1}\quad\rho_{2,2}\,\,\right]$ denote parameter vectors of common seasonal patterns.  
The second-stage parameter set is defined as
\begin{equation}
    \bm{\delta} = \left[\,\, \bm{\delta}^{\text{H}} \quad \bm{\delta}^{\text{C}}\,\,\right],
    \label{para_delta}
\end{equation}
where $\bm{\delta}^{\text{H}} = \left\{\delta_r^{\text{H}}(\mathcal G)\right\}_{r\in\mathcal R,\,\mathcal G\in\mathcal{A}}$ and $\bm{\delta}^{\text{C}} = \left\{\delta_r^{\text{C}}(\mathcal G)\right\}_{r\in\mathcal R,\,\mathcal G\in\mathcal{A}}$ are age-group-region-specific loadings for heat and cold effects. These two stages are repeated until convergence.

\paragraph{Initial round.} We first estimate the parameters of the Hermite spline and cross-basis functions using the EM algorithm, subject to the constraints that $\bm{\delta}^{\text{H}} = \bm{\delta}^{\text{C}} = \bm{1}$. The estimated parameter set is denoted as $\hat{\bm{\theta}}^{(1)}$. Then, holding $\bm{\theta} = \hat{\bm{\theta}}^{(1)}$, we estimate $\bm{\delta}^{\text{H}}$ and $\bm{\delta}^{\text{C}}$ by maximizing the Poisson log-likelihood under the model specified in Equation~\eqref{hermite_delta_dlnm}. The estimated parameter set is denoted as $\hat{\bm{\delta}}^{(1)}$.

    \paragraph{Subsequent rounds.} In round $j \geq 2$, by taking $\hat{\bm{\theta}}^{(j-1)}$ as the initial parameter set, we estimate again the parameter set $\bm{\theta}$ using the EM algorithm, subject to $\bm{\delta} = \hat{\bm{\delta}}^{(j-1)}$. The estimated parameter set is denoted as $\hat{\bm{\theta}}^{(j)}$. Then, by fixing $\bm{\theta} = \hat{\bm{\theta}}^{(j)}$ and taking $\hat{\bm{\delta}}^{(j-1)}$ as the initial parameter set, we re-estimate $\bm{\delta}$ by maximizing the conditional log-likelihood. The estimated parameter set is denoted as $\hat{\bm{\delta}}^{(j)}$. 
    The two stages are repeated until the absolute relative change in the log-likelihood between the $(j-1)$th and the $j$th rounds fall below the pre-specified tolerance level $\varepsilon_{\mathcal{L}}$, at which point the iterative procedure is deemed to have converged. In our case study, we set $\varepsilon_\mathcal{L} = 10^{-2}$. Further mathematical details of the estimation procedure are provided in Appendix~\ref{app:Estimation}.

\section{Empirical results}\label{sec:Empirical-results}

In this section, we present the empirical results obtained by fitting the data described in Section \ref{sec:Data} to our proposed temperature-augmented Hermite mortality model in Equation \eqref{hermite_delta_dlnm}. We begin by presenting the estimation results for the Hermite baseline component, followed by an examination of the temperature effects modeled by the DLNM. Finally, we evaluate the in-sample fit of the proposed model to the observed mortality rates.

\subsection{Hermite spline estimation}\label{sec: hermite-component}

We present the estimated Hermite baseline term and trend coefficients in Table \ref{tab:hermite-trend-coef}. The estimated coefficients have consistent signs and broadly comparable magnitudes across regions. Following the Hermite spline interpretation of \cite{richards2020hermite}, $\alpha_r$ and $\gamma_r$ determine the logarithmic baseline mortality levels at the lower and upper age boundaries (ages 65 and 110, respectively), while $m_{0,r}$ and $m_{1,r}$ specify the slopes of the mortality profile at the two boundary ages. We can see that the estimated coefficients vary across regions, indicating regional heterogeneity in both the logarithm of the mortality levels and the gradients of the mortality profile. In particular, $\hat{\alpha}_r$ ranges from $-4.5469$ in ES23 to $-4.2630$ in ES61, and $\hat{\gamma}_r$ ranges from $-2.7515$ in ES22 to $-2.4023$ in ES30, suggesting regional differences in the baseline mortality levels. Regional variation is also evident in the boundary gradients, with $\hat{m}_{0,r}$ ranging from $1.8442$ in ES13 to $2.8711$ in ES30, and $\hat{m}_{1,r}$ ranging from $-9.2994$ in ES24 to $-6.7903$ in ES30.

\begin{table}[ht!]
\caption{\textbf{Estimated region-specific Hermite term and trend coefficients.}}
{\fontsize{9.5}{10}\selectfont
\centering
\begin{tabularx}{\linewidth}{X *{6}{d{4.7}}}
\toprule

& \multicolumn{4}{c}{Hermite term}
& \multicolumn{2}{c}{Trend} \\
\cmidrule(lr){2-5}\cmidrule(lr){6-7}
\multicolumn{1}{l}{NUTS-2} & \multicolumn{1}{c}{$\hat{\alpha}_r$}
& \multicolumn{1}{c}{$\hat{m}_{0,r}$}
& \multicolumn{1}{c}{$\hat{\gamma}_r$}
& \multicolumn{1}{c}{$\hat{m}_{1,r}$}
& \multicolumn{1}{c}{$\hat{\phi}_r$}
& \multicolumn{1}{c}{$\hat{\kappa}_r$} \\
\cmidrule(lr){1-1} \cmidrule(lr){2-2} \cmidrule(lr){3-3} \cmidrule(lr){4-4} \cmidrule(lr){5-5} \cmidrule(lr){6-6} \cmidrule(lr){7-7}
ES11 & -4.4634 & 2.2417 & -2.5386 & -7.8014 & -0.0209 & -0.0104 \\
ES12 & -4.4041 & 2.4597 & -2.4844 & -7.7406 & -0.0227 & -0.0115 \\
ES13 & -4.4086 & 1.8442 & -2.7226 & -8.9966 & -0.0228 & -0.0077 \\
ES21 & -4.4396 & 2.3950 & -2.5770 & -8.0234 & -0.0266 & -0.0106 \\
ES22 & -4.4907 & 2.0631 & -2.7515 & -9.0659 & -0.0254 & -0.0051 \\
ES23 & -4.5469 & 2.1096 & -2.6408 & -8.8890 & -0.0205 & -0.0109 \\
ES24 & -4.4400 & 2.0160 & -2.6792 & -9.2994 & -0.0223 & -0.0110 \\
ES30 & -4.4992 & 2.8711 & -2.4023 & -6.7903 & -0.0304 & -0.0138 \\
ES41 & -4.5284 & 2.0389 & -2.5561 & -7.8141 & -0.0209 & -0.0086 \\
ES42 & -4.5363 & 2.7909 & -2.5777 & -8.6230 & -0.0233 & -0.0117 \\
ES43 & -4.3280 & 2.4412 & -2.5876 & -8.3521 & -0.0240 & -0.0102 \\
ES51 & -4.4055 & 2.3365 & -2.5876 & -8.3848 & -0.0262 & -0.0106 \\
ES52 & -4.3536 & 2.7822 & -2.5650 & -8.6568 & -0.0273 & -0.0123 \\
ES61 & -4.2630 & 2.8276 & -2.6181 & -8.9803 & -0.0260 & -0.0116 \\
ES62 & -4.3664 & 2.8374 & -2.5560 & -8.7708 & -0.0282 & -0.0111 \\
\bottomrule
\end{tabularx}%
}
\par\smallskip\footnotesize
\emph{Notes}: This table reports the estimated region-specific Hermite term and trend coefficients in the temperature-augmented Hermite mortality model.
\label{tab:hermite-trend-coef}
\end{table}

The trend coefficients $\hat{\phi}_r$ and $\hat{\kappa}_r$ are negative across all regions, indicating a downward trend in mortality rates over time due to mortality improvements. ES30 (Madrid, a highly urbanized region) exhibited the largest decline in the logarithm of the mortality rates, with approximate reductions of $3.04\%$  per annum at age 65 and $1.38\%$ per annum at age 110. By comparison, the estimated declines are more gradual in the less urbanized regions of ES11 (Galicia, $2.09\%$ per annum at age 65 and $1.04\%$ per annum at age 110) and ES13 (Cantabria, $2.28\%$ per annum at age 65 and $0.77\%$ per annum at age 110). This contrast may be attributable to better healthcare infrastructure and greater accessibility in highly urbanized areas, compared with relatively limited access to healthcare services in less urbanized regions.

\begin{figure}[ht!]
    \centering
    \includegraphics[width=0.9\linewidth]{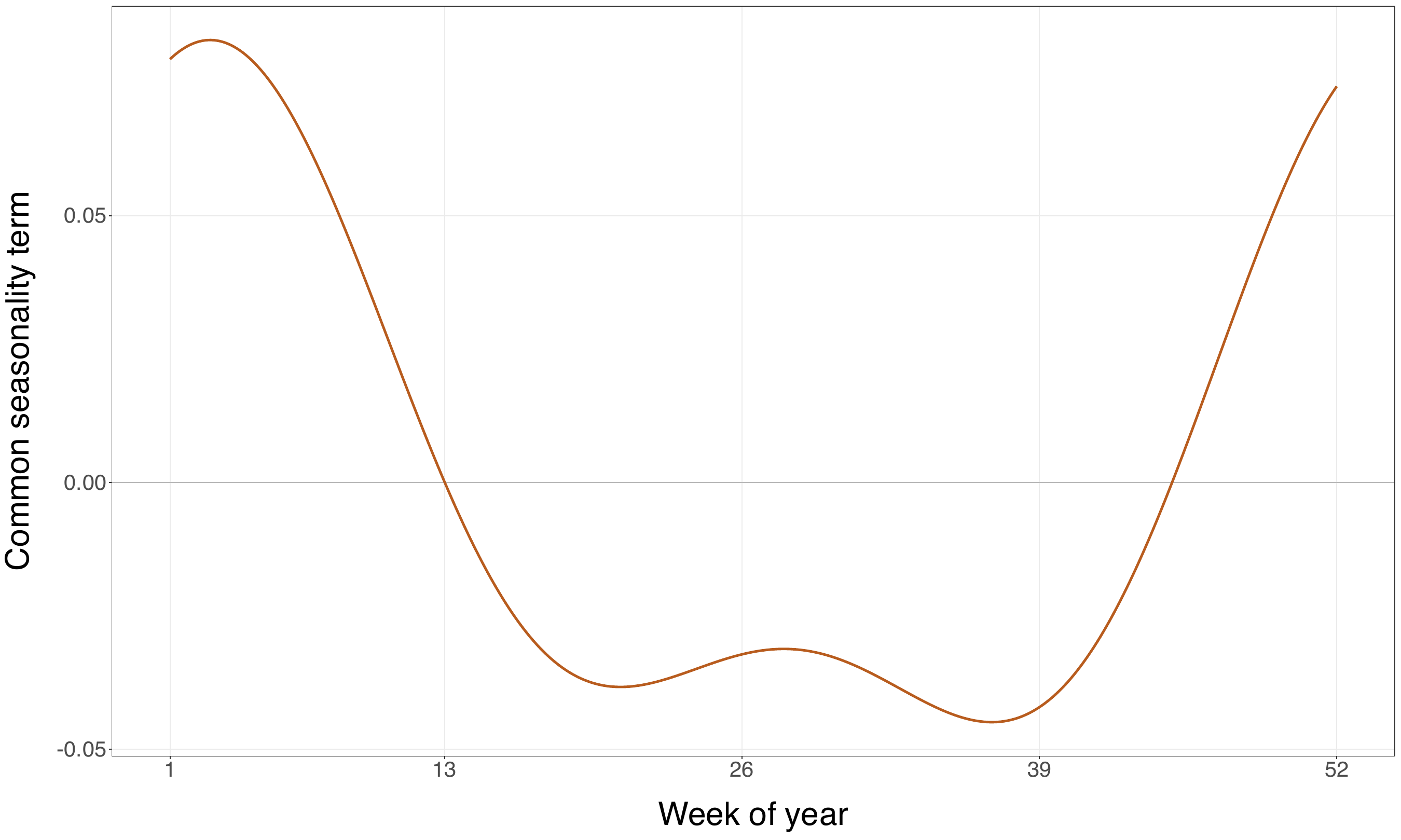}
    \caption{\textbf{Estimated seasonal patterns in the temperature-augmented Hermite mortality model} \newline \footnotesize
    \emph{Notes}: This figure displays the estimated seasonal patterns in the temperature-augmented Hermite mortality model. The estimated seasonality coefficients are $\hat{\rho}_{1,1} = 0.021$, $\hat{\rho}_{1,2} = 0.015$, $\hat{\rho}_{2,1} = 0.053$, and $\hat{\rho}_{2,2} = 0.021$.}
    \label{fig:common_seasonality}
\end{figure}

We then present the seasonality terms in Figure \ref{fig:common_seasonality} with estimated coefficients $\hat{\rho}_{1,1} = 0.021$, $\hat{\rho}_{1,2} = 0.015$, $\hat{\rho}_{2,1} = 0.053$, and $\hat{\rho}_{2,2} = 0.021$. The common seasonality across regions exhibits a pronounced W-shaped pattern over the course of the year. 
Mortality rates are elevated during the winter months, while a smaller secondary peak occurs in summer. These patterns capture seasonality in mortality beyond the effects of temperature. 

\subsection{Temperature effects via DLNM} \label{sec: temperature-effect-DLNM}
\subsubsection{Overall and lagged effects across ages and regions} \label{sec:cumulative-overall-effects}

In the proposed model, we have two DLNM components that separately capture heat- and cold-related temperature effects. It should be noted that these DLNM models are estimated using all available data and thus describe the common temperature--mortality relationship across age groups and regions. They are then scaled by the estimated loadings to capture differences across age groups and heterogeneity across regions.

\begin{figure}[ht!]
    \centering
    \includegraphics[width=0.95\linewidth]{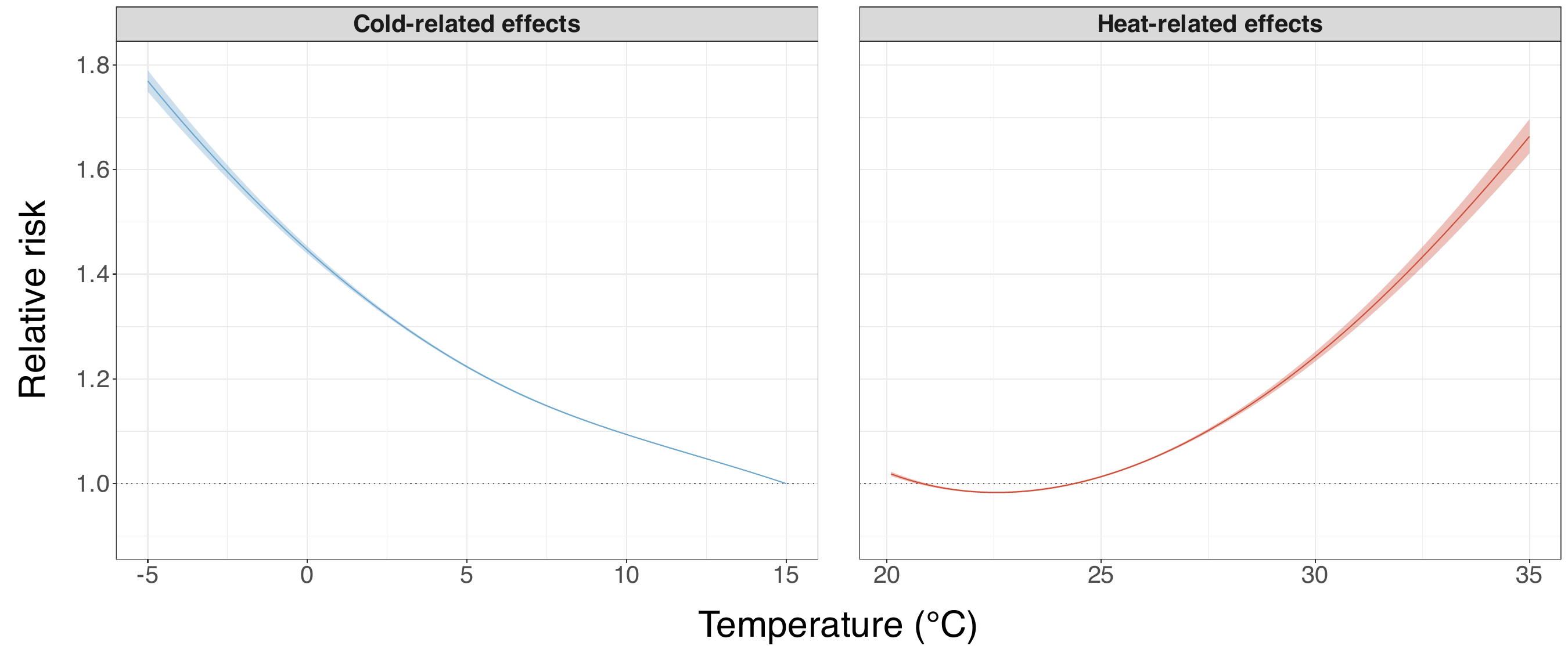}
    \caption{\textbf{Overall cumulative effects of temperature on mortality rates} \newline \footnotesize
    \emph{Notes}: This figure illustrates the overall cumulative effects of temperature on mortality rates derived from the cold-related and heat-related DLNM components. The cold and heat thresholds are set at $15\,\,{}^{\circ}\text{C}$ and $20\,\,{}^{\circ}\text{C}$, respectively. The temperature effects within the range of $15$--$20\,\,{}^{\circ}\text{C}$ are assumed to be negligible, with the relative risk fixed at one.}
    \label{fig:overall_cumulative_effects}
\end{figure}

Figure~\ref{fig:overall_cumulative_effects} presents the estimated cumulative relative risks associated with the heat- and cold-related cross-basis functions. {It should be noted that in our study, the reference temperatures for heat and cold have been set to be $y_{\text{ref}}^{\text{C}} = 15\,\,{}^{\circ}\text{C}$ and $y_{\text{ref}}^{\text{H}} = 20\,\,{}^{\circ}\text{C}$.\footnote{The reference temperatures are selected based on DLNM models estimated separately for each region across all age groups, with the estimated relative risk generally close to 1 over the range of $15$--$20\,\,{}^{\circ}\text{C}$.}} For temperatures between the cold threshold of $15\,\,{}^{\circ}\text{C}$ and the heat threshold of $20\,\,{}^{\circ}\text{C}$, the proposed model assumes that temperature-related effects are negligible, such that the relative risk is fixed at one. Overall, there is a clear U-shaped relationship between relative risk and temperature, with elevated mortality risks under both extreme cold ($-5\,\,{}^{\circ}\text{C}$) and extreme heat ($35\,\,{}^{\circ}\text{C}$) conditions. These results are broadly consistent with previous studies \citep[e.g.,][]{gasparrini2010distributed, robben2025penalized, min2026mortality, begin2026modelling}. 

\begin{figure}[ht!]
    \centering
    \includegraphics[width=0.95\linewidth]{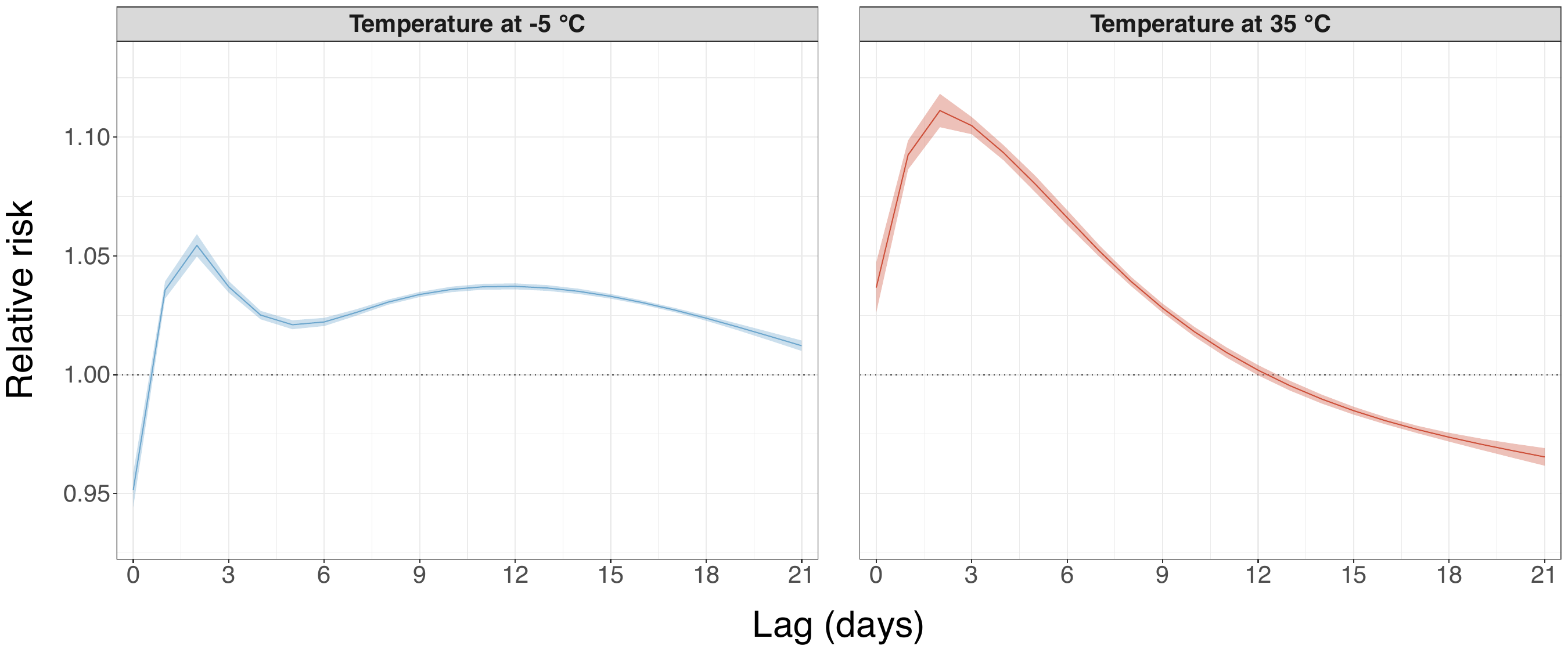}
    \caption{\textbf{Lagged effects for cold- and heat-related mortality risks} \newline \footnotesize
    \emph{Notes}: This figure reports lag-specific relative risks at $-5^\circ\mathrm{C}$ for cold effects and $35^\circ\mathrm{C}$ for heat effects, respectively. The cold-related effects are modest but persistent throughout the 21-day lag period, whereas the heat-related effects are acute, concentrated within the first three days, and decline rapidly thereafter.}
    \label{fig:lagged_effects}
\end{figure}

Figure~\ref{fig:lagged_effects} presents the lag--response relationships for cold-related and heat-related mortality risks at $-5\,\,{}^{\circ}\text{C}$ and $35\,\,{}^{\circ}\text{C}$, respectively. The relative risk, which is below 1 at lag 0, increases over time and peaks at approximately 2--3 days before declining over the subsequent few days. It then remains above 1 throughout the remainder of the 21-day lag period, indicating persistent adverse effects of cold exposure.
In contrast, heat exposure has a more immediate and acute effect on mortality rates. The relative risk is already elevated at lag 0 and rises rapidly to its peak at approximately 2--3 days of lag. It subsequently declines rapidly and falls below 1 after approximately 12 days of lag, indicating a potential \emph{harvesting effect} due to heat waves. Again, these findings are largely consistent with the existing literature.

\subsubsection{Estimated age--group--region-specific loadings} \label{sec:delta-loading}

\begin{figure}[ht!]
    \centering
    \includegraphics[width=0.95\linewidth]{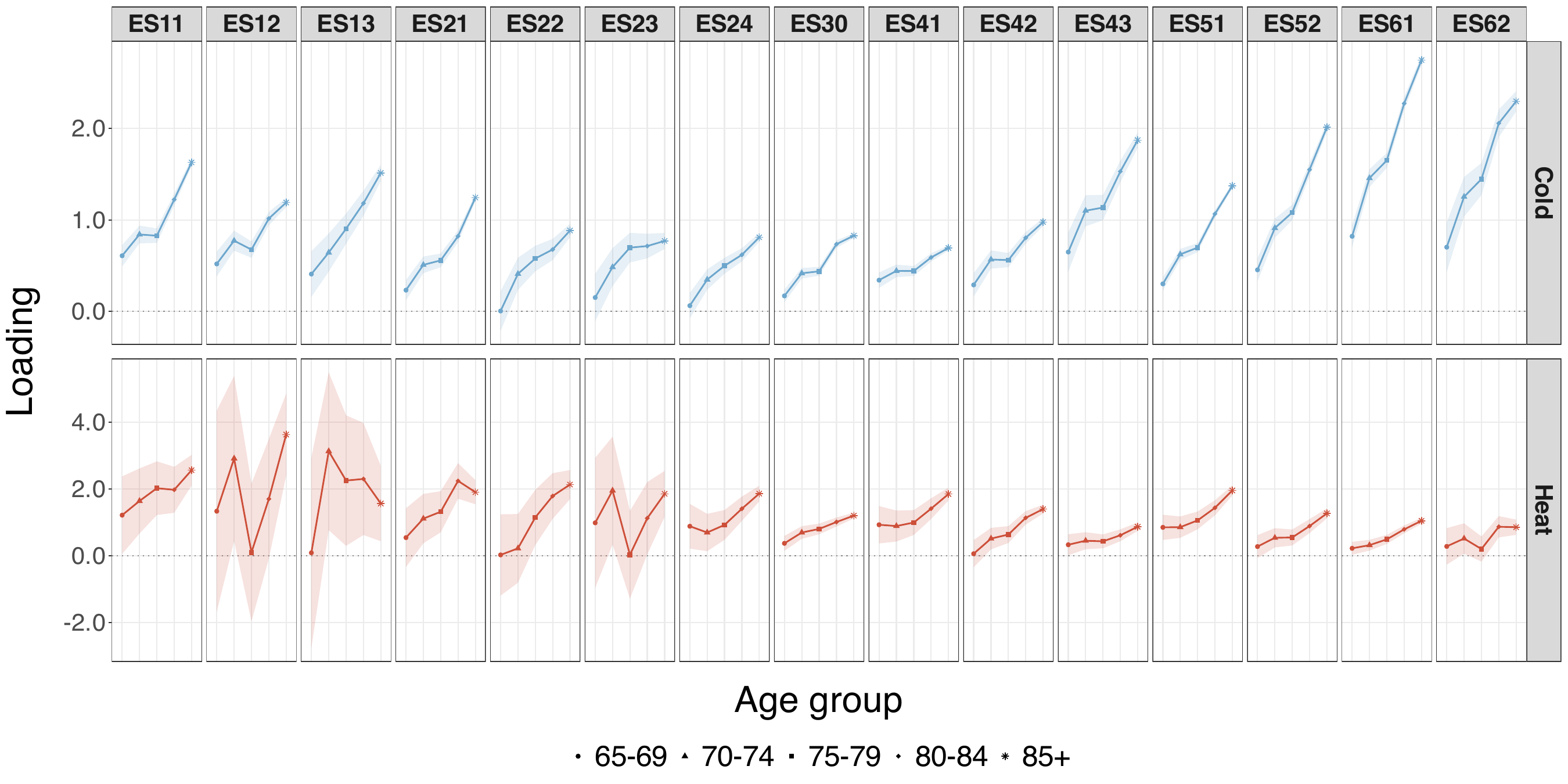}
    \caption{\textbf{Estimated age-group-region-specific loadings for cold and heat effects} \newline \footnotesize
    \emph{Notes}: This figure reports the region-specific loadings for each age group. The upward trend of the loadings for both heat- and cold-related effects in most regions indicates that older age groups are more sensitive and more vulnerable to the heat- and cold-related effects. }
    \label{fig:delta_loading}
\end{figure}

Figure~\ref{fig:delta_loading} illustrates the estimated age--group--region-specific loadings, $\delta$, for cold- and heat-related effects across each age group and region combination. The higher the loading, the more vulnerable the population is to extreme cold and heat. For each subplot, the $x$-axis represents age, with age group increasing from left to right. For most regions, there is an upward trend in the loadings, indicating that older age groups are more vulnerable to both cold- and heat-related mortality risks. Comparing loadings across regions, we find clear regional heterogeneity in the response to temperature extremes. Warmer regions, particularly those in southern Spain (e.g., ES52, ES61, and ES62), tend to be more resilient to extreme heat than to extreme cold, exhibiting smaller heat-related loadings but larger cold-related $\delta$-loadings. On the other hand, relatively cooler regions in northern Spain (e.g., ES11, ES12, ES13, ES21, and ES22) exhibit greater resilience to cold-related mortality effects than to heat-related effects, in contrast to the warmer regions. These results are not surprising and are consistent with local climatic adaptation: people in warmer regions are more vulnerable to extreme cold, while people in colder regions are more vulnerable to extreme heat. Similar findings have also been reported in \citet{heutel2021adaptation} and \citet{hajdu2026impact}. In addition, more economically affluent regions (e.g., ES30 and ES51) are generally less affected by both hot and cold extremes, suggesting that socioeconomic resources may play an important role in enhancing adaptive capacity, alongside potential adaptation to local climatic conditions.\\

\subsection{In-sample fit of mortality rates}\label{sec:in_sample_fit}

\begin{figure}[ht!]
    \centering
    \includegraphics[page=5, width=1\linewidth]{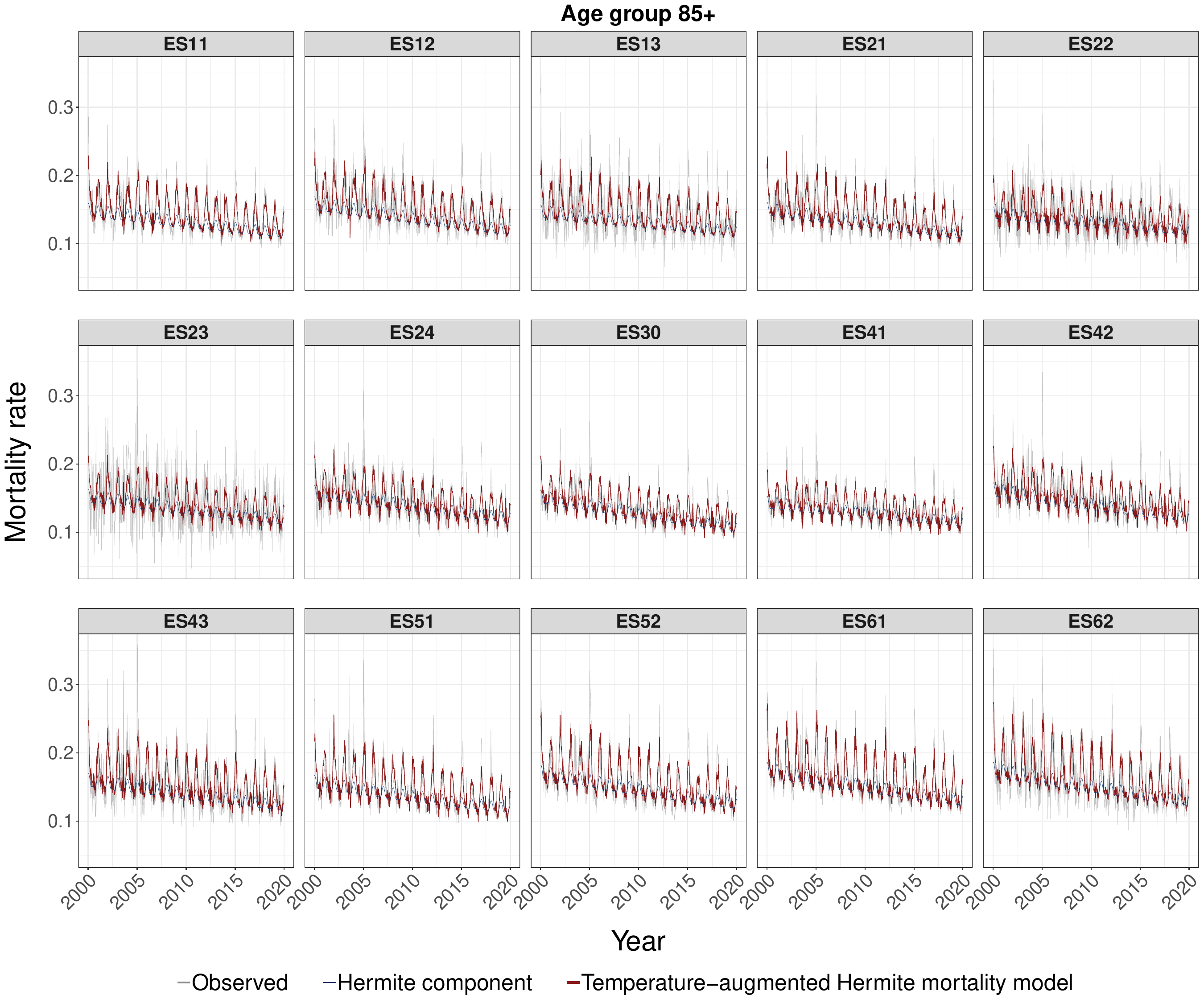}
    \caption{\textbf{Fitted mortality rates over the period 2000--2019 for the age group 85+} \newline \footnotesize
    \emph{Notes}: This figure shows the in-sample fit of weekly mortality rates across the selected mainland Spanish NUTS-2 regions. The gray line represents the observed mortality rates, the blue line represents the Hermite baseline component capturing the baseline seasonal pattern and time trend, and the red line represents the fitted mortality rates from the full temperature-augmented Hermite mortality model, combining the Hermite baseline component and temperature-related effects.}
    \label{fig:fitted_mortality_85}
\end{figure}

Figure~\ref{fig:fitted_mortality_85} compares the observed mortality rates (gray line) for the age group 85 and older with those fitted by the temperature-augmented Hermite mortality model (red line), as well as the fitted Hermite component without adding the DLNM component (blue line), across all regions.\footnote{The corresponding figures for the remaining age groups are presented in Section~\ref{sec:fit} of the Supplementary Material.} The observed mortality rates show clear evidence of mortality improvement, with a pronounced downward trend over the period 2000--2019. This long-term downward trend is adequately captured by the Hermite component and is consequently retained in the temperature-augmented Hermite model. We also see that the observed mortality rates exhibit a pronounced seasonal pattern, characterized by recurrent winter peaks and smaller secondary peaks during summer. The fitted Hermite component would capture the underlying smooth seasonal pattern over the period 2000--2019, while the temperature-augmented Hermite model further adjusts the magnitude of mortality peaks according to cold- and heat-related temperature effects.
Consequently, the temperature-augmented  Hermite model provides a better fit to the historical data, particularly around the winter and summer mortality peaks. The differences between the two fitted lines represent the contribution of temperature effects beyond the long-term trend and regular seasonal pattern captured by the Hermite component.

\section{Scenario-based mortality projection}
\label{sec:Scenario-projection}
This section first introduces the shared socioeconomic pathways (SSPs) used in this study and describes how we obtain the corresponding projected temperature trajectories. These temperature projections capture scenario uncertainty and overcome the limitation that climate models typically provide a single temperature trajectory. We then present projected weekly mortality rates using the proposed temperature-augmented Hermite mortality model, combining the simulated temperature trajectories with bootstrapped model parameters to account for both climate and parameter uncertainty.

\subsection{Projected temperature under SSP scenarios}\label{sec:CMIP6-data}

The SSPs describe alternative socioeconomic development pathways that shape future greenhouse gas emissions through 2100. These pathways underpin the climate projections used in the Intergovernmental Panel on Climate Change (IPCC) Sixth Assessment Report \citep{masson2021climate}. There are five SSP narratives representing sustainable development (SSP1), middle-of-the-road development (SSP2), regional rivalry (SSP3), inequality (SSP4), and fossil-fueled development (SSP5). When combined with different radiative forcing targets, these socioeconomic pathways define specific climate scenarios. For example, SSP1-2.6 combines the SSP1 socioeconomic pathway with a radiative forcing target of 2.6 Wm$^{-2}$ by 2100.\footnote{It should be noted that SSP1-2.6 represents a low-emissions pathway that is increasingly considered unlikely under current policy trajectories.} This paper focuses on three representative climate scenarios, namely SSP2-4.5, SSP3-7.0, and SSP5-8.5.

\begin{figure}[ht!]
    \centering
    \includegraphics[width=1\linewidth]{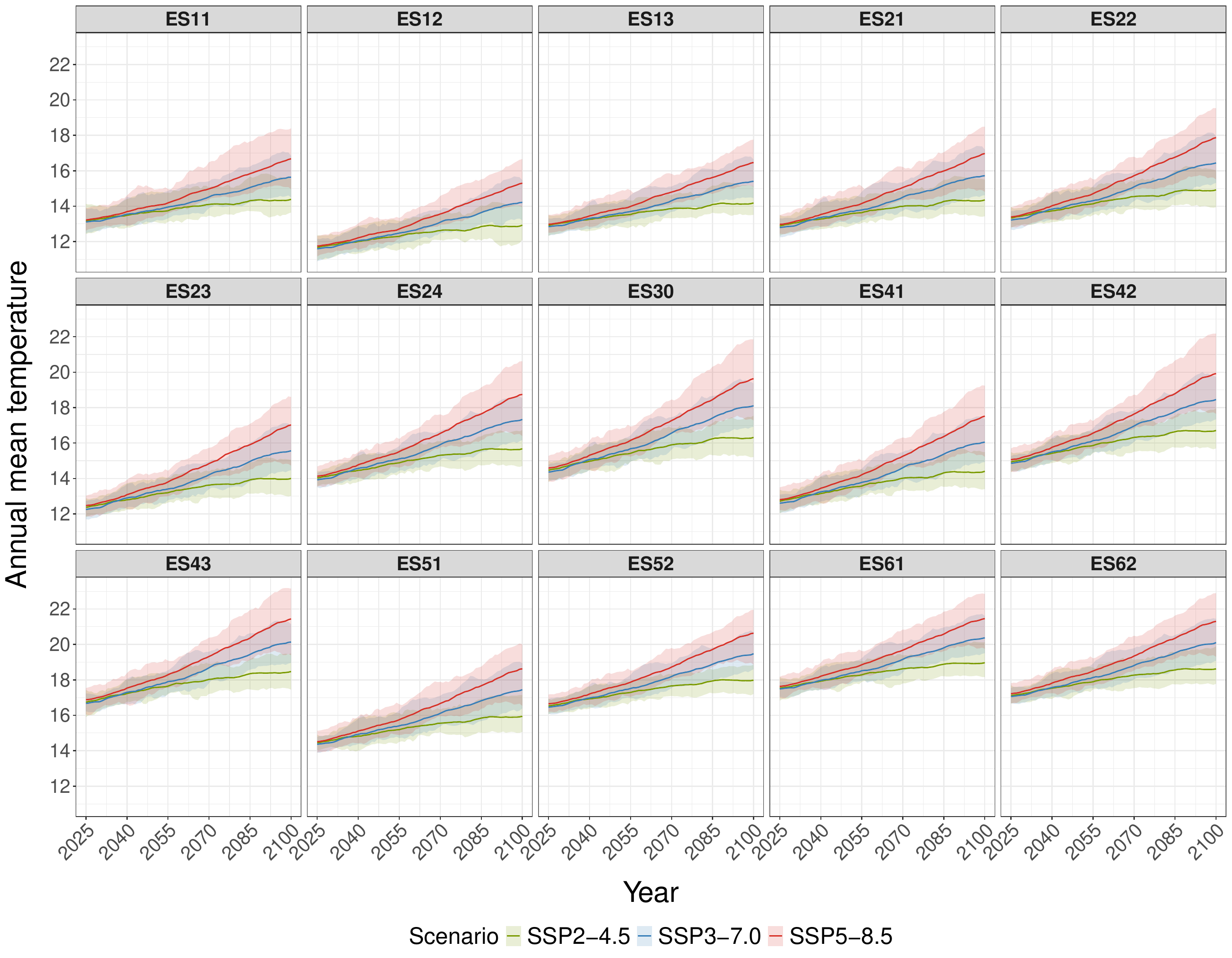}
    \caption{\textbf{Projected annual mean daily temperature under different SSP scenarios} \newline \footnotesize
    \emph{Notes}: This figure displays projected annual mean temperature and the 95\% projection interval under SSP2-4.5, SSP3-7.0, and SSP5-8.5. Annual mean temperature is calculated by averaging bias-corrected daily temperatures within each year. The solid lines show the simulated annual mean temperature, and the shaded bands show the 2.5th and 97.5th percentiles across simulated annual temperature paths. The visualized lines and shaded areas are smoothed using a 10-year centered moving average.}
    \label{fig:ssp_proj}
\end{figure}

The SSP-based temperature projections are obtained from the Coupled Model Intercomparison Project Phase 6 (CMIP6), available from \cite{cds_cmip6_2021}. We consider projections from 15 CMIP6 climate models, under three SSP scenarios, denoted by $s \in \mathcal{S} = \{\text{SSP2-4.5}, \text{SSP3-7.0}, \text{SSP5-8.5}\}$.\footnote{Details of the CMIP6 climate models are provided in Section~\ref{sec:cmip6_model_list} of the Supplementary Material.} For each combination of climate model and SSP scenario, CMIP6 provides a single deterministic temperature trajectory, which limits our ability to quantify uncertainty in future temperature projections. To account for this uncertainty, we follow the approach proposed by \cite{avanzi2025dynamic}, first applying bias correction to the temperature projections and then generating multiple stochastic temperature trajectories. Further details of the methodology are provided in Appendix~\ref{sec:bias_correction}. For each SSP scenario, we generate 1,000 simulated paths for each of the 15 CMIP6 climate models, resulting in 15,000 simulated paths in total. For visualization, we compute the daily mean, $2.5$th percentile, and $97.5$th percentile across the 15,000 simulated paths for each region and scenario. We then average these daily summaries over each year and apply a 10-year moving average to obtain the smoothed annual series shown in Figure~\ref{fig:ssp_proj}.

Across all 15 NUTS-2 regions, the annual mean daily temperature exhibits an upward trend over time under all three SSP scenarios. Under SSP2-4.5, the annual mean temperature generally increases by about $1$--$2\,\,{}^\circ\mathrm{C}$ by the end of the 21st century. SSP3-7.0 shows a stronger warming pattern, while SSP5-8.5 produces the most substantial increase, with the annual mean daily temperature rising by about $4$--$6\,\,{}^\circ\mathrm{C}$ in most regions by 2100. The divergence across scenarios becomes more pronounced after around 2055, as temperatures under SSP5-8.5 rise more rapidly than those under SSP2-4.5 and SSP3-7.0. The wider projection interval under SSP5-8.5 also indicates greater projection uncertainty under the high-emission scenario. Figure~\ref{fig:ssp_proj} further highlights pronounced regional variation in future temperature exposure across mainland Spain. Regions that are already warmer, particularly southern and inland regions such as ES30, ES42, ES43, ES61, and ES62, maintain higher annual mean daily temperatures throughout the projection period, whereas northern regions, including ES11, ES12, ES13, ES21, and ES22, remain comparatively cooler.

\subsection{Mortality projection by temperature-augmented Hermite mortality model}
\label{sec:mu_projection}
Once the projected temperature trajectories under different SSP scenarios for 2025--2100 are obtained, we consider mortality projections incorporating two sources of uncertainty: temperature projection uncertainty and model parameter uncertainty.
Temperature projection uncertainty refers to uncertainty in future temperature inputs, arising from differences across CMIP6 climate models and from the residual distribution estimated from the bias-corrected historical backcasts. This uncertainty is incorporated by generating $Q$ simulated temperature trajectories for each CMIP6 climate model using Equation~\eqref{temperature-simulation}.
Parameter uncertainty refers to the estimation uncertainty associated with the coefficients of the temperature-augmented Hermite mortality model. Following \cite{guibert2025impact} and \cite{min2026mortality}, we account for this uncertainty by applying a parametric bootstrap to the estimated parameter vector $\hat{\bm{\Theta}} = [\, \, \hat{\bm{\theta}} \quad \hat{\bm{\delta}} \, \,]$, repeated $B$ times. Assuming that the sampling distribution of $\hat{\bm{\Theta}}$ can be approximated by a multivariate normal distribution, we generate the bootstrapped parameter vector $\bm{\Theta}^{*}_b = [ \, \, \bm{\theta}^*_b \quad \bm{\delta}^*_b \, \, ]$, for the temperature-augmented Hermite mortality model as 
\begin{equation}
\label{bootstrap_eq}
\bm{\Theta}^{*}_b
\sim
\mathcal{MVN}\left(
\hat{\bm{\Theta}},
\hat{\bm{\Sigma}}_{\bm{\Theta}}
\right),
\quad b = 1, \dots, B,
\end{equation}
where $\hat{\bm{\Theta}}$ denotes the maximum likelihood estimate of the model parameter vector, and $\hat{\bm{\Sigma}}_{\bm{\Theta}}$ is its estimated asymptotic covariance matrix \citep[see Chapter 6 of][]{wood2017generalized}.
To reduce computational complexity, we set $Q = B$. In this case, each simulation draw for model uncertainty is matched to one simulation draw for scenario uncertainty in each CMIP6 climate model.

\begin{figure}[ht!]
    \centering
    \includegraphics[page=3, width=1\linewidth]{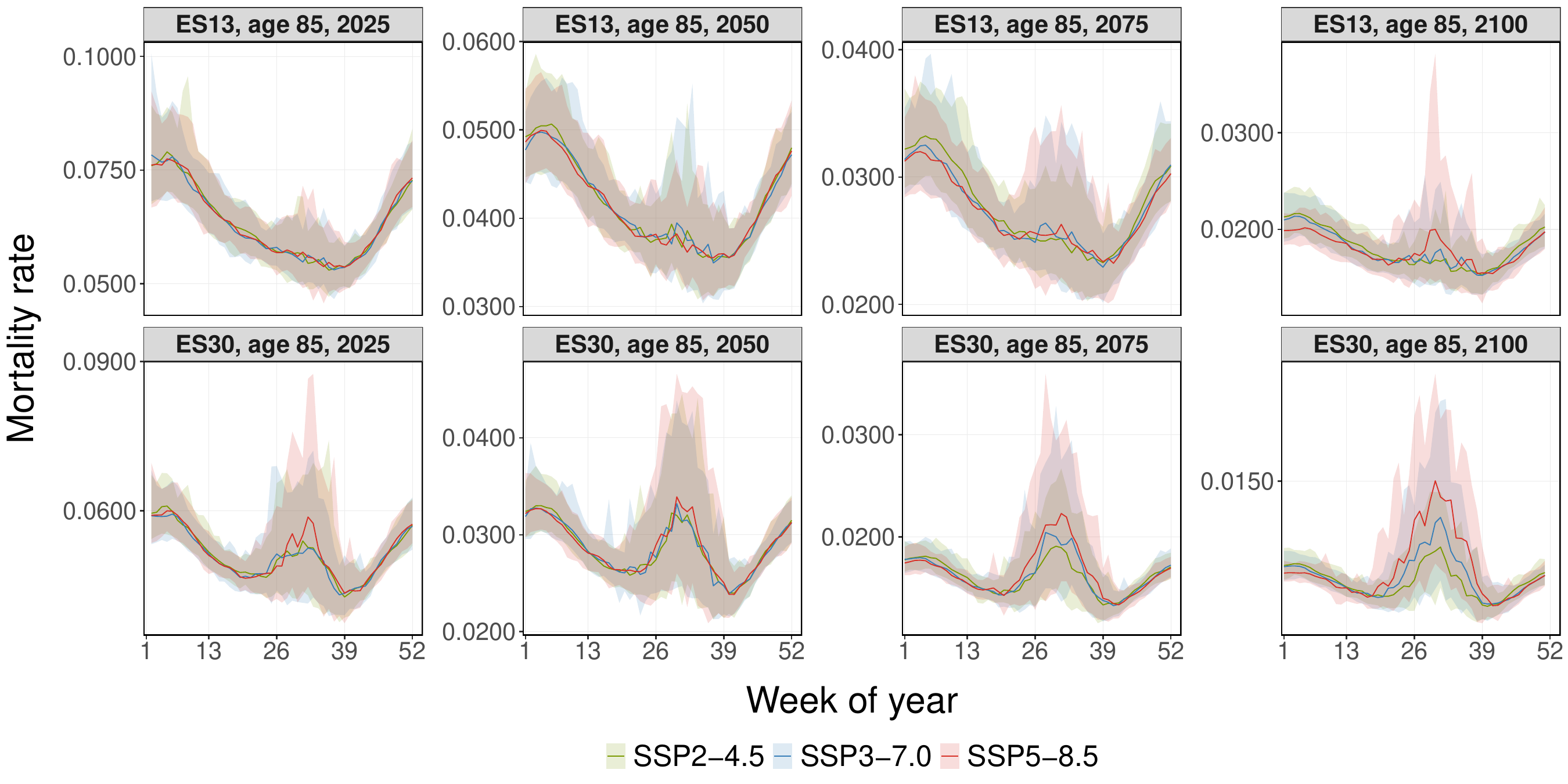}
    \caption{\textbf{Mortality projections for age 85 in ES13 and ES30} \newline \footnotesize
    \emph{Notes}: This figure shows the projected weekly mortality hazards in ES13 and ES30 for the age group 85 under SSP2–4.5, SSP3–7.0, and SSP5–8.5. Each panel presents the seasonal pattern of projected mortality hazards in years 2025, 2050, 2075, and 2100. The solid lines represent the projected mortality hazards, and the shaded areas indicate the corresponding 95\% projection intervals under each SSP scenario.}
    \label{fig:mu_projection}
\end{figure}

Figure~\ref{fig:mu_projection} presents the projected weekly mortality rates at age 85 for Cantabria (ES13) and Madrid (ES30) in 2025, 2050, 2075, and 2100 under the three SSP scenarios. ES13 and ES30 are used as representative examples of relatively less urbanized and more urbanized regions, respectively. For these two regions, the corresponding figures for the remaining age groups are presented in Section~\ref{sec:projection} of the Supplementary Material.\footnote{Results for the other 13 regions across all age groups are available upon request.}

For both regions, we can see that projected mortality at age 85 declines substantially from 2025 to 2100 due to anticipated future mortality improvements. Furthermore, rising temperatures under the SSP scenarios are expected to progressively reshape the seasonal pattern of mortality. Over time, we observe winter compensation effects in both regions: projected mortality during winter becomes relatively lower due to warmer winter conditions under all three SSP scenarios. Conversely, projected mortality during summer becomes relatively higher as temperatures continue to rise, indicating the increasing adverse effects of more frequent and extreme heat waves in summer.

However, it should be noted that the magnitude of this seasonal reshaping differs between the two regions, indicating regional heterogeneity in climate-related mortality risks. Overall, the impact of climate change is more pronounced through the winter compensation effect in ES13, while in ES30, it is more pronounced through the increase in summer mortality.
In ES13, a clear winter compensation effect emerges early in the projection period, with lower winter mortality under SSP5-8.5 than under SSP2-4.5 and SSP3-7.0 due to the warmer winter temperatures. In contrast, the increase in summer mortality is less apparent in the earlier years. As warming continues to intensify toward 2100, however, summer mortality increases more substantially, with SSP5-8.5 eventually producing the highest summer mortality among the three scenarios. A different pattern emerges in ES30. Although we observe a clear winter mortality compensation effect, projected winter mortality remains similar across the three scenarios over most of the projection period, with differences becoming more apparent only by 2100, when SSP5-8.5 results in the lowest winter mortality. The increase in summer mortality emerges more prominently, particularly under SSP5-8.5, and becomes increasingly pronounced over time. The projection intervals also widen substantially around the summer peak, indicating greater uncertainty in summer mortality. By 2075 and 2100, the projected summer mortality peak exceeds the corresponding winter peak under all three scenarios. These observations are largely consistent with our findings in Sections~\ref{sec:Data} and~\ref{sec:Empirical-results}.

\section{Climate impacts on whole life insurance pricing}\label{sec:Pricing}

This section assesses the impact of climate change under the SSP scenarios on a hypothetical whole life insurance portfolio. Based on the projected temperature-related mortality rates in Section~\ref{sec:Scenario-projection}, we evaluate the expected death payoffs for each week of the year and the distribution of expected portfolio P\&L across regions using the simulation results. The analysis provides a quantitative assessment of climate-related mortality risks for insurance pricing.

\subsection{Simulation design}

To evaluate the impact of climate change on a portfolio of whole life insurance policies, we develop a simulation framework that accounts for uncertainty in both temperature projections and model parameter estimates. For each SSP scenario, we generate a total of $N = 15Q$ temperature paths, with $Q$ paths generated for each of the 15 climate models. Each temperature path is then paired one-to-one with a bootstrapped parameter vector generated according to Equation~\eqref{bootstrap_eq}. 

We construct a hypothetical portfolio comprising a total of $M$ whole life insurance policies, which are allocated across the 15 NUTS-2 regions according to their corresponding population weights $w_r$, for $r \in \mathcal{R}$, where $\sum_{r\in\mathcal R} w_r = 1$. The number of policies allocated to region $r$, denoted by $m_r$, is approximately $M w_r$, with minor rounding adjustments to ensure that all allocations are integers. Let $x_0$ denote the age at policy issue and $x_t$ the attained age in week $t$. Since only integer ages are considered in this study, the attained age is given by $x_t=x_0+\lfloor\frac{t-1}{52}\rfloor$.

Table~\ref{tab:assumption} summarizes the key assumptions underlying the simulation design.

\begin{table}[ht!]
\caption{\textbf{Key assumptions for the whole life insurance portfolio}}
{\fontsize{9.5}{10}\selectfont
\centering
\renewcommand{\arraystretch}{1}
\begin{tabularx}{\linewidth}{l X l}
\toprule
Symbol & Definition & Values \\
\midrule
$M$ & Total number of policies across all regions & $1{,}000$ \\
$x_0$ & Age at policy issue & $65$ \\
$\omega$ & Limiting age & $110$ \\
$t_0$& Policy issue year & $2025$ and $2050$ \\
$L$ & Individual death benefit & \$1 million \\
$\iota$ & Annual effective interest rate & 5\% \\
$Q$ & The number of temperature paths for each climate model & $1{,}000$ \\
$N$ & Total number of temperature paths across the 15 climate models& $15{,}000$ \\
\bottomrule
\end{tabularx}%
}
\par\smallskip\footnotesize 
\emph{Notes}: This table lists the key assumptions for the whole life insurance portfolio.
\label{tab:assumption}
\end{table}

\subsection{Expected payoffs}

For $n=1,...,N$, we project the future weekly mortality hazard $\mu_r^{(s,n)}(x_t,t)$ for an individual aged $x_t$ in week $t$ and region $r$ under scenario $s$. The corresponding weekly death probability is then computed as
\begin{equation}
q_r^{(s,n)}(x_t,t) = 1 - \exp\left(-\mu_r^{(s,n)}(x_t,t)\right).
\end{equation}

Based on the projected death probabilities, we calculate the discounted expected death benefit associated with death in week $t$ as
\begin{equation}
l_r^{(s,n)}(x_t,t) = \left(1+\iota\right)^{-\frac{t}{52}} \, L\,  \prod_{j=1}^{t-1} \left(1 - q_r^{(s,n)}(x_j,j)\right) q_r^{(s,n)}(x_t,t),
\end{equation}
where $L$ denotes the death benefits and $\iota$ denotes the annual effective interest rate.

For each temperature path $n$, the expected present value (EPV) of the total portfolio payoff associated with deaths in week $t$ under scenario $s$ is given by
\begin{equation}
L^{(s,n)}(x_t,t) = \sum_{r\in\mathcal{R}}  m_r \, l_r^{(s,n)}(x_t,t).
\end{equation}

To assess the impact of climate change on expected portfolio payoffs, we compare the projected payoffs under each SSP scenario with those obtained from the Hermite baseline model, which does not incorporate temperature information.\footnote{The Hermite baseline model refers to the multi-population Hermite model described in Section \ref{sec:hermite-model}, rather than the Hermite component of the temperature-augmented Hermite model.} Let $L_{\mathcal{H}}^{(n)}(x_t,t)$ denote the corresponding weekly portfolio payoff under the Hermite baseline model for simulation draw $n$.
For scenario $s$, we define the weekly portfolio excess payoff ratio relative to the Hermite baseline model as
\begin{equation}
R^{(s)}(x_t,t) = 100\% \, \frac{\displaystyle \bar{L}^{(s)}(x_t,t)-\displaystyle \bar{L}_{\mathcal{H}}^{}(x_t,t)}{\displaystyle \bar{L}_{\mathcal{H}}^{}(x_t,t)} , 
\end{equation}
where $\bar{L}^{(s)}(x_t,t)$ and $\bar{L}_{\mathcal{H}}^{}(x_t,t)$ denote the average EPVs of the total portfolio payoff across $N$ simulated paths for the temperature-augmented Hermite model under scenario $s$ and the Hermite baseline model, respectively. 

A positive value of $R^{(s)}(x_t,t)$ indicates that, under scenario $s$, the EPV of the death benefit associated with deaths in week $t$ obtained from the temperature-augmented Hermite model exceeds the corresponding value obtained from the Hermite baseline model. Conversely, a negative value indicates a lower EPV relative to the baseline. Note that the Hermite baseline model assumes a continuation of historical mortality trends and seasonal patterns, without explicitly accounting for future changes in climate.

\begin{figure}[ht!]
    \centering
    \includegraphics[width=0.8\linewidth]{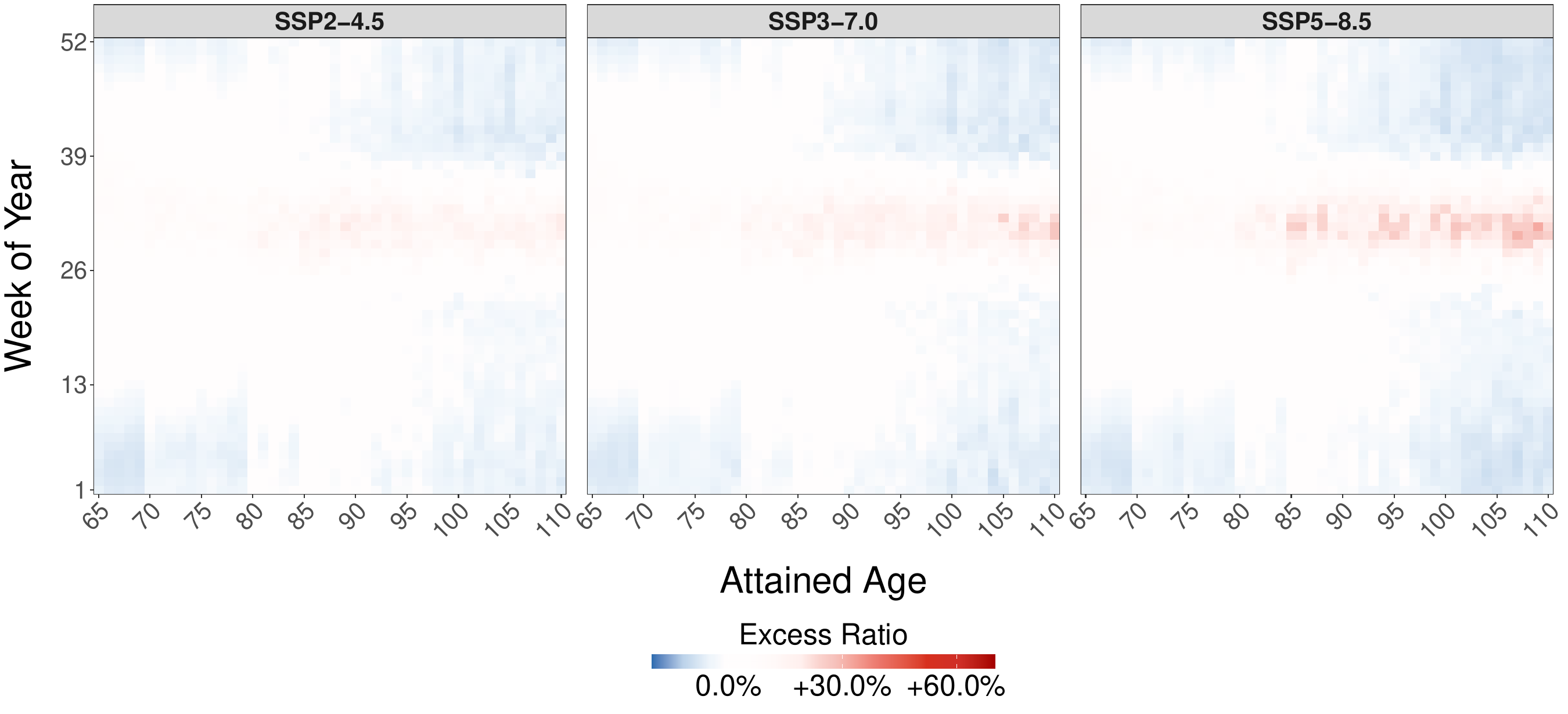}
    \caption{\textbf{Weekly portfolio excess ratio for policies issued in 2025} \newline \footnotesize
    \emph{Notes}: This figure shows the excess payoff ratio by attained age and week of year for policies issued in 2025 under SSP2-4.5, SSP3-7.0, and SSP5-8.5. The excess ratio measures the temperature-related deviation in projected payoff relative to the payoff obtained under the Hermite baseline mortality model. Positive values (red) indicate that the temperature effect increases the payoff relative to the Hermite baseline mortality model, while negative values (blue) indicate that it decreases the payoff.}
    \label{fig:heatmap_2025}
\end{figure}

\begin{figure}[ht!]
    \centering
    \includegraphics[width=0.8\linewidth]{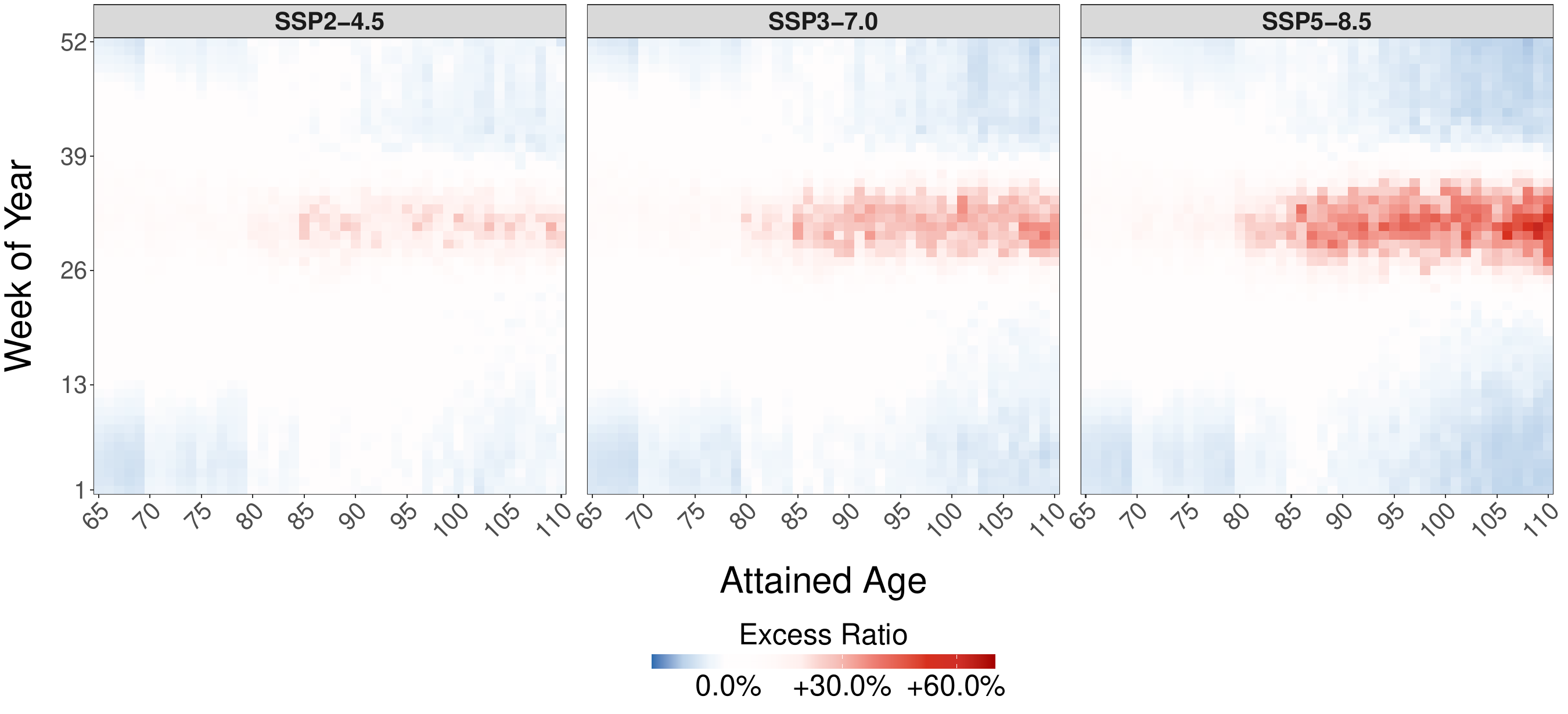}
    \caption{\textbf{Weekly portfolio excess ratio for policy issued in 2050} \newline \footnotesize
    \emph{Notes}: This figure shows the excess payoff ratio by attained age and week of year for policies issued in 2050 under SSP2-4.5, SSP3-7.0, and SSP5-8.5. The excess ratio significantly increases across three SSP scenarios from 2050 due to higher-forcing climate scenarios. }
    \label{fig:heatmap_2050}
\end{figure}

Figures \ref{fig:heatmap_2025} and \ref{fig:heatmap_2050} present the weekly portfolio excess payoff ratios for policies issued in 2025 and 2050, respectively. The heatmaps show that deviations from the Hermite baseline model generally become larger as the climate scenario becomes more severe, progressing from SSP2-4.5 to SSP3-7.0 and SSP5-8.5. A pronounced seasonal pattern is also evident. Positive excess payoff ratios are concentrated around weeks 32--38, particularly at older attained ages, whereas negative ratios are generally observed near the beginning and end of each year.

The positive excess payoff ratios during the warmer part of the year indicate higher projected mortality and, consequently, larger EPVs of portfolio payoffs under the temperature-augmented model relative to the Hermite baseline model. Conversely, the negative excess payoff ratios during the colder part of the year indicate lower projected mortality and correspondingly smaller EPVs of portfolio payoffs.
These seasonal deviations are more pronounced for policies issued in 2050 than for those issued in 2025, particularly under SSP3-7.0 and SSP5-8.5. This is expected because policies issued in 2050 are more exposed to the impact of climate change over their duration.
 The differences across SSP scenarios also become more pronounced over time.
During winter, the negative excess payoff ratios generally increase in magnitude as temperatures rise, reflecting the beneficial effect of milder winter conditions on mortality. In contrast, the positive excess payoff ratios during summer generally increase as the climate scenarios become more severe, reflecting greater adverse effects of higher temperatures on mortality.

These results highlight the importance of incorporating temperature-related mortality effects when projecting future mortality and associated life insurance payoffs. A mortality model that does not incorporate future climate information may overlook important changes in the magnitude and seasonal pattern of projected portfolio payoffs, thereby providing an incomplete assessment of future mortality risk. To formally assess the impact of climate change on portfolio performance, we next conduct a P\&L analysis based on the simulation results.

\subsection{Profit and loss analysis}

This study assumes that insurers determine the portfolio premium according to the pure premium principle. For each SSP scenario $s$, the premium is set equal to the average portfolio payoff EPV across the $N$ simulated paths. This scenario-specific premium is then held fixed and compared with the portfolio payoff EPV under each individual path $n$, generating a distribution of portfolio P\&L across the $N$ paths.

The EPV at issuance of the payoff from an individual policy in region $r$ is given by
\begin{equation}
l_r^{(s,n)}(x_0) = \sum_{t \in \mathcal{T}} l_r^{(s,n)}(x_t,t).
\end{equation}

We further compute the EPV of the total life insurance portfolio payoff across all regions as
\begin{equation}
L^{(s,n)}(x_0) = \sum_{r \in \mathcal{R}} m_r l_r^{(s,n)}(x_0).
\end{equation}

The scenario-specific portfolio pure premium is then obtained by averaging the portfolio payoff EPVs across $N$ paths under scenario $s$, and is given by
\begin{equation}
P^{(s)}(x_0) = \frac{1}{N} \sum_{n=1}^N L^{(s,n)}(x_0)=\bar{L}^{(s,n)}(x_0).
\end{equation}

Using this scenario-specific portfolio pure premium, insurers can evaluate the distribution of portfolio P\&L across the $N$ simulated paths. For each path $n$ under scenario $s$, the portfolio P\&L is given by
\begin{equation}
\Pi^{(s,n)}(x_0) = P^{(s)}(x_0) - L^{(s,n)}(x_0).
\end{equation}

\begin{figure}[ht!]
    \centering
    \includegraphics[width=1\linewidth]{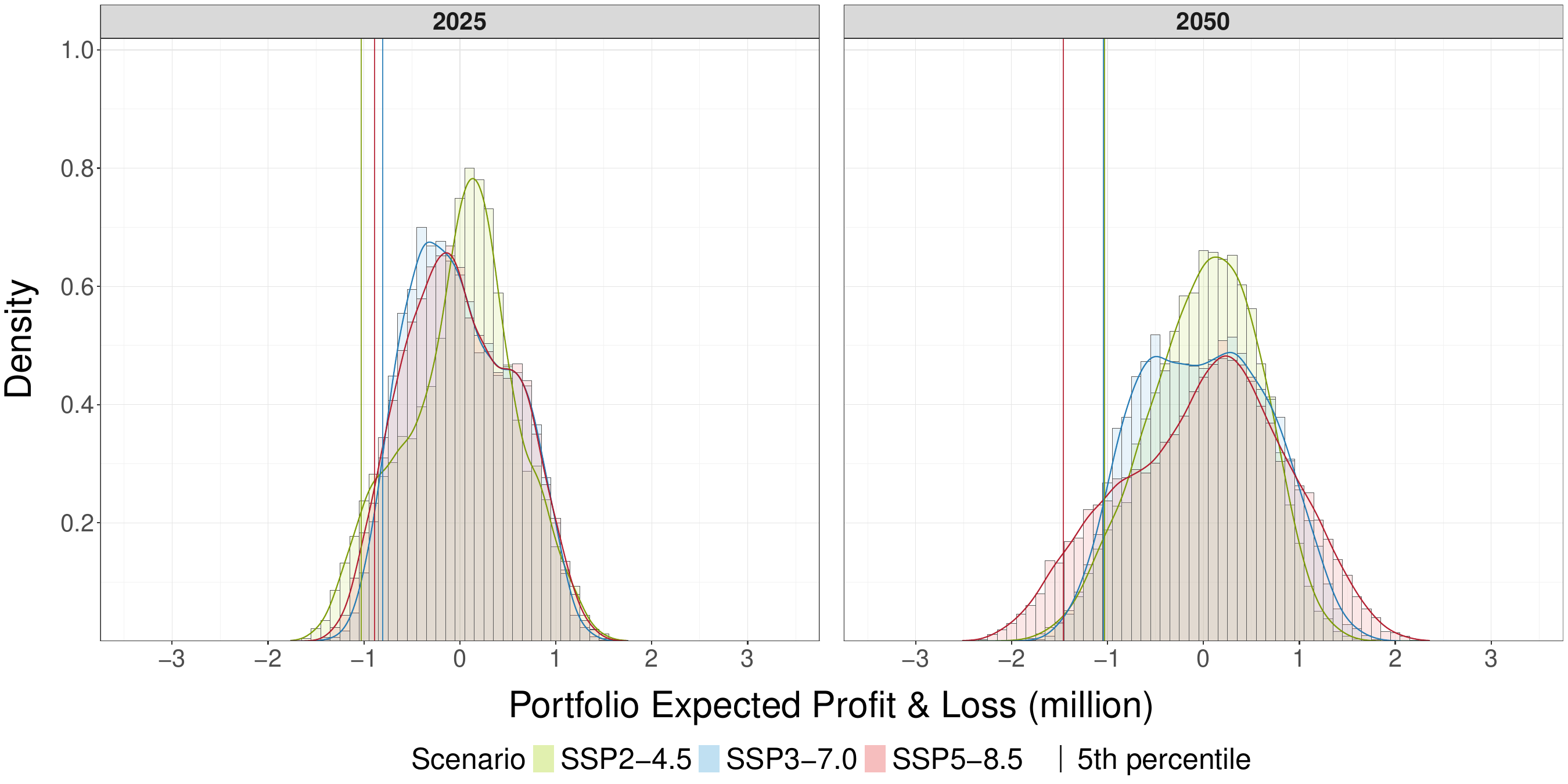}
    \caption{\textbf{The distribution of expected portfolio P\&L in issue year 2025 and 2050} \newline \footnotesize
    \emph{Notes}: This figure shows the simulated portfolio P\&L present value distributions for issue age 65 in issue year 2025 and 2050 under the three SSP scenarios. The histograms show the empirical simulated distributions, the smooth curves show the corresponding kernel density estimates, and the vertical lines indicate the 5th percentile.}
    \label{fig:histogram}
\end{figure}

Since the pure premium is determined by averaging the portfolio payoff EPVs across the same $N$ paths, the average portfolio P\&L across these paths is zero by construction. The P\&L analysis therefore focuses on the downside and upside in portfolio performance across the simulated paths. Figure \ref{fig:histogram} presents the distributions of portfolio P\&L based on the simulation results, together with their 5th percentiles, for policies issued in 2025 and 2050 under the three SSP scenarios.

For policies issued in 2025, the three P\&L distributions overlap substantially, indicating relatively modest differences across the SSP scenarios. However, the 5th percentile under SSP2-4.5 corresponds to a larger portfolio loss than those under SSP3-7.0 and SSP5-8.5. This may be explained by the winter compensation effect. In the near term, the reduction in winter mortality under warmer climate scenarios is expected to outweigh the increase in mortality associated with summer heat. As a result, projected mortality and the EPV of death benefit payments are lower under SSP3-7.0 and SSP5-8.5. The loss at the 5th percentile is also slightly smaller under SSP3-7.0 than under SSP5-8.5, although the difference is modest. This may reflect the stronger increase in summer mortality under SSP5-8.5, which partially offsets the benefit of lower winter mortality.

For policies issued in 2050, the differences among the three P\&L distributions become more evident. As the projection horizon approaches 2100, the increase in summer mortality appears to outweigh the reduction in winter mortality associated with warmer conditions. This is consistent with the changing seasonal patterns of mortality shown earlier in Figure \ref{fig:mu_projection}. The mortality projections also become more uncertain under SSP5-8.5.
These features translate into greater variation in portfolio outcomes. The P\&L distribution under SSP5-8.5 is noticeably wider and flatter than those under SSP2-4.5 and SSP3-7.0, indicating greater dispersion and more probability mass away from the center of the distribution. The distribution also exhibits slight negative skewness and the largest loss at the 5th percentile. Taken together, these features indicate greater tail risk, particularly on the loss side, and a higher potential for severe portfolio losses under SSP5-8.5.

\begin{table}[ht!]
\caption{\textbf{Summary statistics of portfolio expected P\&L across systematic paths}}
{\fontsize{9.5}{10}\selectfont
\centering
\begin{tabularx}{\linewidth}{X *{6}{d{4.7}}}
\toprule
\multirow{2}{*}{Statistic}
& \multicolumn{3}{c}{2025}
& \multicolumn{3}{c}{2050} \\
\cmidrule(lr){2-4}
\cmidrule(lr){5-7}
& \multicolumn{1}{c}{SSP2-4.5} & \multicolumn{1}{c}{SSP3-7.0} & \multicolumn{1}{c}{SSP5-8.5}
& \multicolumn{1}{c}{SSP2-4.5} & \multicolumn{1}{c}{SSP3-7.0} & \multicolumn{1}{c}{SSP5-8.5} \\

\midrule
Variance
& 0.3363 & 0.2890 & 0.3191
& 0.3401 & 0.4453 & 0.7122 \\[2pt]

$Q_{1}$
& -1.2954 & -1.0198 & -1.1118
& -1.3898 & -1.3312 & -1.8639 \\[2pt]

$Q_{2.5}$
& -1.1679 & -0.9123 & -1.0051
& -1.1981 & -1.1904 & -1.6491 \\[2pt]

$Q_{5}$
& -1.0287 & -0.8067 & -0.8887
& -1.0289 & -1.0427 & -1.4607 \\[2pt]

$Q_{10}$
& -0.8413 & -0.6800 & -0.7356
& -0.7970 & -0.8748 & -1.1889 \\
\bottomrule
\end{tabularx}%
}
\par\smallskip\footnotesize
\emph{Notes}: This table reports the variance, and the $p$th percentiles $Q_{p}$ for $p = \{1,\ 2.5, \ 5, \ 10 \}$ of portfolio expected P\&L across 15,000 systematic paths under each SSP scenario. 
\label{tab:summary_stat}
\end{table}

Table~\ref{tab:summary_stat} complements Figure~\ref{fig:histogram} by reporting the variance and percentiles $Q_p$, where $p=\{1,2.5,5,10\}$, of the portfolio P\&L distribution. For policies issued in 2025, SSP2-4.5 has the highest variance ($0.3363$) and the largest losses across these percentiles. SSP3-7.0 and SSP5-8.5 show lower variance and smaller losses, consistent with the winter compensation effect moderating downside risk under the warmer scenarios.
For policies issued in 2050, the differences across SSP scenarios are more pronounced. The variance increases from $0.3401$ under SSP2-4.5 to $0.4453$ under SSP3-7.0 and $0.7122$ under SSP5-8.5, indicating greater dispersion as the climate scenario becomes more severe. SSP5-8.5 also produces the largest losses across all reported percentiles. This is consistent with the increasing dominance of adverse summer mortality effects over winter compensation, resulting in greater downside risk under SSP5-8.5.

Overall, the P\&L analysis shows that climate change can materially affect the distribution and downside risk of life insurance portfolio outcomes, even when premiums are set conditional on the same climate scenario used to generate the future mortality outcomes. These effects become increasingly important under higher-forcing climate scenarios, where greater uncertainty in future mortality translates into wider P\&L distributions and more adverse loss outcomes. The differences are also larger for policies issued in later years, highlighting the increasing relevance of climate-related mortality risk for longer-term life insurance liabilities. From an actuarial perspective, these findings highlight the importance of incorporating climate-related mortality effects into long-term liability projections, pricing, and risk assessment, particularly when evaluating exposure to adverse tail outcomes.

\section{Concluding remarks}\label{sec:Conclusion}

Climate change poses an important challenge for life insurers because long-term insurance liabilities depend on mortality assumptions that may themselves evolve as temperatures rise. Assessing this risk requires more than projecting aggregate mortality: insurers need mortality estimates at sufficiently granular ages and frequencies to translate climate scenarios into the timing and magnitude of future benefit payments. In this article, we develop a temperature-augmented Hermite mortality model for this purpose and use it to assess the implications of climate-related mortality changes for a whole life insurance portfolio.

Our empirical analysis highlights an important interaction between rising temperatures and the seasonality of mortality. Warmer conditions reduce cold-related mortality during winter but increase heat-related mortality during summer, with the relative importance of these two effects changing over time and across climate scenarios. The proposed framework captures these effects while allowing their magnitude to vary across ages and regions. This granularity is particularly relevant for life insurance applications, since changes in the timing of deaths directly affect the present value of future death-benefit payments.

The life insurance application shows that these climate-related changes in mortality can have material consequences for portfolio outcomes. In the near term, reductions in winter mortality can partly offset, and in some cases dominate, increases in summer mortality. Over longer horizons, however, the adverse effects of higher summer temperatures become increasingly important. For policies issued in 2050, differences across climate scenarios are substantially more pronounced, and the most severe scenario produces a wider P\&L distribution and greater downside risk. Importantly, this additional risk remains even when the insurer sets the pure premium using the same climate scenario from which future mortality outcomes are generated. Climate change therefore affects not only the expected level and seasonal timing of benefit payments, but also the uncertainty surrounding those payments.

These findings have direct implications for actuarial practice. Mortality projections that extrapolate historical trends and regular seasonal patterns without explicitly accounting for future temperature changes may provide an incomplete picture of the risks embedded in long-duration life insurance liabilities. Climate-related mortality risk should therefore be considered not only when constructing long-term mortality assumptions, but also when pricing life insurance products, assessing portfolio risk, and evaluating exposure to adverse outcomes. The increasing differences observed for policies issued at later dates further suggest that this issue is likely to become more important as the climate evolves and insurers write liabilities extending further into the future.

Several extensions would further strengthen this actuarial assessment. Future research could incorporate additional climate variables and extreme-weather measures, allow socioeconomic characteristics and adaptation to influence future mortality responses, and investigate how climate-related mortality uncertainty interacts with reserving, capital requirements, reinsurance, and other risk-management strategies. More broadly, our results illustrate how climate-sensitive mortality modeling can provide a bridge between physical climate scenarios and the financial risks ultimately borne by life insurers.

\section*{Data and code availability statement}

The datasets used in this article are publicly available. We collected weekly deaths by 5-year age group and selected NUTS-2 regions in Spain from \href{https://ec.europa.eu/eurostat/databrowser/view/demo_r_mwk2_05/default/table}{Eurostat}, downloaded on November 21, 2025. Annual population data by 5-year age group and selected NUTS 2 regions in Spain were also collected from \href{https://ec.europa.eu/eurostat/databrowser/view/demo_r_d2jan/default/table}{Eurostat}, downloaded on November 21, 2025. We retrieved historical temperature data from ERA5 reanalysis through the \href{https://cds.climate.copernicus.eu/datasets/derived-era5-single-levels-daily-statistics}{Copernicus Climate Data Store}, downloaded on November 25, 2025, and future SSP scenario-based temperature data from the CMIP6 model through the \href{https://cds.climate.copernicus.eu/datasets/projections-cmip6}{Copernicus Climate Data Store}, downloaded on July 6, 2026. All analyses were carried out using \texttt{R}.

\vspace{0.25cm}
\setlength{\bibsep}{6pt}
\titlespacing*{\section}{0pt}{6pt}{12pt}
\begin{spacing}{0.0}
\bibliographystyle{apa-good}
\bibliography{reference}

@article{dong2022air,
  title={Air pollution and mortality impacts},
  author={Dong, Zhe Michelle and Shang, Han Lin and Bruhn, Aaron},
  journal={Risks},
  volume={10},
  number={6},
  pages={126},
  year={2022},
  publisher={MDPI}
}

@article{li2025beyond,
  title={Beyond annual data: Mortality forecasting with mixed frequency data},
  author={Li, Runze and Zhou, Rui and Pitt, David},
  journal={Insurance: Mathematics and Economics},

  year={2025},
  publisher={Elsevier},
  volume = {126},
  pages = {103172},
}

@article{arik2025measuring,
  title={On measuring {COVID-19} excess mortality: Insights and challenges},
  author={Arik, Ayse and Klein, Allen and Li, Han},
  journal={Insurance: Mathematics and Economics},
  pages={103199},
  year={2025},
  publisher={Elsevier}
}

@article{hanebeck2026dependence,
  title={Dependence modelling across major causes of death via time-varying copula state space models},
  author={Hanebeck, Ariane and Li, Han and Czado, Claudia},
  journal={Journal of the Royal Statistical Society Series C},
  volume={75},
  number={2},
  pages={407--430},
  year={2026},
  publisher={Oxford University Press UK}
}

@article{miao2026gradient,
  title={Gradient boosted multi-population mortality modeling with high-frequency data},
  author={Miao, Ziting and Li, Han and Chen, Yuyu},
  journal={ASTIN Bulletin},
  pages={1--26},
  year={2026},
  publisher={Cambridge University Press}
}

@article{ofosuhene2026modelling,
  author = {Ofosu-Hene, Eric Dei and Carr, Peter and Harris, Jonathan and Morris, Richard and Tavener, Chris and Horvat-Gitsels, Lisanne},
  title = {{Modelling climate-related mortality and morbidity risks in the {UK}: Insights from a practitioner survey}},
  journal = {British Actuarial Journal},
  year = {2026},
  volume = {31},
  eid = {e16},
  pages = {1--31},
  doi = {10.1017/S1357321726100476}
}

@techreport{bankofengland2021a,
  author = {{Bank of England}},
  title = {{Climate Biennial Exploratory Scenario 2021: Financial risks from climate change}},
  year = {2021},
  institution = {Prudential Regulation Authority}
}

@article{boudreault2023changing,
  author = {Boudreault, M. and Clacher, I. and {Siu-Hang, Li} and Pigott, C. and Zhou, R.},
  title = {{A changing climate for actuarial science}},
  journal = {Annals of Actuarial Science},
  year = {2023},
  volume = {17},
  pages = {415--419},
  doi = {10.1017/S1748499523000222}
}

@article{miljkovic2018examining,
  author = {Miljkovic, T. and Miljkovic, D. and Maurer, K.},
  title = {Examining the impact on mortality arising from climate change: {I}mportant findings for the insurance industry},
  journal = {European Actuarial Journal},
  year = {2018},
  volume = {8},
  pages = {363--381},
  doi = {10.1007/s13385-018-0178-2}
}

@article{romanello2024countdown,
  title={The 2024 report of the Lancet Countdown on health and climate change: facing record-breaking threats from delayed action},
  author={Romanello, Marina and Walawender, Maria and Hsu, Shih-Che and Moskeland, Annalyse and Palmeiro-Silva, Yasna and Scamman, Daniel and Ali, Zakari and Ameli, Nadia and Angelova, Denitsa and Ayeb-Karlsson, Sonja and others},
  journal={The Lancet},
  volume={404},
  number={10465},
  pages={1847--1896},
  year={2024},
  publisher={Elsevier}
}

@article{heutel2021adaptation,
  title={Adaptation and the mortality effects of temperature across {US} climate regions},
  author={Heutel, Garth and Miller, Nolan H and Molitor, David},
  journal={Review of Economics and Statistics},
  volume={103},
  number={4},
  pages={740--753},
  year={2021},
  publisher={MIT Press One Rogers Street, Cambridge, MA 02142-1209, USA journals-info~…}
}

@article{hajdu2026impact,
  title={The impact of heat and cold on mortality and life expectancy in {E}urope, 2015--2024},
  author={Hajdu, Tam{\'a}s},
  journal={Economics \& Human Biology},
  pages={101607},
  volume={61},
  year={2026},
  publisher={Elsevier}
}

@article{tillett1983excess,
  title={Excess deaths attributable to influenza in {E}ngland and {Wales}: {A}ge at death and certified cause},
  author={Tillett, HILARY E and Smith, JWG and Gooch, CD},
  journal={International Journal of Epidemiology},
  volume={12},
  number={3},
  pages={344--352},
  year={1983},
  publisher={Oxford University Press}
}

@article{aastrom2011heat,
  title={Heat wave impact on morbidity and mortality in the elderly population: A review of recent studies},
  author={{\AA}str{\"o}m, Daniel Oudin and Bertil, Forsberg and Joacim, Rockl{\"o}v},
  journal={Maturitas},
  volume={69},
  number={2},
  pages={99--105},
  year={2011},
  publisher={Elsevier}
}

@article{hermite1878sur,
  author  = {Hermite, Charles},
  title   = {Sur la formule d'interpolation de {L}agrange},
  journal = {Journal f{\"u}r die reine und angewandte Mathematik},
  volume  = {1878},
  number  = {84},
  pages   = {70--79},
  year    = {1878},
  doi     = {10.1515/crelle-1878-18788405}
}

@book{kreyszig1999advanced,
  author    = {Kreyszig, Erwin},
  title     = {Advanced Engineering Mathematics},
  publisher = {John Wiley \& Sons},
  address   = {New York, NY, United States of America},
  year      = {1999}
}

@article{schoenberg1973cardinal,
  title={Cardinal interpolation and spline functions V. The {B}-splines for cardinal {H}ermite interpolation},
  author={Schoenberg, IJ and Sharma, Ambikeshwar},
  journal={Linear Algebra and its Applications},
  volume={7},
  number={1},
  pages={1--42},
  year={1973},
  publisher={Elsevier}
}

@article{hartley1958maximum,
  title={Maximum likelihood estimation from incomplete data},
  author={Hartley, Herman O},
  journal={Biometrics},
  volume={14},
  number={2},
  pages={174--194},
  year={1958},
  publisher={JSTOR}
}

@article{akaike1973information,
  title={A new look at the statistical model identification},
  author={Akaike, Hirotugu},
  journal={IEEE Transactions on Automatic Control},
  volume={19},
  number={6},
  pages={716--723},
  year={1974},
  publisher={Ieee}
}

@article{dempster1977maximum,
  title={Maximum likelihood from incomplete data via the {EM} algorithm},
  author={Dempster, Arthur P and Laird, Nan M and Rubin, Donald B},
  journal={Journal of the Royal Statistical Society Series B},
  volume={39},
  number={1},
  pages={1--22},
  year={1977},
  publisher={Wiley Online Library}
}

@article{schwarz1978estimating,
  title={Estimating the dimension of a model},
  author={Schwarz, Gideon},
  journal={Annals of Statistics},
  pages={461--464},
  year={1978},
  publisher={JSTOR}, 
  volume={6},
  number={2}
}

@article{dominici2002air,
  title={Air pollution and mortality: Estimating regional and national dose-response relationships},
  author={Dominici, Francesca and others},
  journal={Journal of the American Statistical Association},
  volume={97},
  number={457},
  pages={100--111},
  year={2002},
  publisher={Taylor \& Francis}
}

@article{piani2010statistical,
  title={Statistical bias correction of global simulated daily precipitation and temperature for the application of hydrological models},
  author={Piani, Claudio and Weedon, GP and Best, M and Gomes, SM and Viterbo, Pedro and Hagemann, Stefan and Haerter, JO},
  journal={Journal of Hydrology},
  volume={395},
  number={3-4},
  pages={199--215},
  year={2010},
  publisher={Elsevier}
}

@article{gasparrini2010distributed,
  title={Distributed lag non-linear models},
  author={Gasparrini, Antonio and Armstrong, Ben and Kenward, Mike G},
  journal={Statistics in Medicine},
  volume={29},
  number={21},
  pages={2224--2234},
  year={2010},
  publisher={Wiley Online Library}
}

@article{gasparrini2011impact,
  title={The impact of heat waves on mortality},
  author={Gasparrini, Antonio and Armstrong, Ben},
  journal={Epidemiology},
  volume={22},
  number={1},
  pages={68--73},
  year={2011},
  publisher={LWW}
}

@article{gasparrini2011distributed,
  title={Distributed lag linear and non-linear models in \texttt{R}: The package \texttt{dlnm}},
  author={Gasparrini, Antonio},
  journal={Journal of Statistical Software},
  volume={43},
  pages={1--20},
  year={2011}
}

@article{cheng2014temperature,
  title={Temperature variation between neighboring days and mortality: A distributed lag non-linear analysis},
  author={Cheng, Jian and Zhu, Rui and Xu, Zhiwei and Xu, Xiangqing and Wang, Xu and Li, Kesheng and Su, Hong},
  journal={International Journal of Public Health},
  volume={59},
  pages={923--931},
  year={2014},
  publisher={Springer}
}

@article{armstrong2006models,
  title={Models for the relationship between ambient temperature and daily mortality},
  author={Armstrong, Ben},
  journal={Epidemiology},
  volume={17},
  number={6},
  pages={624--631},
  year={2006},
  publisher={LWW}
}

@book{wood2017generalized,
  title={Generalized Additive Models: {An} Introduction with R},
  author={Wood, Simon N},
  year={2017},
  publisher={CRC Press},
  address={New York, NY, United States of America}
}

@article{guo2018quantifying,
  title={Quantifying excess deaths related to heatwaves under climate change scenarios: A multicountry time series modelling study},
  author={Guo, Yuming and others},
  journal={PLoS Medicine},
  volume={15},
  number={7},
  pages={e1002629},
  year={2018},
  publisher={Public Library of Science San Francisco, CA USA}
}

@article{armstrong2019role,
  author = {Armstrong, Ben and others},
  title = {The role of humidity in associations of high temperature with mortality: A multicountry, multicity study},
  journal = {Environmental Health Perspectives},
  year    = {2019},
  volume  = {127},
  number  = {9},
  pages   = {1--8}
}

@article{richards2020modelling,
  title={Modelling seasonal mortality with individual data},
  author={Richards, Stephen J and Ramonat, Stefan J and Vesper, Gregory T and Kleinow, Torsten},
  journal={Scandinavian Actuarial Journal},
  volume={2020},
  number={10},
  pages={864--878},
  year={2020},
  publisher={Taylor \& Francis}
}

@article{richards2020hermite,
  title={A {H}ermite-spline model of post-retirement mortality},
  author={Richards, Stephen J},
  journal={Scandinavian Actuarial Journal},
  volume={2020},
  number={2},
  pages={110--127},
  year={2020},
  publisher={Taylor \& Francis}
}

@article{masson2021climate,
  title={Climate change 2021: the physical science basis},
  author={Masson-Delmotte, Val{\'e}rie and Zhai, Panmao and Pirani, Anna and Connors, Sarah L and P{\'e}an, Clotilde and Berger, Sophie and Caud, Nada and Chen, Yang and Goldfarb, Leah and Gomis, Melissa I and others},
  journal={Contribution of working group I to the sixth assessment report of the intergovernmental panel on climate change},
  volume={2},
  number={1},
  pages={2391},
  year={2021},
  publisher={Geneva, Switzerland}
}

@article{qian2021projecting,
  title={Projecting health impacts of future temperature: A comparison of quantile-mapping bias-correction methods},
  author={Qian, Weijia and Chang, Howard H},
  journal={International Journal of Environmental Research and Public Health},
  volume={18},
  number={4},
  pages={1992},
  year={2021},
  publisher={MDPI}
}

@misc{cds_cmip6_2021,
  author       = {{Copernicus Climate Change Service}},
  year         = {2021},
  title        = {{CMIP6} Climate Projections},
  howpublished = {Climate Data Store},
  note         = {available at \url{https://doi.org/10.24381/cds.c866074c}. Accessed July 6, 2026}
}

@article{li2022joint,
  title={Joint extremes in temperature and mortality: A bivariate {POT} approach},
  author={Li, Han and Tang, Qihe},
  journal={North American Actuarial Journal},
  volume={26},
  number={1},
  pages={43--63},
  year={2022},
  publisher={Taylor \& Francis}
}

@article{tang2023hermite,
  title={A {H}ermite spline approach for modelling population mortality},
  author={Tang, Sixian and Li, Jackie and Tickle, Leonie},
  journal={Annals of Actuarial Science},
  volume={17},
  number={2},
  pages={243--284},
  year={2023}
}

@article{madaniyazi2022assessing,
  title={Assessing seasonality and the role of its potential drivers in environmental epidemiology: A tutorial},
  author={Madaniyazi, Lina and others},
  journal={International Journal of Epidemiology},
  volume={51},
  number={5},
  pages={1677--1686},
  year={2022},
  publisher={Oxford University Press}
}

@article{wen2023new,
  title={A new method to separate the impacts of interday and intraday temperature variability on mortality},
  author={Wen, Bo and Wu, Yao and Guo, Yuming and Li, Shanshan},
  journal={BMC Medical Research Methodology},
  volume={23},
  number={1},
  pages={92},
  year={2023},
  publisher={Springer}
}

@article{li2023pricing,
  title={Pricing extreme mortality risk in the wake of the COVID-19 pandemic},
  author={Li, Han and Liu, Haibo and Tang, Qihe and Yuan, Zhongyi},
  journal={Insurance: Mathematics and Economics},
  volume={108},
  pages={84--106},
  year={2023},
  publisher={Elsevier}
}

@article{madaniyazi2024seasonality,
  title={Seasonality of mortality under climate change: A multicountry projection study},
  author={Madaniyazi, Lina and others},
  journal={The Lancet Planetary Health},
  volume={8},
  number={2},
  pages={e86--e94},
  year={2024},
  publisher={Elsevier}
}

@article{brouhns2002poisson,
  title={A {P}oisson log-bilinear regression approach to the construction of projected lifetables},
  author={Brouhns, Natacha and Denuit, Michel and Vermunt, Jeroen K},
  journal={Insurance: Mathematics and Economics},
  volume={31},
  number={3},
  pages={373--393},
  year={2002},
  publisher={Elsevier}
}

@article{renshaw2006cohort,
  title={A cohort-based extension to the {L}ee--{C}arter model for mortality reduction factors},
  author={Renshaw, Arthur E and Haberman, Steven},
  journal={Insurance: Mathematics and Economics},
  volume={38},
  number={3},
  pages={556--570},
  year={2006},
  publisher={Elsevier}
}

@article{guibert2025impact,
  title={Impact of climate change on mortality: An extrapolation of temperature effects based on time series data in {F}rance},
  author={Guibert, Quentin and Pincemin, Ga{\"e}lle and Planchet, Fr{\'e}d{\'e}ric},
  journal={International Journal of Forecasting},
  year    = {2026},
  volume   = {42},
  number   = {2},
  pages    = {359--413},
  doi      = {10.1016/j.ijforecast.2025.07.004}
}

@misc{era5_2025,
  author       = {{Copernicus Climate Change Service}},
  year         = {2025},
  title        = {{ERA5} Hourly Time-Series Data on Single Levels from 1940 to Present},
  howpublished = {Climate Data Store},
  note         = {Available at \url{https://doi.org/10.24381/1cf1ad76}. Accessed November 25, 2025}
}

@article{robben2025penalized,
  title={A penalized distributed lag non-linear {L}ee--{C}arter framework for regional weekly mortality forecasting},
  author={Robben, Jens and Barigou, Karim},
  journal={Working Paper, \emph{available at \url{https://arxiv.org/abs/2509.24087}}},
  year={2025}
}

@article{begin2025modelling,
  title={Modelling seasonal mortality: An age--period--cohort approach},
  author={B{\'e}gin, Jean-Fran{\c{c}}ois and Boudreault, Mathieu and Landry, Thomas},
  journal={Insurance: Mathematics and Economics},
  pages={103162},
  year={2025},
  volume={125},
  publisher={Elsevier}
}

@article{begin2026modelling,
  title={Modelling the impacts of climate change on deaths caused by heat and cold waves with age–period–cohort models},
  author={B{\'e}gin, Jean-Fran{\c{c}}ois and Boudreault, Mathieu and Landry, Thomas},
  journal={Working Paper, \emph{available at \url{https://doi.org/10.13140/RG.2.2.20313.89442}}},
  year={2026}
}

@article{wangevaluating,
  title={Evaluating Climate Change Impacts on Mortality, Life Insurance, and Annuities},
  author={Wang, Lintao and Zhou, Rui and King, Andrew and Chen, Ping},
  journal={Working Paper, \emph{available at \url{https://papers.ssrn.com/sol3/papers.cfm?abstract_id=5253347}}},
  year={2025}
}

@article{avanzi2025dynamic,
  title={Dynamic Financial Analysis ({DFA}) of general insurers under climate change},
  author={Avanzi, Benjamin and Li, Yanfeng and Taylor, Greg and Wong, Bernard},
  journal={ASTIN Bulletin},
  volume={Forthcoming},
  year={2026},
  publisher={Cambridge University Press}
}

@article{li2026modeling,
  title={Modeling cold-related excess deaths via stationary vine copulas},
  author={Li, Han and Nagler, Thomas and Czado, Claudia},
  journal={Scandinavian Actuarial Journal},
  volume={Forthcoming},
  year={2026},
  publisher={Taylor \& Francis}
}

@article{arandjelovic2026impact,
  title={The impact of climate change on reserves in life insurance},
  author={Arandjelovi{\'c}, Aleksandar and Shevchenko, Pavel V},
  journal={European Actuarial Journal},
  volume={\emph{Forthcoming}},
  year={2026},
  publisher={Springer}
}

@article{min2026mortality,
  title={Mortality forecasting under climate risk: A stochastic approach with distributed lag nonlinear models},
  author={Min, Jiacheng and Li, Han and Nagler, Thomas and Li, Shuanming},
  journal={Journal of the Royal Statistical Society Series A},
  volume={\emph{Forthcoming}},
  year={2026},
  publisher={Oxford University Press UK}
}

@article{kouton2026pricing,
  title={Pricing life insurance using bivariate temperature-mortality seasonal hidden Markov models},
  author={Kouton, Samuel Darwin Dona and Barigou, Karim and Nguyen, Thai},
  journal={Working Paper, \emph{available at \url{https://papers.ssrn.com/sol3/papers.cfm?abstract_id=6837878}}},
  year={2026}
}

@article{so2026climate,
  title={Climate-Driven Mortality Forecasting Using Deep Learning},
  author={So, Kenrick and Barigou, Karim and Robben, Jens},
  journal={Working Paper, \emph{available at \url{https://arxiv.org/abs/2606.26980}}},
  year={2026}
}

@article{begin2024modeling,
  title={Modeling and forecasting subnational mortality in the presence of aggregated data},
  author={B{\'e}gin, Jean-Fran{\c{c}}ois and Sanders, Barbara and Xu, Xueyi},
  journal={North American Actuarial Journal},
  volume={28},
  number={4},
  pages={882--908},
  year={2024},
  publisher={Taylor \& Francis}
}
\end{spacing}

\titleformat{\section}{\normalfont\bfseries}{\thesection}{1em}{}
\titleformat{\subsection}{\normalfont\bfseries}{\thesubsection}{1em}{}
\titleformat{\subsubsection}{\normalfont\bfseries}{\thesubsubsection}{1em}{}

\titlespacing*{\section}{0pt}{6pt}{6pt}
\titlespacing*{\subsection}{0pt}{6pt}{6pt}
\titlespacing*{\subsubsection}{0pt}{6pt}{6pt}

\newpage
\section*{Appendices}
\setcounter{section}{0}
\renewcommand{\thesection}{\Alph{section}}
\renewcommand{\theHsection}{appendix.\Alph{section}}
\section{Estimation of temperature-augmented Hermite mortality model}\label{app:Estimation}

This appendix discusses the mathematical derivations of the iterative estimation procedure of the temperature-augmented Hermite mortality model. As explained in Section~\ref{sec:estimation}, we estimate the model using multiple rounds to ensure convergence, similar to the methods used in \citet{robben2025penalized} and \citet{begin2026modelling}. 

\subsection{Overview of the estimation}

We estimate the parameters using a two-stage approach: in the first stage, we estimate the parameter set $\bm{\theta}$, subject to $\bm{\delta} = \bm{1}$. In the second stage, we estimate $\bm{\delta}$ by maximizing the Poisson log-likelihood under the model specified in Equation~\eqref{hermite_delta_dlnm} by fixing the estimated parameter set from the first stage \citep[see, e.g.,][for brief discussion on two-stage estimation techniques]{begin2024modeling}. These two stages constitute one round of the estimation.

To ensure the convergence of the model estimation, we further repeat the two-stage procedure by updating
$\bm{\theta}$ and $\bm{\delta}$ in each subsequent round. In the initial round, we estimate the parameters of the Hermite spline and cross-basis functions, subject to the constraints that $\bm{\delta}^{\text{H}} = \bm{\delta}^{\text{C}} = \bm{1}$. Then, we estimate $\bm{\delta}^{\text{H}}$ and $\bm{\delta}^{\text{C}}$ by maximizing the Poisson log-likelihood and fixing $\bm{\theta} = \hat{\bm{\theta}}^{(1)}$. The estimated parameter set is denoted as $\hat{\bm{\delta}}^{(1)}$.
In the subsequent round $j = 2, ..., J$, by taking $\hat{\bm{\theta}}^{(j-1)}$ as the initial parameter set, we estimate again the parameter set $\bm{\theta}$ using the EM algorithm, subject to $\bm{\delta} = \hat{\bm{\delta}}^{(j-1)}$. The estimated parameter set is denoted as $\hat{\bm{\theta}}^{(j)}$. Then, by fixing $\bm{\theta} = \hat{\bm{\theta}}^{(j)}$ and taking $\hat{\bm{\delta}}^{(j-1)}$ as the initial parameter set, we re-estimate $\bm{\delta}$ by maximizing the Poisson log-likelihood.

Since only weekly death counts by age group, $D_r(\mathcal{G},t)$, are observed, whereas the age-specific death counts, $D_r(x,t)$, remain unobserved, we formulate the model estimation as an incomplete-data maximum-likelihood problem.
Specifically, $D_r(x,t)$ is treated as a latent variable to estimate mortality at individual ages. This incomplete-data structure naturally motivates the use of the EM algorithm.

The EM algorithm provides an iterative likelihood-based procedure for incomplete-data problems.\footnote{\cite{hartley1958maximum} first proposed an iterative method for maximum likelihood estimation with incomplete data, and \cite{dempster1977maximum} further generalized and formalized this method as the EM algorithm, consisting of the expectation step (E step) and the maximization step (M step).} 
In the estimation (E) step, the conditional expectation of the complete-data log-likelihood is computed given the observed data and the current parameter estimates. In the maximization (M) step, the model parameters are updated by maximizing the expected complete-data log-likelihood.

\subsection{The mathematical derivation of the estimation}
In the first stage of each round, we estimate $\bm{\theta}$ by fixing $\bm{\delta}$. To simplify the notation, we drop $\bm{\delta}$ in the following derivation. The complete-data log-likelihood is given by
\begin{align}
        \log\left(\mathcal{L}(\bm{\theta}) \right) &= \sum_{x \in \mathcal{X}}\sum_{t \in \mathcal{T}} \sum_{r \in \mathcal{R}} \big(D_r(x,t) \log(E_r(x,t)\mu_r(x,t;\bm{\theta})) \!-\! E_r(x,t)\mu_r(x,t;\bm{\theta}) \!-\!\log (D_r(x,t)!)\big) \notag \\
        &\propto \sum_{x \in \mathcal{X}}\sum_{t \in \mathcal{T}} \sum_{r \in \mathcal{R}} \big(D_r(x,t)\log \left( \mu_r(x,t;\bm{\theta}) \right) - E_r(x,t)\mu_r(x,t;\bm{\theta})\big), \label{complete-ll}
\end{align}
where the parameter set $\bm{\theta}$ is specified as in Equation~\eqref{para_theta}.
Since we compute the conditional expectation of the complete-data log-likelihood in the E step, we need to express the conditional expectation of the individual age death counts at EM iteration $n$, for $n \geq 1$. 

Based on Section~\ref{sec:assupmtions}, we assume that 
\begin{equation}
     D_r(x,t;\bm{\theta}) \sim \text{Poi}\left(\lambda_r(x,t;\bm{\theta}) \right), 
    \qquad
    D_r(\mathcal{G},t;\bm{\theta})  \sim \text{Poi}\left(\Lambda_{r}(\mathcal{G},t;\bm{\theta}) \right), \notag
\end{equation}
where $\lambda_r(\mathcal{G},t;\bm{\theta}) = E_r(x,t)\mu_r(x,t;\bm{\theta})$ and $\Lambda_r(\mathcal{G},t;\bm{\theta}) = \sum_{x \in \mathcal{G}}E_r(x,t)\mu_r(x,t;\bm{\theta})$ denote the Poisson intensity for individual age $x$ and age group $\mathcal{G}$ on week $t$ in region $r$, respectively. 

The probability mass function of individual age death count $D_r(x,t;\bm{\theta})$ and age group death count $D_r(\mathcal{G},t;\bm{\theta})$ in step $n$ are
\begin{align*}
    f\left(d_r(x,t);\bm{\theta}^{(n)} \right) = &\, \frac{\left(\lambda_r(x,t;\bm{\theta}^{(n)})\right)^{d_r(x,t)}e^{-\lambda_r(x,t;\bm{\theta}^{(n)})}}{d_r(x,t)!}, \\    
    f\left(d_r(\mathcal{G},t); \bm{\theta}^{(n)} \right) = &\, \frac{\left(\Lambda_r(\mathcal{G},t; \bm{\theta}^{(n)})\right)^{d_r(\mathcal{G},t)}e^{-\Lambda_r(\mathcal{G},t; \bm{\theta}^{(n)})}}{d_r(\mathcal{G},t)!}.
\end{align*}

For all individual ages $x \in \mathcal{G}$, the joint probability mass function can be expressed as 
\begin{equation}
        f \left(\{d_r(x,t)\}_{x \in \mathcal{G}} ; \bm{\theta}^{(n)} \right) 
        = \prod_{x \in \mathcal{G}} \frac{\left(\lambda_r(x,t;\bm{\theta}^{(n)})\right)^{d_r(x,t)}e^{-\lambda_r\left(x,t;\bm{\theta}^{(n)}\right)}}{d_r(x,t)!}. \notag
\end{equation}

The conditional joint probability mass function is given by 
\begin{align}
        & f\left( \{d_r(x,t)\}_{x \in \mathcal{G}} \mid d_r(\mathcal{G},t); \bm{\theta}^{(n)} \right) \notag \\[1ex]
        = & \frac{f \left(\{d_r(x,t)\}_{ x \in \mathcal{G}} ; \bm{\theta}^{(n)} \right)}{f\left(d_r(\mathcal{G},t) ; \bm{\theta}^{(n)} \right)} 
        = \frac{d_r(\mathcal{G},t)!}{\prod_{x \in \mathcal{G}}d_r(x,t)!}\prod_{x \in \mathcal{G}}\left(\frac{\lambda_r(x,t;\bm{\theta}^{(n)})}{\Lambda_r(\mathcal{G},t; \bm{\theta}^{(n)})}\right)^{d_r(x,t)}, \notag
\end{align}
where $\Lambda_r(\mathcal{G},t; \bm{\theta}^{(n)}) = \sum_{x \in \mathcal{G}} \lambda_r(x,t;\bm{\theta}^{(n)})$ and $d_r(\mathcal{G},t) = \sum_{x \in \mathcal{G}}d_r(x,t)$ are assumed. 

Thus, the random variable $(\{D_r(x,t)\}_{x \in \mathcal{G}} \mid D_r(\mathcal{G},t) = d_r(\mathcal{G},t); \bm{\theta}^{(n)})$ follows multinomial distribution such that
\begin{equation}
    \left(\{D_r(x,t)\}_{x \in \mathcal{G}} \mid D_r(\mathcal{G},t) = d_r(\mathcal{G},t); \bm{\theta}^{(n)} \right) \sim \text{Multinomial}\left(d_r(\mathcal{G},t), \left\{p_r\left(x,t;\bm{\theta}^{(n)}\right)\right\}_{x \in \mathcal{G}}\right). \notag
\end{equation}
where $p_r\left(x,t;\bm{\theta}^{(n)}\right) =  \frac{\lambda_r(x,t;\bm{\theta}^{(n)})}{\Lambda_r(\mathcal{G},t; \bm{\theta}^{(n)})}$. The conditional expectation can be expressed as 
\begin{equation}
    \mathbb{E}\left[D_r(x,t) ; D_r(\mathcal{G},t), \bm{\theta}^{(n)} \right] 
    = D_r(\mathcal{G},t) \frac{\lambda_r(x,t;\bm{\theta}^{(n)})}{\Lambda_r(\mathcal{G},t; \bm{\theta}^{(n)})} 
    = D_r(\mathcal{G},t) \frac{E_r(x,t)\mu_r\left(x,t;\bm{\theta}^{(n)}\right)}{\sum_{y \in \mathcal{G}}E_r(y,t)\mu_{r}\left(y,t; \bm{\theta}^{(n)}\right)}, \notag
\end{equation}
and the conditional expectation of the complete-data log-likelihood in the E step is given by
\begin{align*}
        & \, \mathbb{E}\left[\log \left(\mathcal{L}\left(\bm{\theta}\right)\right)\mid D_r(\mathcal{G},t); \bm{\theta}^{(n)} \right] \notag
        \\[1ex]
        \propto & \sum_{\mathcal{G} \in \mathcal{A}}\sum_{x \in \mathcal{G}}\sum_{t \in \mathcal{T}} \sum_{r \in \mathcal{R}} \left(\mathbb{E}\left[D_r(x,t) \mid D_r(\mathcal{G},t); \bm{\theta}^{(n)} \right] \log \left( \mu_r(x,t;\bm{\theta})\right) - E_r(x,t) \mu_r(x,t;\bm{\theta})\right). \notag
\end{align*}

In the M step, we update the parameter vector from $\hat{\bm{\theta}}^{(n)}$ to $\hat{\bm{\theta}}^{(n+1)}$ by numerically solving
\begin{equation}
        \hat{\bm{\theta}}^{(n+1)} = \underset{\bm{\theta}}{\arg\!\max} \ 
        \mathbb{E}\left[\log \left( \mathcal{L}\left(\bm{\theta}\right)\right) \mid D_r(\mathcal{G},t); \bm{\theta}^{(n)} \right]. \notag
\end{equation}

The E and M steps are repeated until $\left\|\bm{\theta}^{(n+1)}-\bm{\theta}^{(n)}\right\|_{\infty}<\varepsilon_{\text{EM}}$, for a pre-specified tolerance $\varepsilon_{\text{EM}}$, set to $10^{-2}$ in this article.

Once the parameter set $\hat{\bm{\theta}}^{(n)}$ is obtained from the EM algorithm, we estimate the loadings in the second stage while keeping $\hat{\bm{\theta}}^{(n)}$ fixed. In this stage, the fitted Hermite splines $\hat{\mathcal{H}}_r(x,t)$, and heat and cold effects, $\bar{\bm{C}}_{r}^{\text{H}}(t)\hat{\bm{\eta}}^{\text{H}}$ and $\bar{\bm{C}}_{r}^{\text{C}}(t)\hat{\bm{\eta}}^{\text{C}}$, are treated as fixed
offset terms. Then, the loadings $\bm{\delta}^{(n)}$ are estimated by maximizing the log-likelihood function in Equation \eqref{complete-ll}.

Following the initial round $j = 1$, we repeat the two-stage estimation procedure to update $\bm{\theta}$ and $\bm{\delta}$ in subsequent rounds $j = 2, ... , J$. We use the following stopping criterion:
\begin{equation}
        \Delta_{\mathcal{L}}^{(j)} = \left|  \frac{ \log \left( \mathcal{L}\left(\hat{\bm{\theta}}^{(j)}, \hat{\bm{\delta}}^{(j)} \right) \right)- \log \left( \mathcal{L}\left(\hat{\bm{\theta}}^{(j-1)}, \hat{\bm{\delta}}^{(j-1)} \right) \right)}{\log \left( \mathcal{L}\left(\hat{\bm{\theta}}^{(j-1)}, \hat{\bm{\delta}}^{(j-1)} \right) \right)} \right| < \varepsilon_\mathcal{L},
\end{equation}
where $\log(\mathcal{L}(\hat{\bm{\theta}}^{(j)}, \hat{\bm{\delta}}^{(j)} ))$ denotes the observed-data Poisson log-likelihood evaluated based on $\hat{\bm{\theta}}^{(j)}$ and $\hat{\bm{\delta}}^{(j)}$ in the $j$th round, and $\varepsilon_\mathcal{L}$ is pre-specified tolerance level, set to $10^{-2}$ in this article.
\section{Model selection}
\label{sec:model_selection}
\begin{figure}[ht!]
    \centering
    \includegraphics[width=0.9\linewidth]{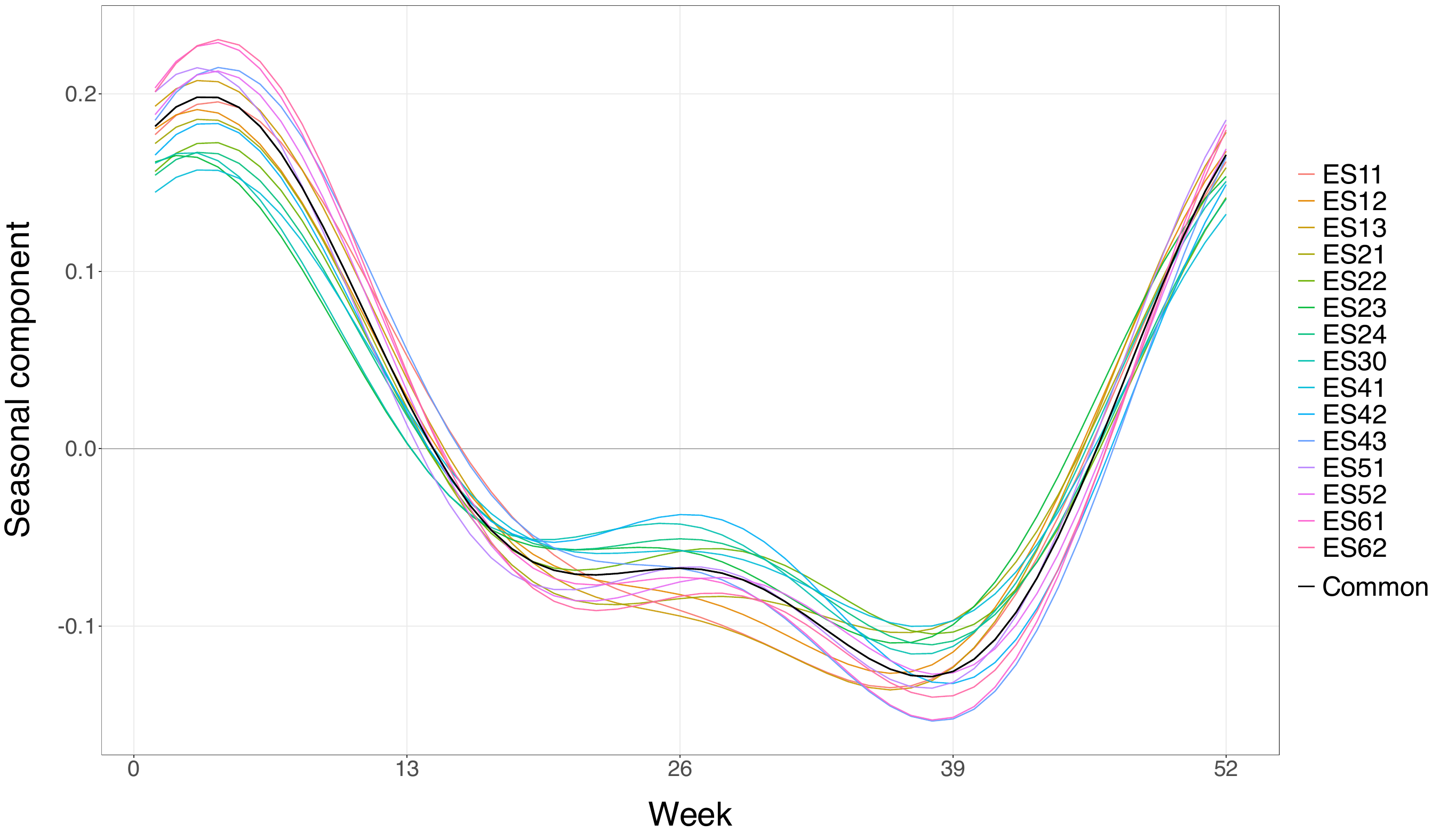}
    \caption{Estimated common seasonal pattern in region-specific Hermite baseline model. \newline \footnotesize
    \emph{Notes}: This figure shows the common and region-specific seasonal pattern in Hermite baseline model. The black line shows the common seasonal pattern, while the coloured lines show the region-specific seasonal patterns for the 15 NUTS-2 regions.}
    \label{fig:common_seasonality_selection}
\end{figure}

We first fit a region-specific Hermite baseline model for each NUTS-2 region. The model is expressed as follows:
\begin{align}
\log\left(\mu_r(x,t)\right) = &\, \underbrace{\alpha_r h_{00}(u) + m_{0,r} h_{10}(u) +\gamma_r h_{01}(u) + m_{1,r} h_{11}(u)}_{\text{Region-specific Hermite spline term}} + \underbrace{(\phi_r h_{00}(u) + \kappa_r h_{01}(u)) \frac{t}{52}}_{\text{Region-specific trend}} \nonumber \\
&+ \underbrace{\sum_{j=1}^2 \rho_{1,j,r}\sin\left(\frac{2\pi j t}{52}\right) + \sum_{k=1}^2 \rho_{2,k,r}\cos\left(\frac{2\pi k t}{52}\right)}_{\text{Region-specific seasonal pattern}},
\end{align}
where all parameters are region-specific and depend on $r \in \mathcal{R}$. This baseline specification is used to select the optimal Hermite spline component before incorporating the cross-basis components. For further interpretation and model selection, the model can be decomposed into three components: Hermite spline terms, trend terms, and seasonal terms. Figure \ref{fig:common_seasonality_selection} shows that the seasonal patterns are very similar across all regions. This motivates us to consider the common seasonality pattern when we construct the temperature-augmented Hermite mortality model. 

After fixing the seasonal terms as common across regions, we then examine whether the Hermite spline and trend components should be specified as region-specific or common. Now we construct the Hermite baseline model by pooling data from all regions; we denote four potential models $\mathcal{M}_0$, $\mathcal{M}_1$, $\mathcal{M}_2$, and $\mathcal{M}_3$ as follows:
\begin{itemize}
    \item [$\mathcal{M}_0$:] Common Hermite spline term with common trend and  common seasonal pattern, such that
    \begin{align}
        \log\left(\mu_r(x,t)\right) =&\, \underbrace{\alpha h_{00}(u) + m_{0} h_{10}(u) +\gamma h_{01}(u) + m_{1} h_{11}(u)}_{\text{Common Hermite spline term}} + \underbrace{(\phi h_{00}(u) + \kappa h_{01}(u)) \frac{t}{52}}_{\text{Common trend}} \nonumber \\ 
        &+ \underbrace{\sum_{j=1}^2 \rho_{1,j}\sin\left(\frac{2\pi j t}{52}\right) + \sum_{k=1}^2 \rho_{2,k}\cos\left(\frac{2\pi k t}{52}\right)}_{\text{Common seasonal pattern}}.
    \label{baseline_m0}
    \end{align}
    \item [$\mathcal{M}_1$:]  Region-specific Hermite spline with common trend and common seasonal pattern, such that
    \begin{align}
        \log\left(\mu_r(x,t)\right)= & \, \underbrace{\alpha_r h_{00}(u) + m_{0,r} h_{10}(u) +\gamma_r h_{01}(u) + m_{1,r} h_{11}(u)}_{\text{Region-specific Hermite spline term}} + \underbrace{(\phi h_{00}(u) + \kappa h_{01}(u)) \frac{t}{52}}_{\text{Common trend}} \nonumber \\ 
        &+ \underbrace{\sum_{j=1}^2 \rho_{1,j}\sin\left(\frac{2\pi j t}{52}\right) + \sum_{k=1}^2 \rho_{2,k}\cos\left(\frac{2\pi k t}{52}\right)}_{\text{Common seasonal pattern}}.
    \label{baseline_m1}
    \end{align}
    \item [$\mathcal{M}_2$:] Common Hermite spline  with region-specific trend and common seasonal pattern, such that
    \begin{align}
        \log\left(\mu_r(x,t)\right)= & \underbrace{\alpha h_{00}(u) + m_{0} h_{10}(u) +\gamma h_{01}(u) + m_{1} h_{11}(u)}_{\text{Common Hermite spline term}} + \underbrace{(\phi_r h_{00}(u) + \kappa_r h_{01}(u)) \frac{t}{52}}_{\text{Region-specific trend}} \nonumber \\ 
        &+ \underbrace{\sum_{j=1}^2 \rho_{1,j}\sin\left(\frac{2\pi j t}{52}\right) + \sum_{k=1}^2 \rho_{2,k}\cos\left(\frac{2\pi k t}{52}\right)}_{\text{Common seasonal pattern}}.
    \label{baseline_m2}
    \end{align}
    \item [$\mathcal{M}_3$:] Region-specific Hermite spline with region-specific trend and common seasonal pattern, such that
    \begin{align}
        \log\left(\mu_r(x,t)\right) =&\, \underbrace{\alpha_r h_{00}(u) + m_{0,r} h_{10}(u) +\gamma_r h_{01}(u) + m_{1,r} h_{11}(u)}_{\text{Region-specific Hermite spline term}}  \nonumber \\ 
        &+ \underbrace{(\phi_r h_{00}(u) + \kappa_r h_{01}(u)) \frac{t}{52}}_{\text{Region-specific trend}} \nonumber \\ 
        &+ \underbrace{\sum_{j=1}^2 \rho_{1,j}\sin\left(\frac{2\pi j t}{52}\right) + \sum_{k=1}^2 \rho_{2,k}\cos\left(\frac{2\pi k t}{52}\right)}_{\text{Common seasonal pattern}}.
    \label{baseline_m3}
    \end{align}
\end{itemize}

We compare the four candidate models using AIC and BIC, as reported in Table \ref{tab:baseline_model_comparison}. Since $\mathcal{M}_3$ has the lowest AIC and BIC, we select it as the best model and use it as the Hermite baseline component of our proposed modeling framework. This model allows both the Hermite spline and trend components to vary by region.

\begin{table}[H]
\centering
\resizebox{\linewidth}{!}{
\begin{tabular}{@{}ccccccccc@{}}
\toprule
\multirow{2.5}{*}{Model} & \multicolumn{2}{c}{Hermite spline} & \multicolumn{2}{c}{Trend} & \multicolumn{2}{c}{Seasonal pattern} & \multirow{2.5}{*}{AIC} & \multirow{2.5}{*}{BIC}\\
\cmidrule(lr){2-3}\cmidrule(lr){4-5}\cmidrule(lr){6-7}
& Common & Region-specific & Common & Region-specific & Common & Region-specific & &\\
\cmidrule(lr){2-2} \cmidrule(lr){3-3} \cmidrule(lr){4-4} \cmidrule(lr){5-5} \cmidrule(lr){6-6} \cmidrule(lr){7-7} \cmidrule(lr){8-8} \cmidrule(lr){9-9} 
$\mathcal{M}_0$ & \checkmark &  & \checkmark &  & \checkmark &  & 610,415.4 & 610,508.1\\
$\mathcal{M}_1$ &  & \checkmark & \checkmark &  & \checkmark &  & 562,258.8 & 562,870.4\\
$\mathcal{M}_2$ & \checkmark &  &  & \checkmark & \checkmark &  & 575,536.0 & 575,888.2\\
$\mathcal{M}_3$ &  & \checkmark &  & \checkmark & \checkmark &  & \textbf{561,403.9} & \textbf{562,275.0}\\
\bottomrule
\end{tabular}
}
\caption{\textbf{Hermite baseline model comparison via AIC and BIC.} \newline \footnotesize
\emph{Notes}: This table shows the model selection results based on AIC and BIC. A checkmark indicates the specification used for each model component. ``Common'' denotes a component shared across regions, whereas ``Region-specific'' denotes a component allowed to vary by region. Bold values indicate the lowest AIC and BIC. }
\label{tab:baseline_model_comparison}
\end{table}

\section{Bias-corrected temperature projections}\label{sec:bias_correction}
This appendix describes the procedure used to generate stochastic temperature projections.
For region $r$, we first apply a quantile mapping approach  \citep{piani2010statistical,qian2021projecting} to estimate bias-correction coefficients as
\begin{equation}
    \label{bias-corrected}
    {y}_{r,q} = {\nu}_{0,m,r} + {\nu}_{1,m,r} \, \hat{{y}}_{m,r,q},
\end{equation}
where ${y}_{r,q}$ and $\hat{y}_{m,r,q}$ denote the $q$th quantiles of the observed historical temperature data and the backcasts from CMIP6 climate model $m$, respectively, over the reference period 2000--2014. The estimated coefficients $\hat{\nu}_{0,m,r}$ and $\hat{\nu}_{1,m,r}$ are then used in bias-corrected temperature projections. 

Let $y_{r}(t)$ be historical observed temperature and $\hat{y}_{m,r}^{*}(t)$ be the bias-corrected model backcasts, computed by $\hat{y}_{m,r}^{*}(t)=\hat{\nu}_{0,m,r} + \hat{\nu}_{1,m,r} \hat{{y}}_{m,r}(t)$, for $t\in \tau_{\text{Historical}}=\{2000, ..., 2014\}$. We further compute residuals $z_{m,r}(t) = y_{r}(t) - \hat{y}_{m,r}^{*}(t)$ and assume that they follow a normal distribution with zero mean and standard deviation $\sigma_{m,r}$ such that
\begin{equation}
    \label{residual}
    z_{m,r}(t) \sim \mathcal{N}\left(0,\sigma_{m,r}^{2} \right).
\end{equation}

We thus simulate the bias-corrected future temperature projections for 2025--2100 as follows:
\begin{equation}
    \label{temperature-simulation}
    \tilde{y}_{m,r}(t) = \hat{y}_{m,r}^{*}(t) + \tilde z_{m,r}(t),
\end{equation}
where $\tilde{y}_{m,r}(t)$ denotes a simulated bias-corrected temperature trajectory at future time $t$ for region $r$ under climate model $m$, and $\hat{y}_{m,r}^{*}(t)$ is the corresponding bias-corrected model forecasts, for $t\in \tau_{\text{Future}}=\{2025, ..., 2100\}$. The random variable $\tilde z_{m,r}(t)$ denotes a simulated value from the residual distribution of climate model $m$.

\clearpage
\section*{Supplementary material}
\setcounter{section}{0}
\renewcommand{\thesection}{\Alph{section}}
\renewcommand{\theHsection}{supplement.\Alph{section}}

\section{In-sample fit of mortality rates}
\label{sec:fit}
\begin{figure}[H]
    \centering
    \includegraphics[page=1, width=0.65\linewidth]{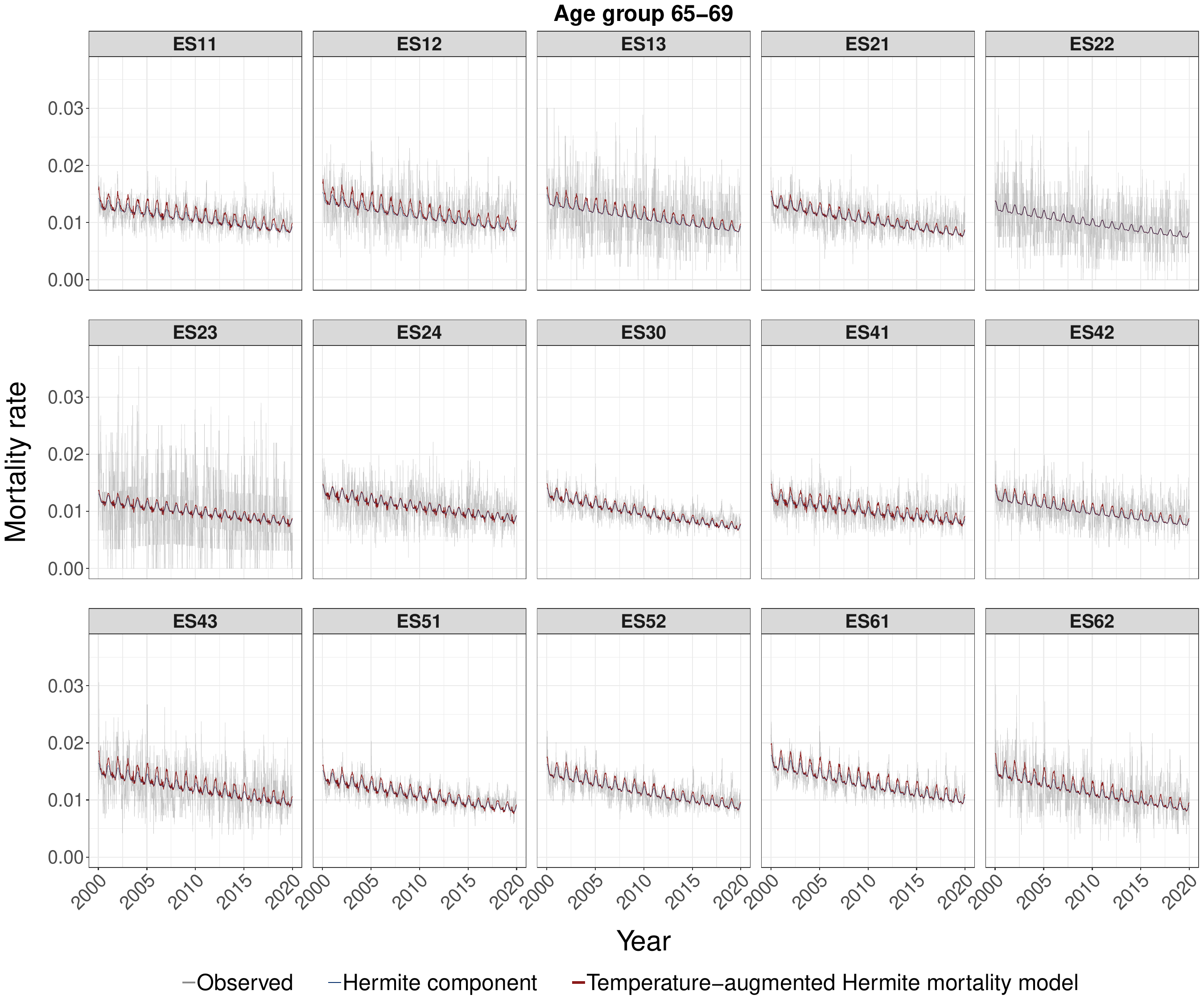}
    \vspace{-0.1in}
    \caption{\textbf{Fitted mortality rates over the period 2000--2019 for the ages 65--69} \newline \footnotesize
    \emph{Notes}: This figure shows the in-sample fit of weekly mortality rates for age 65--69 across the selected regions.}
    \label{fig:fitted_mortality_65}
\end{figure}

\begin{figure}[H]
    \centering
    \includegraphics[page=2, width=0.65\linewidth]{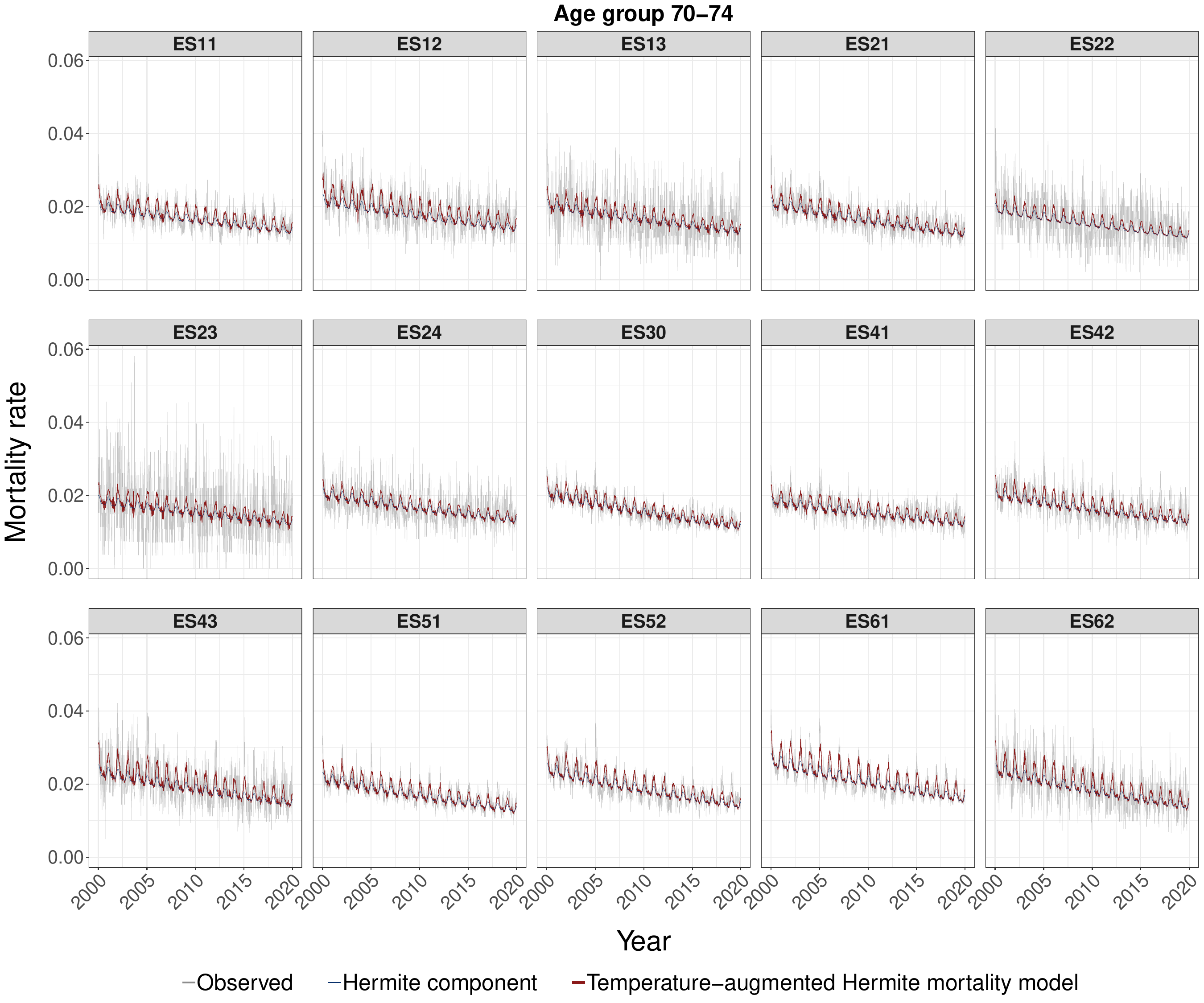}
    \vspace{-0.1in}
    \caption{\textbf{Fitted mortality rates over the period 2000--2019 for the ages 70--74} \newline \footnotesize
    \emph{Notes}: This figure shows the in-sample fit of weekly mortality rates for age 70--74 across the selected regions.}
    \label{fig:fitted_mortality_70}
\end{figure}

\begin{figure}[H]
    \centering
    \includegraphics[page=3, width=0.65\linewidth]{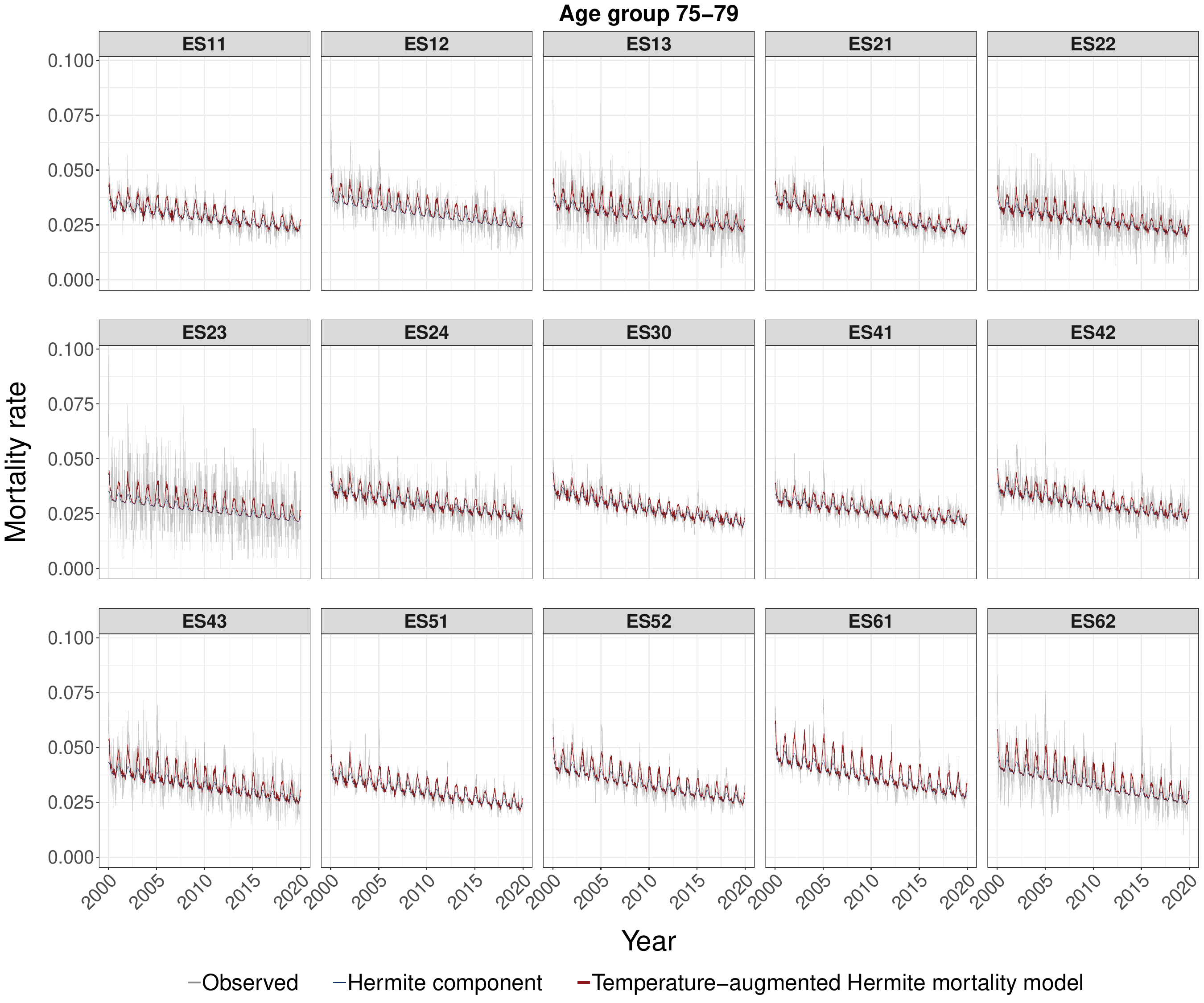}
    \caption{\textbf{Fitted mortality rates over the period 2000--2019 for the ages 75--79} \newline \footnotesize
    \emph{Notes}: This figure shows the in-sample fit of weekly mortality rates for age 75--79 across the selected regions.} 
    \label{fig:fitted_mortality_75}
\end{figure}

\begin{figure}[H]
    \centering
    \includegraphics[page=4, width=0.65\linewidth]{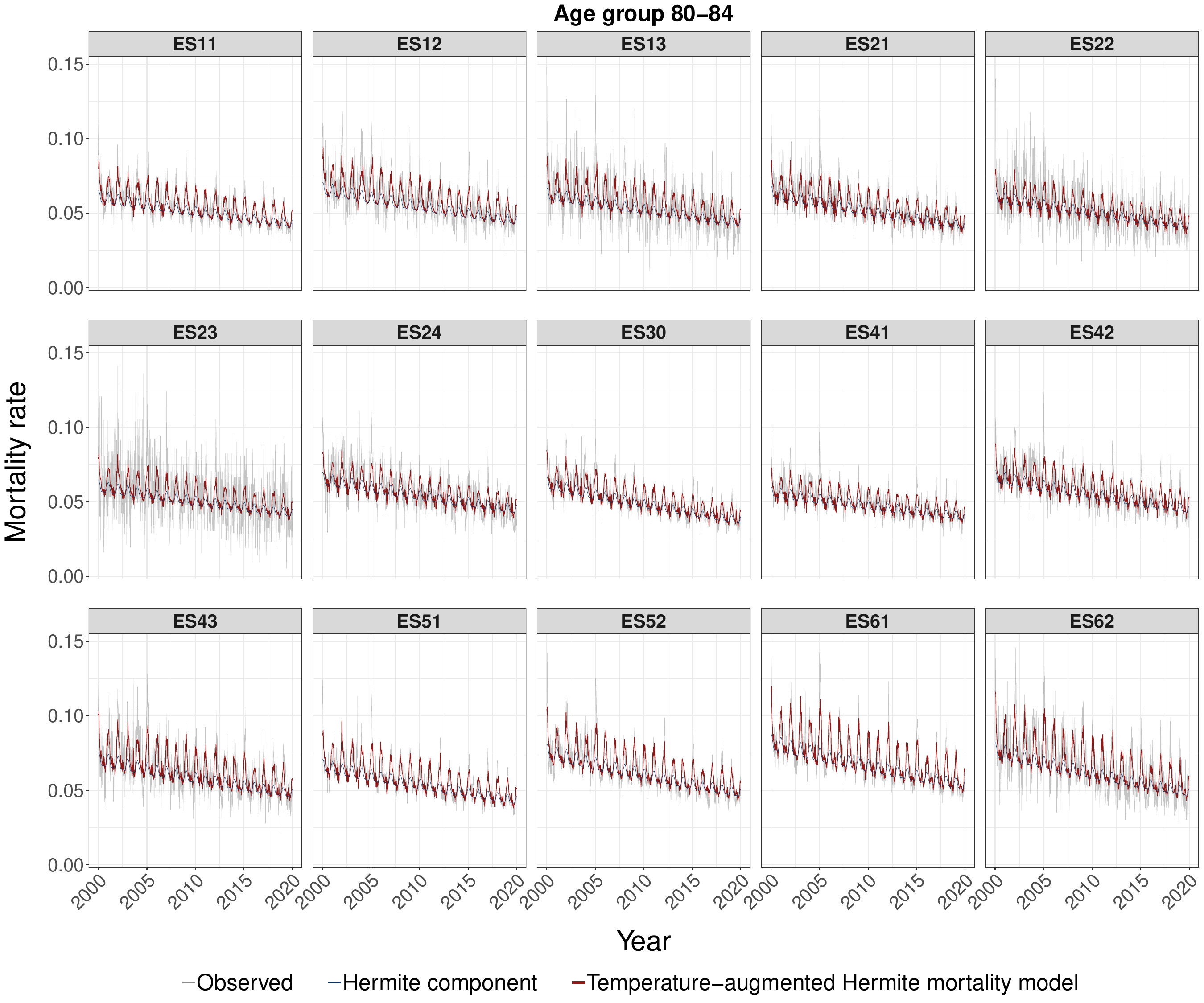}
    \caption{\textbf{Fitted mortality rates over the period 2000--2019 for the ages 80--84} \newline \footnotesize
    \emph{Notes}: This figure shows the in-sample fit of weekly mortality rates for age 80--84 across the selected egions.} 
    \label{fig:fitted_mortality_80}
\end{figure}

To supplement Section~\ref{sec:in_sample_fit}, Figures~\ref{fig:fitted_mortality_65} to \ref{fig:fitted_mortality_80} compare the observed and fitted mortality rates for the age groups 65--69, 70--74, 75--79, and 80--84, respectively.

\vspace{-0.25in}

\section{Mortality projection}\label{sec:projection}

\begin{figure}[H]
    \centering
    \includegraphics[page=1, width=0.85\linewidth]{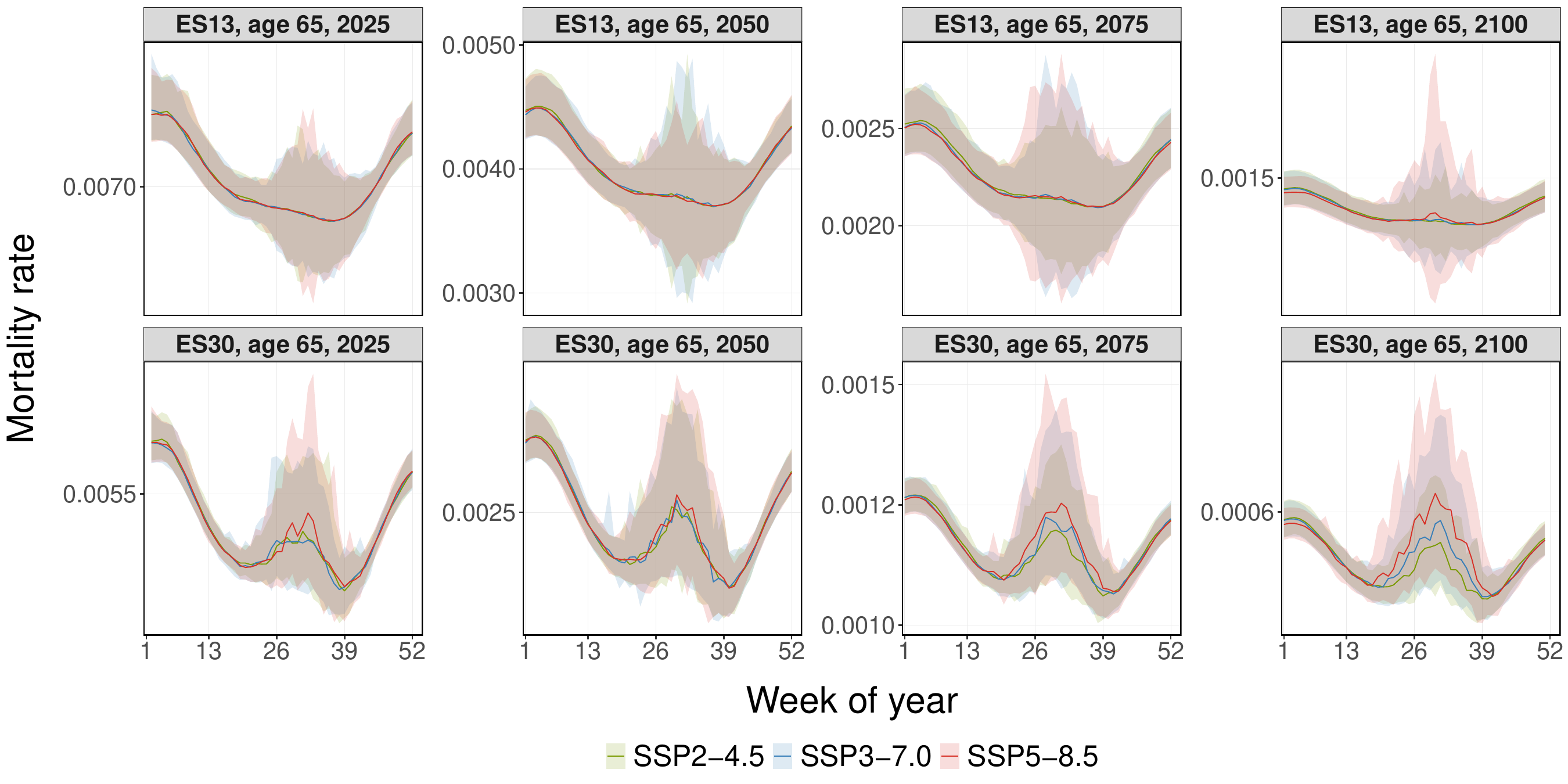}
    \caption{\textbf{Mortality projections for age 65 in ES13 and ES30} \newline \footnotesize
    \emph{Notes}: This figure shows the projected weekly mortality hazards in ES13 and ES30 for the age group 65 under SSP2–4.5, SSP3–7.0, and SSP5–8.5. Each panel presents the seasonal pattern of projected mortality hazards in years 2025, 2050, 2075, and 2100.} 
    \label{fig:mu_projection_65}
\end{figure}

\begin{figure}[H]
    \centering
    \includegraphics[page=2, width=0.85\linewidth]{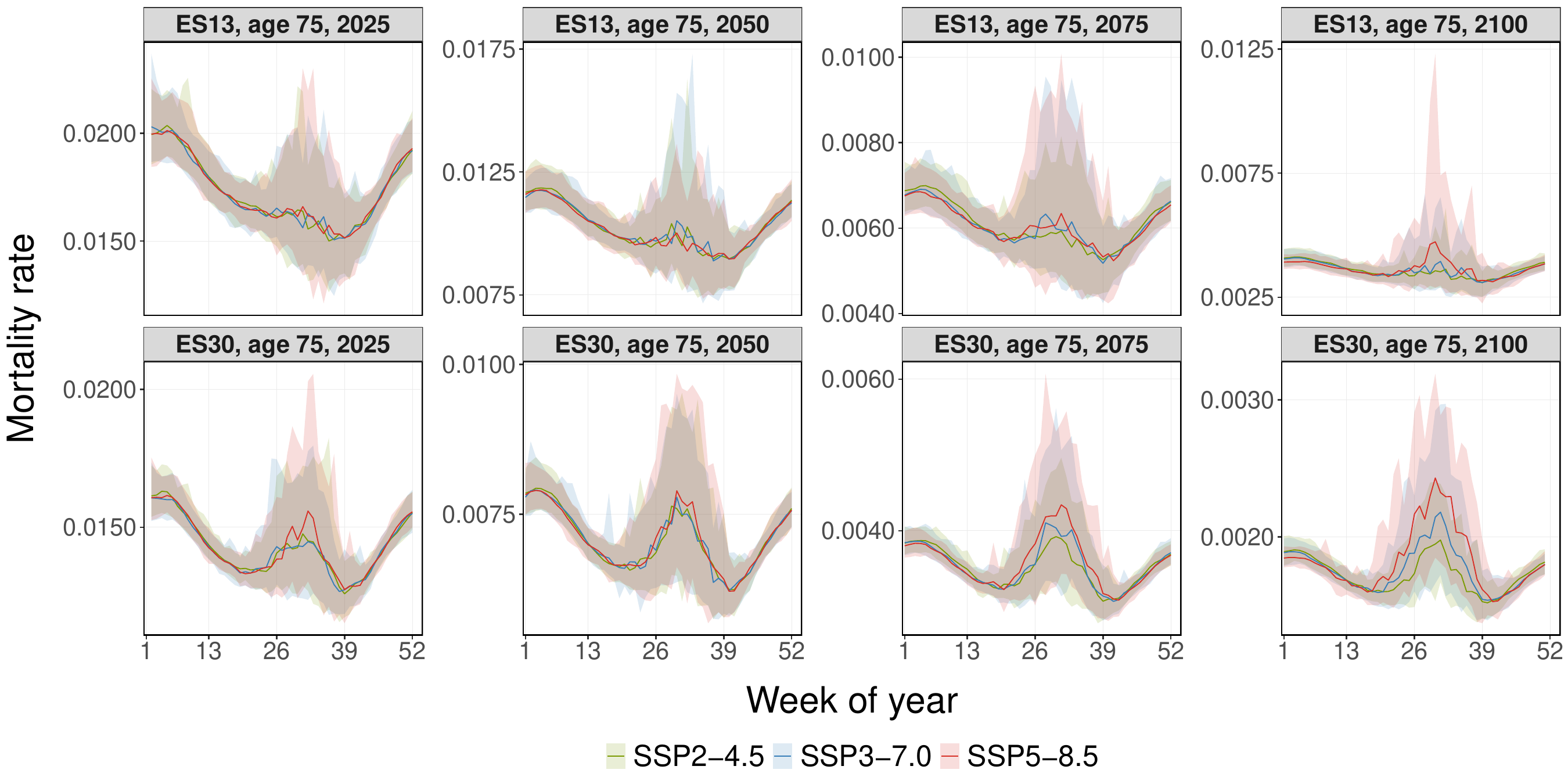}
    \caption{\textbf{Mortality projections for age 75 in ES13 and ES30} \newline \footnotesize
    \emph{Notes}: This figure shows the projected weekly mortality hazards in ES13 and ES30 for the age group 75 under SSP2–4.5, SSP3–7.0, and SSP5–8.5. Each panel presents the seasonal pattern of projected mortality hazards in years 2025, 2050, 2075, and 2100.} 
    \label{fig:mu_projection_75}
\end{figure}

To supplement the results presented in Section~\ref{sec:mu_projection}, Figure~\ref{fig:mu_projection_65} and \ref{fig:mu_projection_75} illustrate the projected weekly mortality rates at age 65 and 75, respectively, for ES13 (Cantabria) and ES30 (Madrid) in 2025, 2050, 2075, and 2100 under the three SSP scenarios.
\vspace{-1.5em}

\section{List of CMIP6 climate models}\label{sec:cmip6_model_list}

In this section, we present the 15 CMIP6 climate models used in our case study, as listed in Table~\ref{tab:cmip6_models} and obtain the corresponding simulated temperature projections from the \href{https://cds.climate.copernicus.eu/datasets/projections-cmip6}{Copernicus Climate Data Store}. Each CMIP6 climate model provides projections for the period 2025--2100 under SSP1-2.6, SSP2-4.5, and SSP5-8.5, using the ensemble member specified in the table.

\begin{table}[ht!]
\caption{\textbf{List of CMIP6 climate models used in mortality projection}}
{\fontsize{10}{10}\selectfont
\centering
\renewcommand{\arraystretch}{1.2}
\begin{tabularx}{\linewidth}{l X l}
\toprule
\textbf{Climate model} & \textbf{Country (modeling centre)} & \textbf{Ensemble member} \\
\midrule
ACCESS-CM2 & Australia (CSIRO-ARCCSS) & \texttt{r1i1p1f1} \\
AWI-CM-1-1-MR & Germany (AWI) & \texttt{r1i1p1f1} \\
BCC-CSM2-MR & China (BCC) & \texttt{r1i1p1f1} \\
CESM2 & USA (NCAR) & \texttt{r4i1p1f1} \\
CNRM-CM6-1 & France (CNRM-CERFACS) & \texttt{r1i1p1f2} \\
CNRM-ESM2-1 & France (CNRM-CERFACS) & \texttt{r1i1p1f2} \\
GFDL-ESM4 & USA (NOAA-GFDL) & \texttt{r1i1p1f1} \\
INM-CM4-8 & Russia (INM) & \texttt{r1i1p1f1} \\
INM-CM5-0 & Russia (INM) & \texttt{r1i1p1f1} \\
IPSL-CM6A-LR & France (IPSL) & \texttt{r1i1p1f1} \\
MIROC6 & Japan (MIROC) & \texttt{r1i1p1f1} \\
MIROC-ES2L & Japan (MIROC) & \texttt{r1i1p1f2} \\
MPI-ESM1-2-LR & Germany (MPI-M) & \texttt{r1i1p1f1} \\
MRI-ESM2-0 & Japan (MRI) & \texttt{r1i1p1f1} \\
NorESM2-MM & Norway (NCC) & \texttt{r1i1p1f1} \\
\bottomrule
\end{tabularx}%
}
\par\smallskip\footnotesize
\emph{Notes}: This table lists the modeling centre and selected ensemble member for each climate model.
\label{tab:cmip6_models}
\end{table}

\end{document}